\RequirePackage{fix-cm}
\documentclass[natbib,smallextended]{svjour3}       
\smartqed  
\usepackage{graphicx}
\usepackage{amsmath}
\usepackage{amssymb}
\usepackage{color}
\usepackage[citecolor=blue,urlcolor=blue,breaklinks]{hyperref}
\newcommand{\be}{\begin{equation}}
\newcommand{\ee}{\end{equation}}
\journalname{Living Rev. Relativ.}
\newcommand       \bea          {\begin{eqnarray}}
\newcommand       \eea          {\end{eqnarray}}

\begin{document}

\title{Kilonovae}

\author{Brian D.~Metzger}

\institute{B.~D.\ Metzger%
\at
Department of Physics\\
Columbia Astrophysics Laboratory\\
Columbia University\\
\email{bmetzger@phys.columbia.edu}
}

\date{Received: date / Accepted: date}

\maketitle

\begin{abstract}
The mergers of double neutron star (NS-NS) and black hole (BH)-NS binaries are promising gravitational wave (GW) sources for Advanced LIGO and future GW detectors.  The neutron-rich ejecta from such merger events undergoes rapid neutron capture ($r$-process) nucleosynthesis, enriching our Galaxy with rare heavy elements like gold and platinum.  The radioactive decay of these unstable nuclei also powers a rapidly evolving, supernova-like transient known as a ``kilonova'' (also known as ``macronova").  Kilonovae are an approximately isotropic electromagnetic counterpart to the GW signal, which also provides a unique and direct probe of an important, if not dominant, $r$-process site.  I review the history and physics of kilonovae, leading to the current paradigm of week-long emission with a spectral peak at near-infrared wavelengths.  Using a simple light curve model to illustrate the basic physics, I introduce potentially important variations on this canonical picture, including: $\sim$day-long optical (``blue'') emission from lanthanide-free components of the ejecta; $\sim$hour-long precursor UV/blue emission, powered by the decay of free neutrons in the outermost ejecta layers; and enhanced emission due to energy input from a long-lived central engine, such as an accreting BH or millisecond magnetar.  I assess the prospects of kilonova detection following future GW detections of NS-NS/BH-NS mergers in light of the recent follow-up campaign of the LIGO binary BH-BH mergers.
\end{abstract}

\newpage

\setcounter{tocdepth}{3}
\tableofcontents


\section{Introduction}

The discovery of gravitational waves (GW) from the inspiral and coalescence of binary black holes (BH) by the Laser Interferometer Gravitational Wave Observatory (LIGO) has opened a fresh window on the cosmos \citep{LIGO+16}.  Even the limited sample of BH-BH mergers discovered thus far is already placing stringent constraints on the formation channels of compact object binaries \citep{LIGO+16}, as well as more fundamental predictions of general relativity in the strong field regime \citep{Miller16}.  We are fortunate witnesses to the birth of a new field of research: GW astronomy.  

Beyond information encoded in the GW strain data alone, the discovery of an electromagnetic (EM) counterpart in coincidence with the GW chirp could reveal a much richer picture of these events \citep{Bloom+09GW}.  By identifying the host galaxies of the merging systems, and their locations within or around their hosts, we would obtain valuable information on the binary formation channels, age of the stellar population, evidence for dynamical formation channels in dense stellar systems, or displacement due to supernova [SN] birth kicks), similar to as has been done in the past for gamma-ray bursts (GRBs) and SNe \citep{Fruchter+06, Fong&Berger13}.    By measuring the redshifts of their host galaxies, we could determine the distances to the GW sources, thus reducing degeneracies in the GW parameter estimation, especially of the binary inclination with respect to the line of sight.  Redshift measurements might also enable the use of a large sample of GW events as standard rulers to probe the cosmic expansion history \citep{Holz&Hughes05, Nissanke+13}.  

Except perhaps in rare circumstances, the merger of stellar mass BH-BH binaries are not expected to produce luminous EM emission due to the absence of baryonic matter in these systems.  Thus, despite the large sample of BH-BH mergers which we expect to accumulate over the next few years, a full synthesis of the GW and EM skies will probably require the discovery of GWs from merging binaries containing neutron stars (NS), of either the NS-NS or BH-NS varieties.  Population synthesis models of field binaries predict GW detection rates of NS-NS/BH-NS mergers of $\sim 0.2\mbox{\,--\,}300$ per year, once Advanced LIGO/Virgo reach their full design sensitivities near the end of this decade \citep{Abadie+10,Dominik+14}.  Empirical rates based on observed binary pulsar systems in our galaxy predict a comparable range, with a best bet rate of $\approx 8$ NS-NS mergers per year \citep{Kalogera+04,Kim+15}.

Among the greatest challenges of GW astronomy are the large uncertainties in the measured sky positions of the GW sources, which are primarily determined by triangulating the GW arrival times with an array of detectors.  With just the two North American LIGO detectors now operational, current sky error regions are very large (initially $\approx 850\mathrm{\ deg}^{2}$ for GW150914, though later improved to $\approx 250 \mathrm{\ deg}^{2}$; \citealt{Abbott+16, Abbott+16PRX}).  Once Virgo in Italy, and eventually KAGRA \citep{KAGRA} in Japan and LIGO-India join the network, these will be reduced to more manageable values of 10\,--\,100 square degrees or less \citep{Fairhurst11,Nissanke+13,Rodriguez+14}.  However, even in the best cases, these sky areas still greatly exceed the fields of view of most radio, optical, and X-ray telescopes, especially those with the required sensitivity to detect the potentially dim EM counterparts of NS-NS and BH-NS mergers \citep{Metzger&Berger12}.  

Several lines of evidence, both observational \citep{Fong+13} and theoretical\footnote{One of the strongest theoretical arguments linking short GRBs and NS-NS/BH-NS mergers is the lack of viable alternative models.  The accretion-induced collapse (AIC) of a NS to a BH was once considered a promising model \citep{MacFadyen+05,Dermer&Atoyan06}.  However,  \cite{Margalit+15} show that the collapse of a NS rotating as a solid body (as would be expected in such evolved systems) is unlikely to produce an accretion disk around the newly-formed BH for the range of nuclear density equations of state consistent with observational constraints on the maximum NS mass \cite[see also][]{Shibata03}.} \citep{Eichler+89,Narayan+92}, support an association between NS-NS or BH-NS mergers and the ``short duration'' class of GRBs (those bursts with durations in the gamma-ray band less than about 2 seconds; \citealp{Nakar07,Berger14}).  At typical LIGO source distances of hundreds of Mpc, a GRB should be easily bright enough to be detected by the \textit{Fermi} and \textit{Swift} satellites, or even with the less sensitive gamma-ray satellites which comprise the Interplanetary Network \citep{Hurley10}.  

Short GRBs are commonly believed to be powered by the accretion of a massive remnant disk onto the compact BH or NS remnant following the merger.  This is typically expected to occur within seconds of the GW chirp, making their temporal association with the GWs unambiguous (the gamma-ray sky is otherwise quiet).  Once a GRB is detected, its associated afterglow can be identified by promptly slewing a sensitive X-ray telescope to the location of the burst.  This exercise is now routine with \textit{Swift}, but may become less so in the next decade without a suitable replacement mission.  Although gamma-ray detectors themselves typically provide poor localizations, the higher angular resolution of the X-ray telescope allows for the discovery of the optical or radio afterglow; this in turn provides an even more precise position, which can allow the host galaxy to be identified.

Although short GRBs are arguably the cleanest EM counterparts, their measured rate within the Advanced LIGO detection volume\footnote{Some subtlety is required here.  Mergers for which the binary plane is viewed face-on (the configuration required to produce a GRB jet pointed towards Earth) are moderately brighter in GWs, and hence detectable to larger distances, than those of typical inclination \citep{Schutz11}.  The time coincidence between the GRB and the NS merger chirp would also increase the effective detection volume modestly by reducing the number of search templates, thus increasing the significance of the signal \citep{Kochanek&Piran93}.  The net effect of these is a factor of $\sim 2$ times larger detection distance for face-on sources, increasing the detection volume by roughly an order of magnitude.} is low, probably less than once per year all-sky if they result from NS-NS mergers \citep{Metzger&Berger12}.  This can be reconciled with the much higher predicted GW event rate cited above \citep{Abadie+10} if the gamma-ray emission is beamed into a narrow solid angle by the bulk relativistic motion of the GRB jet \citep{Fong+15,Troja+16}.  Every effort should be made to guarantee the presence of an all-sky gamma-ray monitor in space throughout the next decade.  However, we should not expect the first---or even the first several dozen---GW chirps from NS-NS/BH-NS mergers to be accompanied by a GRB.  

For the majority of GW-detected mergers, the jetted GRB emission will be relativistically beamed out of our line of sight.  However, as the jet material slows down by shocking the interstellar medium, even off-axis viewers eventually enter the causal emission region of the synchrotron afterglow \citep{Totani&Panaitescu02}.  At X-ray wavelengths, such `orphan afterglow' emission evolves rapidly and only reaches detectable luminosities for viewing angles close to the jet axis.  At optical frequencies, the orphan afterglow is bright enough to be detected within about twice the jet opening angle \citep[][their Figs.~3\,--\,5]{Metzger&Berger12}.  Thus, at least for standard jet structures,\footnote{An exception may occur if the GRB jet is `structured' in its geometry.  Even if gamma-ray emission is confined to the central regions of the jet, larger angles may still contain less- (though still ultra-)relativistic ejecta, producing a more luminous optical synchrotron afterglow than predicted by standard `top hat' jet models \citep{Perna+03,Lamb&Kobayashi16}.} the off-axis afterglow probably does not provide a promising counterpart for most observers.  More isotropic emission could originate from the mildly relativistic `cocoon' of shocked jet material.  However, the cocoon luminosity depends sensitively on how efficiently the shocked jet material mixes with the more heavily baryon-loaded shocked ejecta needed to provide jet collimation \citep{Lazzati+16,Nakar&Piran16}. 

NS-NS/BH-NS mergers are also predicted to be accompanied by a more isotropic counterpart, commonly known as a `kilonova' (also known as `macronova').  Kilonovae are day to week-long thermal, supernova-like transients, which are powered by the radioactive decay of heavy, neutron-rich elements synthesized in the expanding merger ejecta \citep{Li&Paczynski98}.  They provide both a robust EM counterpart to the GW chirp, which is expected to accompany a fraction of BH-NS mergers and essentially all NS-NS mergers, as well as a direct probe of the unknown astrophysical origin of the heaviest elements \citep{Metzger+10}.  

This article provides a pedagogical review of kilonovae, including a brief historical background and recent developments in this rapidly evolving field (Sect.~\ref{sec:history}).  In Sect.~\ref{sec:basics}, I describe the basic physical ingredients relevant to kilonovae, including the key input from numerical simulations of the merger and its aftermath.  For pedagogical reasons, the discussion is organized around a simple toy model for the kilonova light curve (Sect.~\ref{sec:model}), which synthesizes most of the relevant ingredients within a common and easy-to-interpret framework.  My goal is to make the basic results accessible to anyone with the ability to solve a set of coupled ordinary differential equations on their laptop computer.  

I begin by introducing the `vanilla' model of lanthanide-rich ejecta heated by radioactivity, which produces a week-long near-infrared transient (Sect.~\ref{sec:vanilla}).  We then explore several variations on this canonical picture, some more speculative than others.  These include early optical-wavelength (`blue') emission due to Lanthanide-free components of the ejecta (Sect.~\ref{sec:blue}) and the speculative possibility of an early UV-bright `precursor' powered by the decay of free neutrons in the outermost layers of the ejecta (Sect.~\ref{sec:neutrons}).  I also consider the impact on the kilonova signal of energy input from a long-lived accreting BH or magnetar engine (Sect.~\ref{sec:engine}).  In Sect.~\ref{sec:discussion} I assess the prospects for discovering kilonovae following short GRBs or GW-triggers of NS-NS/BH-NS mergers in light of the recent follow-up of the LIGO BH-BH mergers \citep{Abbott+16}.  I use this opportunity to speculate on the promising future years or decades ahead, once kilonovae are routinely discovered in coincidence with a large sample of GW-detected merger events.  I conclude with some personal thoughts and avenues for future progress in Sect.~\ref{sec:conclusions}. 

Although I have attempted to make this review self-contained, the material covered is necessarily limited in scope and reflects my own opinions and biases.  I refer the reader to a number of other excellent recent reviews, which cover some of the topics discussed briefly here in greater detail: \citep{Nakar07,Faber&Rasio12,Berger14,Rosswog15,Fan&Hendry15,Baiotti&Rezzolla16}, including another review dedicated exclusively to kilonovae by \cite{Tanaka16}.  I encourage the reader to consult \cite{Fernandez&Metzger16} for a review of the broader range of EM counterparts of NS-NS/BH-NS mergers.

\begin{table}
\caption{Timeline of major developments in kilonova research}
\centering
\begin{tabular}{r || l}
1974 & Lattimer \& Schramm: $r$-process from BH-NS mergers \\
1975 & Hulse \& Taylor: discovery of binary pulsar system PSR 1913+16\\
1982 & Symbalisty \& Schramm: $r$-process from NS-NS mergers\\
1989 & Eichler et al.: GRBs from NS-NS mergers \\
1994 & Davies et al.: first numerical simulation of mass ejection from NS-NS mergers\\
1998 & Li \& Paczynski: first kilonova model, with parametrized heating \\
1999 & Freiburghaus et al.: NS-NS dynamical ejecta $\Rightarrow$ r-process abundances \\
2005 & Kulkarni: kilonova powered by free neutron-decay (``macronova''), central engine\\
2009 & Perley et al.: optical kilonova candidate following GRB 080503 (Fig.~\ref{fig:magnetarjet})\\
2010 & Metzger et al., Roberts et al., Goriely et al.: kilonova powered by $r$-process heating\\
2013 & Barnes \& Kasen, Tanaka \& Hotokezaka: La/Ac opacities $\Rightarrow$ NIR spectral peak\\
2013 & Tanvir et al., Berger et al.: NIR kilonova candidate following GRB 130603B\\
2013 & Yu, Zhang, Gao: magnetar-boosted kilonova (``merger-nova'')\\
2014 & Metzger \& Fernandez, Kasen et al.: blue kilonova from post-merger remnant disk winds\\
\end{tabular}
\end{table}


\begin{figure}[!t]
\includegraphics[width=1.0\textwidth]{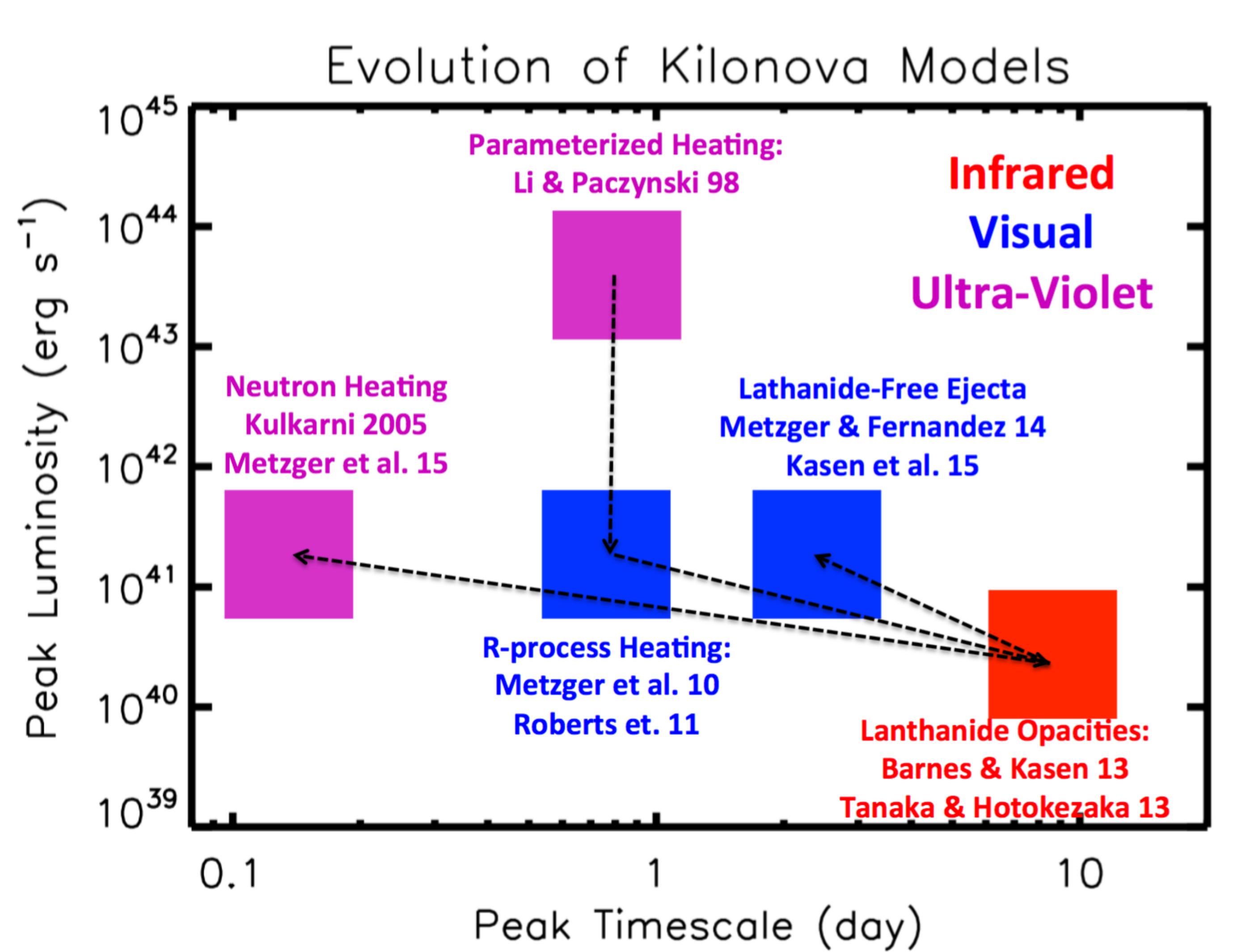}
\caption{Timeline of the development kilonova models in the space of peak luminosity and peak timescale.  The wavelength of the predicted spectral peak are indicated by color as marked in the figure. }
\label{fig:timeline}
\end{figure}

\section{Historical Background }  
\label{sec:history}

\subsection{NS mergers as sources of the $r$-process}

\cite{Burbidge+57} and \cite{Cameron57} realized that approximately half of the elements heavier than iron are synthesized via the capture of neutrons onto lighter seed nuclei like iron) in a dense neutron-rich environment in which the timescale for neutron capture is shorter than the $\beta-$decay timescale.  This `rapid neutron-capture process', or $r$-process, occurs along a nuclear path which resides far on the neutron-rich side of the valley of stable isotopes.  Despite these seminal works occurring almost 70 years ago, the astrophysical environments giving rise to the $r$-process remains an enduring mystery, among the greatest in nuclear astrophysics \citep[e.g.,][for contemporary reviews]{Qian&Wasserburg07,Arnould+07,Thielemann+11}.  

Core collapse SNe have long been considered promising $r$-process sources.  This is in part due to their short delays following star formation, which allows even the earliest generations of metal-poor stars in our Galaxy to be polluted with $r$-process elements, as is observed \citep{Mathews+92,Sneden+08}.  Throughout the 1990s, the high entropy\footnote{A high entropy (low density) results in an $\alpha$-rich freeze-out of the 3 and effective 4-body reactions responsible for forming seed nuclei in the wind, similar to big bang nucleosynthesis.  The resulting higher ratio of neutrons to seed nuclei (for fixed $Y_e$) then allows the $r$-process to proceed to heavier elements.} neutrino-heated winds from proto-neutron stars \citep{Duncan+86,Qian&Woosley96}, which emerge on a timescale of seconds after a successful explosion, were considered the most likely $r$-process site\footnote{Another $r$-process mechanism in the core collapse environment results from $\nu-$induced spallation in the He shell \citep{Banerjee+11}.  This channel is limited to very low metallicity $Z \lesssim 10^{-3}$ and thus cannot represent the dominant $r$-process source over the age of the galaxy (though it could be important for the first generations of stars).  } within the core collapse environment \citep{Woosley+94, Takahashi+94}.  However, more detailed calculations of the wind properties \citep{Thompson+01, Arcones+07, Fischer+10, Hudepohl+10, Roberts+10, MartinezPinedo+12, Roberts+12} later showed that the requisite combination of neutron-rich conditions (electron fraction\footnote{The electron fraction is defined as the ratio of protons to total baryons (neutrons + protons).  Thus, matter with $Y_e < 0.5$ has more neutrons than protons.} $Y_e \lesssim  0.5$) and high entropy were unlikely to obtain.  Possible exceptions include the rare case of a very massive proto-NS \citep{Cardall&Fuller97}, or in the presence of non-standard physics such as an eV-mass sterile neutrino \citep{Tamborra+12,Wu+14}.  

Another exception to this canonical picture may occur if the proto-NS is formed rapidly rotating, with an ultra-strong magnetic field $B \gtrsim 10^{14}\mbox{\,--\,}10^{15}$ G, similar to those which characterize Galactic magnetars.  Magneto-centrifugal acceleration within such a wind \citep{Thompson+04} can act to lower its electron fraction or reduce the number of seed nuclei formed, both during the SN explosion phase \citep{Winteler+12} and during the subsequent proto-NS cooling phase \citep{Thompson03,Metzger+07,Vlasov+14}.  Despite the promise of such models, simulations of MHD-SNe are still in a preliminary state, especially in three dimensions crucial to capturing the growth of non-axisymmetric magnetic kink or sausage mode) instabilities, which can disrupt MHD jet-like structures \citep{Mosta+14}.  The observed rate of hyper-energetic supernovae, which are commonly believed to require an MHD-powered mechanism, is also low compared to the total core collapse rate \citep{Podsiadlowski+04}.  Thus, a a higher $r$-process yield per event is required to explain the Galactic abundances through this channel alone.

Nearly simultaneous with the discovery of the first binary pulsar \citep{Hulse&Taylor75}, \cite{Lattimer&Schramm74, Lattimer&Schramm76} proposed that the merger of compact star binaries---in particular the collision of BH-NS systems---could give rise to the $r$-process by the decompression of highly neutron-rich ejecta \citep{Meyer89}.  \cite{Symbalisty&Schramm82} were the first to suggest NS-NS mergers as the site of the $r$-process.  \cite{Blinnikov+84} and \cite{Paczynski86} first suggested a connection between NS-NS mergers and GRBs.  \cite{Eichler+89} presented a more detailed model for how this environment could give rise to a GRB (albeit one which differs significantly from the current view).  \cite{Davies+94} performed the first numerical simulations of mass ejection from merging neutron stars, finding that $\sim 2\%$ of the binary mass was unbound during the process.  \cite{Freiburghaus+99} presented the first explicit calculations showing that the ejecta properties extracted from a hydrodynamical simulation of a NS-NS merger \citep{Rosswog+99b} indeed produces abundance patterns in basic accord with the solar system $r$-process.    

The neutrino-driven wind following a SN explosion accelerates matter from the proto-NS surface relatively gradually, in which case neutrino absorption reactions on nucleons have time to appreciably raise the electron fraction of the wind from its initial low value near the NS surface.  By contrast, in NS-NS/BH-NS mergers the  different geometry and more dynamical nature of the system allows at least a fraction of the unbound ejecta (tidal tails and disk winds) to avoid strong neutrino irradiation, maintaining a much lower value of $Y_e \lesssim 0.2$ (Sect.~\ref{sec:ejecta}).  

When averaged over the age of the Galaxy, the required production rate of heavy $r$-process nuclei of mass number $A > 140$ is $\sim 2\times 10^{-7} M_{\odot}$ yr$^{-1}$ \citep{Qian00}, although this number comes with large uncertainties \citep{Bauswein+13}.  Given a measured NS-NS merger detection rate by Advanced LIGO/Virgo of $\mathcal{R}_{\rm NS-NS}$, the required $r$-process mass yield per merger event is then approximately \citep[e.g.,][]{Metzger+09,Vangioni+16}
\be
\langle M_{r} \rangle \sim 10^{-2}M_{\odot}\left(\frac{\mathcal{R}_{\rm NS-NS}}{10\,{\rm yr^{-1}}}\right)^{-1}.
\label{eq:Mr}
\ee
As described in Sect.~\ref{sec:ejecta}, numerical simulations of NS-NS/BH-NS mergers find total ejecta masses of $\langle M_{r} \rangle \sim 10^{-3}-10^{-1} M_{\odot}$, consistent (again, with large uncertainties) with NS mergers being a major source of the Galactic $r$-process.  

Several additional lines of evidence support `high yield' $r$-process events like NS-NS/BH-NS mergers being common in our Galaxy, both now and in its early history.  These include the detection of $^{244}$Pu on the ocean floor at abundances roughly 2 orders lower than that expected if the source were frequent, low-yield events like normal SNe \citep{Wallner+15,Hotokezaka+15}.  A large fraction of the stars in the dwarf galaxy Reticulum II are highly enriched in $r$-process elements, indicating that this galaxy was polluted early in its history by a single $r$-process event with a yield much higher than the neutrino-driven wind of a single, non-MHD SN \citep{Ji+16}.  Given the extremely low escape speed of a dwarf galaxy of $\sim 10\mathrm{\ km\ s}^{-1}$, even a moderate SN birth kick would have removed any NS binary from the galaxy prior to merger; on the other hand, a sub-population of the Galactic NS-NS binaries have low proper motions and are indeed inferred to have experienced very low SN kicks \citep{Beniamini+16}.  

It has also been realized that there may exist channels for NS-NS mergers which occur with short delays after star formation \citep{Belczynski+02,Voss&Tauris03,Ramirez-Ruiz+15}.  Depending on the efficiency of compositional mixing between the merger ejecta and the ISM of the Galaxy, realistic delay time distributions for NS-NS/NS-BH mergers within a consistent picture of structure formation via hierarchical growth \citep{Kelley+10} can produce chemical evolution histories which are consistent with observations of the abundances of $r$-process elements in metal-poor halo stars as a function of their iron abundance \citep{Shen+15,Ramirez-Ruiz+15,vandeVoort+15}.  Given the under-resolved nature of current simulations, it is not yet proven that high-yield $r$-process channels are favored.  However, it has become clear that previous claims ruling out NS-NS/BH-NS mergers with closed-box chemical evolution models \citep{Argast+04} were likely premature.

\subsection{A Brief History of Kilonovae}

\cite{Li&Paczynski98} first showed that the radioactive ejecta from a NS-NS or BH-NS merger provides a source for powering transient emission, in analogy with Type Ia SNe.  They developed a toy model for the light curve, similar to that we describe in Sect.~\ref{sec:model}.  Given the low mass and high velocity of the ejecta from a NS-NS/BH-NS merger, they concluded that the ejecta will become transparent to its own radiation quickly, producing emission which peaks on a timescale of about one day, much faster than for normal SNe (which instead peak on a timescale of weeks or longer).  

Lacking a model for the nucleosynthesis, \cite{Li&Paczynski98} parametrized the radioactive heating rate of the ejecta at time $t$ after the merger according to the following prescription,
\be
\dot{Q}_{\rm LP} = \frac{f M c^{2}}{t},
\label{eq:LP98}
\ee
where $M$ is the ejecta mass and $f$ is a free parameter (see below).  The $\propto 1/t$ time dependence was motivated by the total heating rate which results from the sum of the radioactive decay heating rate $\dot{Q}_i \propto \exp(-t/\tau_i)$ of a large number of isotopes $i$, under the assumption that their half-lives $\tau_i$ are distributed equally per logarithmic time (at any time $t$, the heating rate is dominated by isotopes with half-lives $\tau_i \sim t$).  Contemporary models, which process the thermodynamic history of the expanding ejecta based on numerical simulations of the merger through a detailed nuclear reaction network, show that the heating rate at late times actually approaches a steeper power law decay $\propto t^{-\alpha}$, with $\alpha \approx 1.1\mbox{\,--\,}1.4$ \citep{Metzger+10, Roberts+11, Korobkin+12}, similar to what is found for the decay rate of terrestrial radioactive waste \citep{Way&Wigner48}.  \citet{Metzger+10} and \citet{Hotokezaka+17} describe how this power-law decay can be understood from the basic physics of $\beta-$decay and the properties of nuclei on the neutron-rich valley of stability.

\cite{Li&Paczynski98} also left the normalization of the heating rate $f$, to which the peak luminosity of the kilonova is linearly proportional, as a free parameter, considering a range of models with different values of $f = 10^{-5}\mbox{\,--\,}10^{-3}$.  More recent calculations, described below, show that such high heating rates are extremely optimistic, leading to predicted peak luminosities $\gtrsim 10^{42}\mbox{\,--\,}10^{44}\mathrm{\ erg\ s}^{-1}$ \citep[][their Fig.~2]{Li&Paczynski98} which exceed even those of SNe.  These over-predictions leaked to other works throughout the next decade; for instance, \cite{Rosswog05} predicted that BH-NS mergers are accompanied by transients of luminosity $\gtrsim 10^{44}\mathrm{\ erg\ s}^{-1}$, which would rival the most luminous SNe ever discovered (\citealt{Dong+16}).  This unclear theoretical situation led to observational searches for kilonovae following short GRBs which were inconclusive since they were forced to parametrized their results (usually non-detections) in terms of the allowed range of $f$ \citep{Bloom+06,Kocevski+10} instead of in terms of more meaningful constraints on the ejecta properties.

\cite{Metzger+10} determined the true luminosity scale of the radioactively-powered transients of NS mergers by calculating the first light curve models which used radioactive heating rates derived self-consistently from a nuclear reaction network calculation of the $r$-process, based on the dynamical ejecta trajectories of \cite{Freiburghaus+99}.  Based on their derived peak luminosities being approximately one thousand times brighter than a nova, \cite{Metzger+10} first introduced the term `kilonova' to describe the EM counterparts of NS mergers powered by the decay of $r$-process nuclei.  They showed that the radioactive heating rate was relatively insensitive to the precise electron fraction of the ejecta, and they were the first to consider how efficiently the decay products thermalize their energy in the ejecta.  \cite{Metzger+10} also highlighted the critical four-way connection between kilonovae, short GRBs, GWs from NS-NS/BH-NS mergers, and the astrophysical origin of the $r$-process.

Prior to \citet{Metzger+10}, it was commonly believed that kilonovae were in fact brighter, or much brighter, than supernovae \citep{Li&Paczynski98,Rosswog05}.  One exception is \cite{Kulkarni05}, who assumed that the radioactive power was supplied by the decay of $^{56}$Ni or free neutrons.  However, $^{56}$Ni cannot be produced in the neutron-rich ejecta of a NS merger, while all initially free neutrons are captured into seed nuclei during the $r$-process (except perhaps in the very outermost, fastest expanding layers of the ejecta \citep{Metzger+15}; see Sect.~\ref{sec:neutrons}).  Kulkarni introduced the term ``macronovae'' for such Nickel/neutron-powered events.  Despite its inauspicious physical motivation, many authors continue to use the macronova terminology, in part because this name is not tied to a particular luminosity scale (which may change as our physical models evolve).    

Once the radioactive heating rate was determined, attention turned to the yet thornier issue of the ejecta opacity.  The latter is crucial since it determines at what time and wavelength the ejecta becomes transparent and the light curve peaks.  Given the general lack\footnote{At least of an unclassified nature.} of experimental data or theoretical models for the opacity of heavy $r$-process elements, especially in the first and second ionization states of greatest relevance, \cite{Metzger+10,Roberts+11} adopted grey opacities appropriate to the Fe-rich ejecta in Type Ia SNe.  However, then \cite{Kasen+13} showed that the opacity of $r$-process elements can be significantly higher than that of Fe, due to the high density of line transitions associated with the complex atomic structures of some lanthanide and actinide elements (Sect.~\ref{sec:opacity}).  This finding was subsequently confirmed by \cite{Tanaka&Hotokezaka13}.  As compared to the earlier predictions \citep{Metzger+10}, these higher opacities push the bolometric light curve to peak later in time ($\sim 1$ week instead of a $\sim 1$ day timescale), and at a lower luminosity \citep{Barnes&Kasen13}.  More importantly, the enormous optical wavelength opacity caused by line blanketing moved the spectral peak from optical/UV frequencies to the near-infrared (NIR).  

Later that year, \cite{Tanvir+13} and \cite{Berger+13} presented evidence for excess infrared emission following the short GRB 130603B on a timescale of about one week using the \textit{Hubble Space Telescope}.  If confirmed by future observations, this discovery would be the first evidence directly relating NS mergers to short GRBs, and hence to the direct production of $r$-process nuclei (see Sect.~\ref{sec:candidates} for further discussion of kilonova searches after short GRBs).  As discussed further in Sect.~\ref{sec:detection}, the prediction that kilonova emission peaks in the NIR, with the optical emission highly suppressed, has important implications for the strategies of EM follow-up of future GW bursts.  The timeline of theoretical predictions for the peak luminosities, timescales, and spectrap peak of the kilonova emission are summarized in Fig.~\ref{fig:timeline}.

\section{Basic Ingredients}
\label{sec:basics}

The physics of kilonovae can be understood from basic considerations.  Consider the merger ejecta of total mass $M$, which is expanding at a constant velocity $v$, such that its radius is $R \approx vt$ after a time $t$ following the merger.  We assume spherical symmetry, which, perhaps surprisingly, is a reasonable first-order approximation because the ejecta has a chance to expand laterally over the many orders of magnitude in scale from the merging binary ($R_{0} \sim 10^{6}$ cm) to the much larger radius ($R_{\rm peak} \sim 10^{15}$ cm) at which the kilonova emission peaks \citep{Roberts+11,Grossman+14,Rosswog+14}.  

The ejecta is hot immediately after the merger, especially if it originates from the shocked interface between the colliding NS-NS binary (Sect.~\ref{sec:ejecta}).  This thermal energy cannot, however, initially escape as radiation because of its high optical depth at early times,
\be
\tau \simeq \rho \kappa R = \frac{3M\kappa}{4\pi R^{2}} \simeq 70\left(\frac{M}{10^{-2}M_{\odot}}\right)\left(\frac{\kappa}{\rm 1\,cm^{2}\,g^{-1}}\right)\left(\frac{v}{0.1c}\right)^{-2}\left(\frac{t}{\rm 1\,day}\right)^{-2},
\label{eq:tau}
\ee
and the correspondingly long photon diffusion timescale through the ejecta,
\be
t_{\rm diff} \simeq \frac{R}{c}\tau =  \frac{3M\kappa}{4\pi c R} = \frac{3M\kappa}{4\pi c vt},
\label{eq:tdiff}
\ee
where $\rho = 3M/(4\pi R^{3})$ is the mean density and $\kappa$ is the opacity (cross section per unit mass).  As the ejecta expands, the diffusion time decreases with time $t_{\rm diff} \propto t^{-1}$, until eventually radiation can escape on the expansion timescale, as occurs once $t_{\rm diff} = t$ \citep{Arnett82}.  This condition determines the characteristic timescale at which the light curve peaks,
\be
t_{\rm peak} \equiv \left(\frac{3 M \kappa }{4\pi \beta v c}\right)^{1/2} \approx 1.6\,{\rm d}\,\,\left(\frac{M}{10^{-2}M_{\odot}}\right)^{1/2}\left(\frac{v}{0.1c}\right)^{-1/2}\left(\frac{\kappa }{1\,{\rm cm^{2}\,g^{-1}}}\right)^{1/2},
\label{eq:tpeak}
\ee
where the constant $\beta \approx 3$ depends on the precise density profile of the ejecta (see Sect.~\ref{sec:model}).  For values of the opacity $\kappa \sim 1\mbox{\,--\,}100\mathrm{\ cm^{2}\ g^{-1}}$ which characterize the range from Lanthanide-free and Lanthanide-rich matter (Fig.~\ref{fig:opacities}), respectively, Eq.~(\ref{eq:tpeak}) predicts characteristic durations $\sim 1$ day\,--\,1 week.

The temperature of matter freshly ejected at the radius of the merger $R_0 \lesssim 100$ km exceeds billions of degrees.  However, absent a source of persistent heating, this matter will cool through adiabatic expansion, losing all but a fraction $\sim R_0/R_{\rm peak} \sim 10^{-9}$ of its initial thermal energy before reaching the radius $R_{\rm peak} = vt_{\rm peak}$ at which the ejecta becomes transparent (Eq.~\ref{eq:tpeak}).  Such adiabatic losses would leave the ejecta so cold as to be effectively invisible.  

\begin{figure}[!t]
\includegraphics[width=0.5\textwidth]{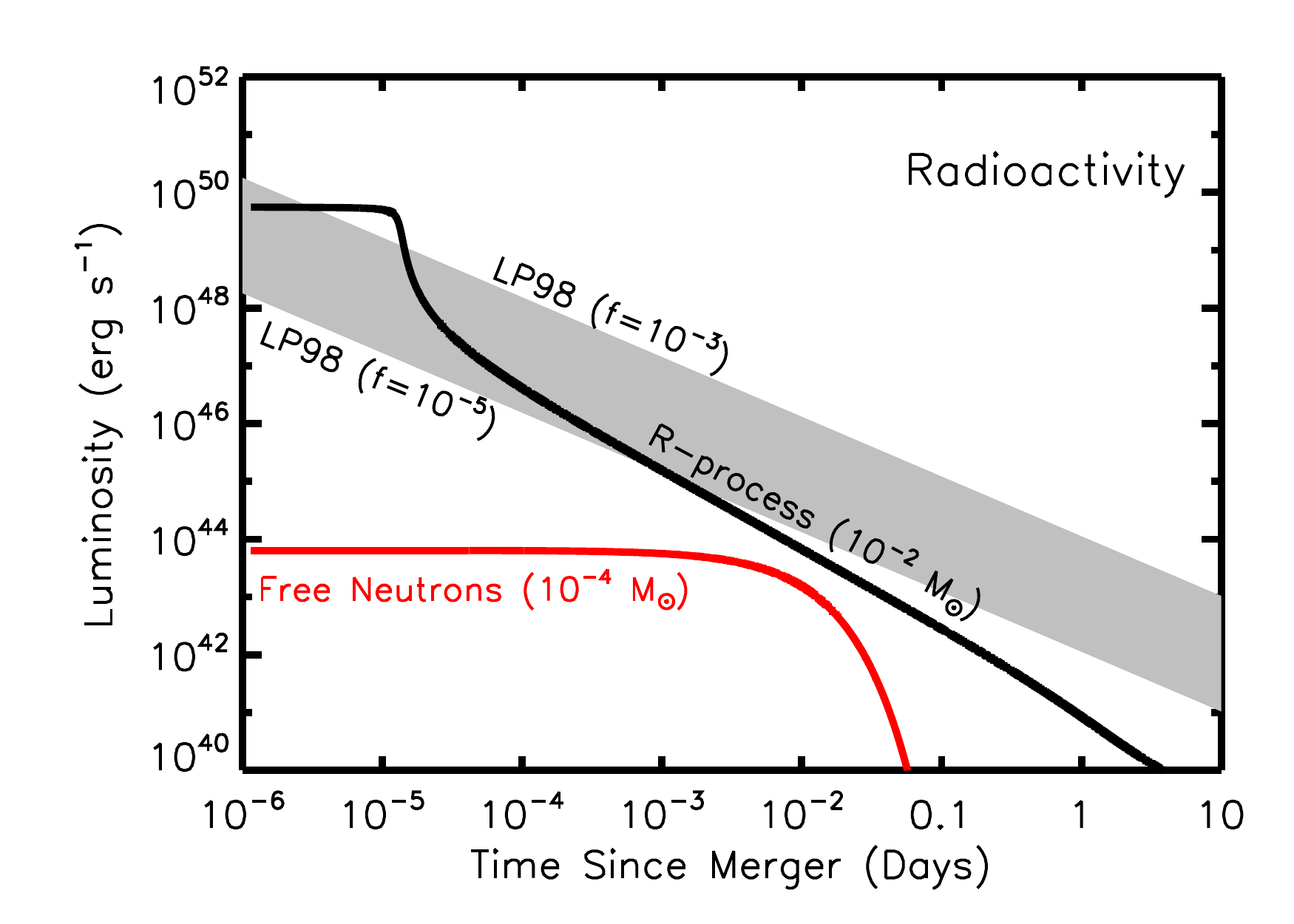}
\includegraphics[width=0.5\textwidth]{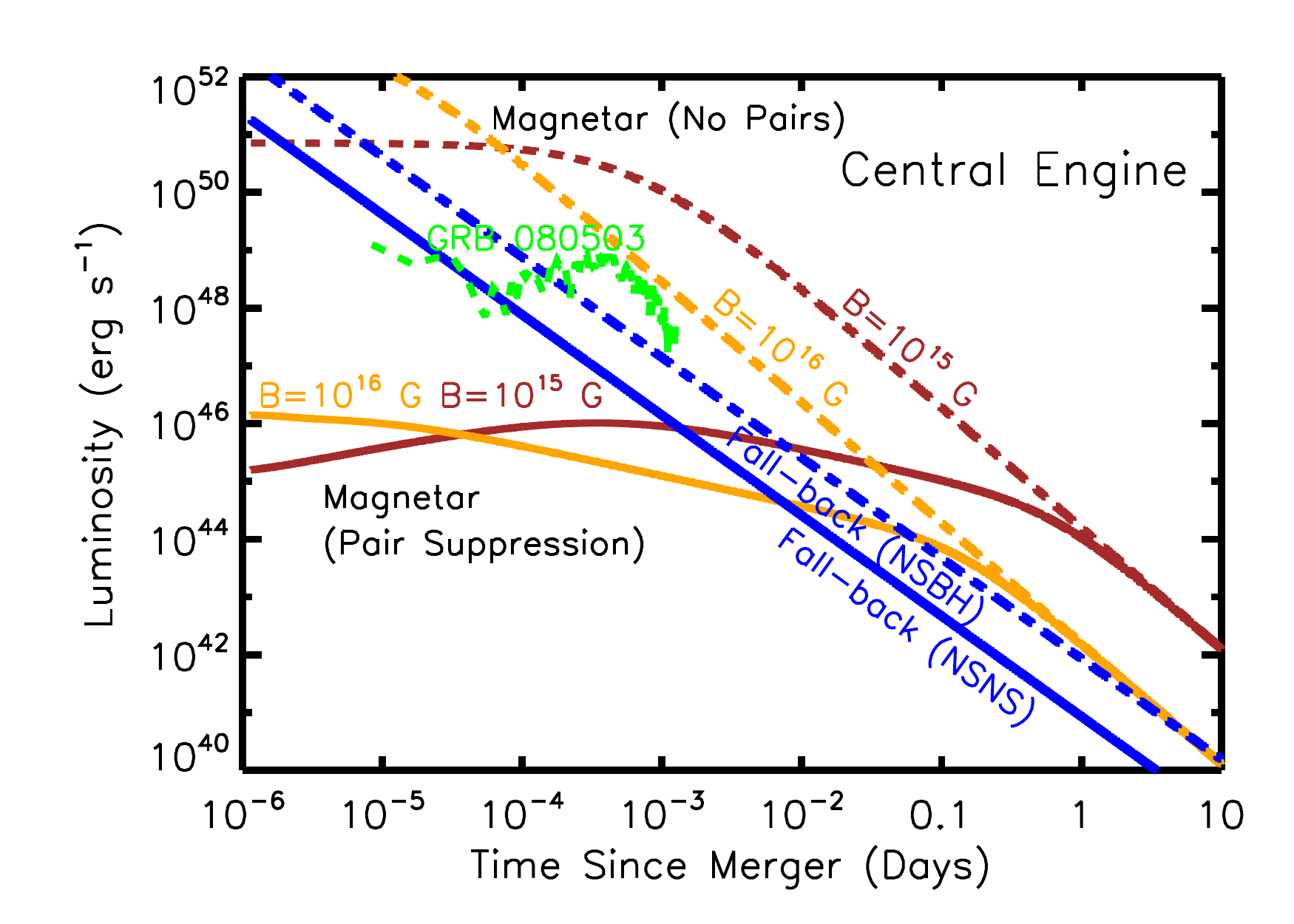}
\caption{Luminosity versus time after the merger of a range of heating sources relevant to powering kilonovae.  LEFT: Sources of radioactive heating include the decay of $\sim 10 ^{-2}M_{\odot}$ of r-process nuclei, as first modeled in a parametrized way by \cite{Li&Paczynski98} (Eq.~\ref{eq:LP98}, grey band) and then more accurately by \cite{Metzger+10}, plotted here using the analytic fit of \cite{Korobkin+12} (Eq.~\ref{eq:edotr}, black line) and applying the thermalization efficiency of \cite{Barnes+16} (Eq.~\ref{eq:eth}).  The outer layers of the merger ejecta may contain $10^{-4}M_{\odot}$ free neutrons (red line), which due to their anomalously long half-life produce significant heating on a timescale of tens of minutes if they exist in the ejecta (Sect.~\ref{sec:neutrons}).  RIGHT: Sources of central engine heating.  These include fall-back accretion (blue lines), shown separately for the case of a NS-NS merger (solid line) and BH-NS merger (dashed line), based on the SPH simulations of \cite{Rosswog07} for an assumed jet efficiency $\epsilon_j = 0.1$ (Eq.~\ref{eq:Lxfb}).  Also shown is the energy input due to the spin-down of a stable central magnetar remnant with an initial spin period of $P = 0.7$ ms dipole field strengths of $B = 10^{15}$ G (brown lines) and $B = 10^{16}$ G (orange lines).  We show separately the total spin-down luminosity $L_{\rm sd}$ (dashed lines; Eq.~\ref{eq:Lsd}), as well as the effective luminosity  accounting also for the suppression of thermalization of the magnetar energy by the high opacity of $e^{\pm}$ pairs in the nebula (solid lines; see Eq.~\ref{eq:Lobs} and surrounding discussion; \citealp{Metzger&Piro14}).  The isotropic X-ray luminosity of the extended emission following the short GRB 080503 is shown with a green line for an assumed redshift $z = 0.3$; \citealp{Perley+09} (see also bottom panel of Fig.~\ref{fig:magnetarjet}). }
\label{fig:heating}
\end{figure}

In reality, the ejecta will be continuously heated by a combination of sources, at a total rate $\dot{Q}(t)$ (Fig.~\ref{fig:heating}).  At a minimum, this heating includes contributions from radioactivity due to $r$-process nuclei and, possibly, free neutrons.  More speculatively, the ejecta can also be heated from within by a central engine, such as a long-lived magnetar or accreting BH.  In most cases of relevance, $\dot{Q}(t)$ is constant or decreasing with time less steeply than $\propto t^{-2}$.  The peak luminosity of the observed emission then equals the heating rate at the peak time ($t = t_{\rm peak}$), i.e.,
\be
L_{\rm peak} \approx \dot{Q}(t_{\rm peak}),
\label{eq:Arnett}
\ee 
a result commonly known as ``Arnett's Law'' \citep{Arnett82}.  

Equations (\ref{eq:tpeak}) and (\ref{eq:Arnett}) make clear that, in order to quantify the key observables of kilonovae (peak timescale, luminosity, and effective temperature), we must understand three key ingredients: 
\begin{itemize}
\item{The mass and velocity of the ejecta from NS-NS/BH-NS mergers.}
\item{The opacity $\kappa$ of expanding neutron-rich matter.}
\item{The variety of sources which contribute to heating the ejecta $\dot{Q}(t)$, particularly on timescales of $t_{\rm peak}$, when the ejecta is first becoming transparent.}
\end{itemize}
 The remainder of this section addresses the first two issues.  The range of different heating sources, which give rise to different `flavors' of kilonovae, are covered in Sect.~\ref{sec:model}.

\subsection{Sources of Ejecta in Binary NS Mergers}
\label{sec:ejecta}


Two broad sources of ejecta characterize NS-NS and BH-NS mergers \citep[see][for a recent review]{Fernandez&Metzger16}.  First, there is matter ejected on the dynamical timescale (typically milliseconds), either by tidal forces or due to compression-induced heating at the interface between merging bodies (Sect.~\ref{sec:dynamical}).  Debris from the merger, which is not immediately unbound or incorporated into the central compact object, can possess enough angular momentum to circularize into an accretion disk around the central compact object.  Outflows from this remnant disk, taking place on longer timescales of up to seconds, provide a second important source of  ejecta (Sect.~\ref{sec:diskejecta}).

In a BH-NS merger, significant mass ejection and disk formation occurs only if the BH has a low mass $M_{\bullet}$ and is rapidly spinning; in such cases, the NS is tidally disrupted during the very final stages of the inspiral instead of being swallowed whole.  Roughly speaking, the condition for the latter is that the tidal radius of the NS, $R_{\rm t} \propto M_{\bullet}^{1/3}$, exceed the innermost stable circular orbit of the BH, $R_{\rm isco} \propto M_{\bullet}$ \citep{Foucart12} and references therein).  For a NS of radius $12$ km and mass $1.4M_{\odot}$, this requires a BH of mass $\lesssim 4(12)M_{\odot}$ for a BH Kerr spin parameter of $\chi_{\rm BH} = 0.7(0.95)$.  For a non-spinning BH, the BH mass range giving rise to tidal disruption---and hence a detectable signal---is very small (however, see \citealp{McWilliams&Levin11,DOrazio+16}).

In the case of a NS-NS merger, the ejecta properties depend sensitively on the fate of the massive NS remnant which is created by the coalescence event.  The latter in turn depends sensitively on the total mass of the original NS-NS binary \citep{shibata2000,Shibata&Taniguchi06}.  Above a threshold mass of $M_{\rm crit} \sim 2.6-3.9M_\odot$ (covering a range of soft and stiff nuclear-theory based equations of state [EOS], respectively), the remnant collapses to a BH essentially immediately, on the dynamical time of milliseconds or less \citep{Hotokezaka+11,Bauswein+13}.  

The maximum mass of a NS, though primarily sensitive to the NS EOS, can be increased if the NS is rotating rapidly \citep{Baumgarte+00,Ozel+10,Kaplan+14}.  For remnant masses $\lesssim M_{\rm crit}$, the remnant is supported by rotation, at least for a temporarily period after the merger.  A massive NS remnant, which is supported exclusively by its differential rotation, is known as a \emph{hypermassive NS} (HMNS).  A somewhat less massive NS, which can be supported even by its solid body rotation (i.e.~once differential rotation has been removed), is known as \emph{supramassive}.  A HMNS is unlikely to survive for more than a few tens to hundreds of milliseconds after the merger, before collapsing to a BH due to the loss of its differential rotation by internal hyrdo-magnetic torques and gravitational wave radiation \citep{Shibata&Taniguchi06,Duez+06,Siegel+13}. In contrast, supramassive remnants must spin-down to the point of collapse through less efficient processes, such as magnetic dipole radiation or GW emission due to secular instabilities, and hence can in principle remain stable for minutes or potentially much longer.  Finally, the merger of a particularly low mass binary, with a total mass less than the maximum mass of a non-rotating NS, $M_{\rm max}(\Omega = 0)$, will produce an indefinitely stable remnant, from which a BH never forms \citep{Metzger+08a,Giacomazzo&Perna13}.  

These mass divisions are illustrated in Fig.~\ref{fig:Erot} using an example EOS, for which the maximum non-rotating NS mass is $M_{\rm max}(\Omega = 0) \approx 2.24M_{\odot}$.  This value is consistent with the lower limit of $M_{\rm max}(\Omega = 0) \approx 2M_{\odot}$ set by the discovery of pulsars with similar masses \citep{Demorest+10, Antoniadis+13}.  Unless the value of $M_{\rm max}(\Omega = 0)$ is fine-tuned to be just slightly above current lower limits, the remnants of at least a moderate fraction of NS-NS mergers are likely to be supramassive \citep{Ozel+10}, if not indefinitely stable.  As we discuss in Sect.~\ref{sec:magnetar}, energy input from such long-lived remnants could substantially enhance the kilonova emission.

\subsubsection{Dynamical Ejecta}
\label{sec:dynamical}

NS-NS mergers eject unbound matter through processes that operate on the dynamical time, and which depend primarily on the total  binary mass, the mass ratio, and the EOS.  Total dynamical ejecta masses typically lie in the range $10^{-4}\mbox{\,--\,}10^{-2}M_\odot$ for NS-NS mergers \citep{Hotokezaka+13}, with velocities $0.1\mbox{\,--\,}0.3$ c.   For BH-NS mergers, the ejecta mass can be up to $\sim 0.1M_\odot$ with similar velocities as in the NS-NS case \citep{Kyutoku+13,Kyutoku+15}.  The ejecta mass is typically greater for eccentric binaries \citep{East+12,Gold+12}, although the dynamical interactions giving rise to eccentric mergers require high stellar densities, probably making them rare events compared to circular inspirals \citep{Tsang13}.

Two main ejection processes operate in NS-NS mergers. First, material at the contact interface between the merging stars is squeezed out by hydrodynamic forces and is subsequently expelled by quasi-radial pulsations of the remnant \citep{Oechslin+07,Bauswein+13,Hotokezaka+13}, ejecting shock-heated matter in a broad range of angular directions.  The second process involves spiral arms from tidal interactions during the merger, which expand outwards in the equatorial plane due to angular momentum transport by hydrodynamic processes.  The relative importance of these mechanisms depends on the EOS and the mass ratio of the binary, with higher mass ratio binaries ejecting greater quantities of mass  \citep{bauswein2013,Lehner+16}.  The ejecta mass also depends on the BH formation timescale; for the prompt collapses which characterize massive binaries, mass ejection from the contact interface is suppressed due to prompt swallowing of this region.  

In BH-NS mergers, mass is ejected primarily by tidal forces that disrupt the NS, with the matter emerging primarily in the equatorial plane  \citep{Kawaguchi+15}.  The ejecta from BH-NS mergers also often covers only part of the azimuthal range \citep{Kyutoku+15}, which may introduce a stronger viewing angle dependence on the kilonova emission than for NS-NS mergers. 

A key property of the ejecta, which is at least as important to the kilonova signal as the total mass, is the electron fraction, $Y_e$.  Simulations that do not account for weak interactions find the ejecta from NS-NS mergers to be highly neutron-rich, with an electron fraction $Y_e \lesssim 0.1$, sufficiently low to produce a robust\footnote{This robustness is rooted in `fission recycling' \citep{Goriely+05}: the low initial $Y_e$ results in a large neutron-to-seed ratio, allowing the nuclear flow to reach heavy nuclei for which fission is possible ($A \sim 250$). The fission fragments are then subject to additional neutron captures, generating more heavy nuclei and closing the cycle.} abundance pattern for heavy nuclei with $A \gtrsim 130$ \citep{Goriely+11,Korobkin+12,Bauswein+13,Mendoza-Temis.Wu.ea:2015}.  More recent merger calculations that include the effects of $e^\pm$ captures and neutrino irradiation in full general-relativity have shown that the dynamical ejecta may have a wider electron fraction distribution ($Y_e \sim 0.1-0.4$) than models which neglect weak interactions \citep{sekiguchi2015,radice2016}.  As a result, lighter $r$-process elements with $90 \lesssim A \lesssim 130$ are synthesized in addition to third-peak elements \citep{wanajo2014}.  These high-$Y_e$ ejecta components are distributed in a relatively spherically-symmetric geometry, while the primarily tidally-ejected, lower-$Y_e$ matter is concentrated closer to the equatorial plane and resides outside the higher-$Y_e$ matter (Fig.~\ref{fig:schematic}).


\subsubsection{Disk Wind Ejecta}
\label{sec:diskejecta}

All NS-NS mergers, and those BH-NS mergers which end in NS tidal disruption, result in the formation of an accretion disk around the central NS or BH remnant.  The disk mass is typically $\sim 0.01-0.3M_{\odot}$, depending on the total mass and mass ratio of the binary, the spins of the binary components, and the NS EOS \citep{Oechslin&Janka06}.   Outflows from this disk, over a timescales of seconds or longer, represent an important source of ejecta mass which can rival---or even dominate---that of the dynamical ejecta.

At early times after the disk forms, its mass accretion rate is high and the disk is a copious source of thermal neutrinos \citep{Popham+99}.  During this phase, mass loss is driven from the disk surface by neutrino heating, in a manner analogous to neutrino-driven proto-NS winds in core collapse SNe \citep{Surman+08,Metzger+08neutron}.  Time dependent models of these remnant tori, which include neutrino emission and absorption, indicate that when a BH forms promptly, the amount of mass ejected through this channel is small, contributing at most a few percent of the outflow, because the neutrino luminosity decreases rapidly in time \citep{Fernandez&Metzger13,Just+15}.  However, if the central NS remnant survives for longer than $\sim 50$~ms (as a hypermassive or supramassive NS), then the larger neutrino luminosity from the NS remnant ejects a non-negligible amount of mass ($\sim 10^{-3}M_\odot$, primarily from the NS itself instead of the disk; \citealp{Dessart+09, Perego+14,Martin+15,Richers+15}).

The disk evolves in time due to the outwards transport of angular momentum, as mediated by magnetic stresses created by MHD turbulence generated by the magneto-rotational instability.  Initial time-dependent calculations of this `viscous spreading' followed the disk evolution over several viscous times using one-zone \citep{Metzger+08b} and one-dimensional height-integrated \citep{Metzger+09} models.  These works showed that, as the disk evolves and its accretion rate decreases, the disk transitions from a neutrino-cooled state to a radiatively inefficient (geometrically thick disk) state as the temperature, and hence the neutrino cooling rate, decreases over a timescale of seconds \citep[see also][]{Lee+09,Beloborodov08}.  Significant outflows occur once during the radiative inefficient phase, because viscous turbulent heating and nuclear recombination are unbalanced by neutrino cooling \citep{Kohri+05}.  This state transition is also accompanied by ``freeze-out''\footnote{A useful analogy can be drawn between weak freeze-out in the viscously-expanding accretion disk of a NS merger, and that which occurs in the expanding Universe during the first minutes following the Big Bang.  However, unlike a NS merger, the Universe freezes-out proton-rich, due to the lower densities (which favor proton-forming reactions over the neutron-forming ones instead favored under conditions of high electron degeneracy).} of weak interactions, leading to the winds being neutron-rich \citep{Metzger+08b,Metzger+09}.  Neutron-rich mater is shielded within the degenerate disk midplane, being ejected only once the disk radius has become large enough, and the neutrino luminosity low enough, that weak interactions no longer appreciably raise $Y_e$ in the outflow.  

These early estimates were followed by two-dimensional, axisymmetric hydrodynamical models of the disk evolution, which show that, in the case of prompt BH formation, the electron fraction of the disk outflows lies in the range $Y_e \sim 0.2\mbox{\,--\,}0.4$ \citep{Fernandez&Metzger13,Just+15}, sufficient to produce the entire mass range of $r$-process elements \citep{Just+15,Wu+16}.  The total fraction of the disk mass which is unbound by these ``viscously-driven'' winds ranges from $\sim 5\%$ for a slowly spinning BH, to $\sim 30\%$ for high BH spin $\chi_{\rm BH} \simeq 0.95$ \citep{Just+15,Fernandez+15b}; see also \cite{Kiuchi+15}, who simulated the long-term evolution of BH-NS disks but without following the electron fraction evolution.  An even larger fraction of the disk mass (up to $\sim 90\%$) is unbound when the central remnant is a long-lived hypermassive or supramassive NS instead of a BH, due to the presence of a hard surface and the higher level of neutrino irradiation from the central remnant \citep{Metzger&Fernandez14}.  A longer-lived remnant also increases the electron fraction of the ejecta, which increases monotonically with the lifetime of the HMNS.  Most of the ejecta is lanthanide-free ($Y_e \gtrsim 0.3$) if the NS survives longer than about 300 ms \citep{Kasen+15}.  

\begin{figure}[!t]

\includegraphics[width=0.5\textwidth]{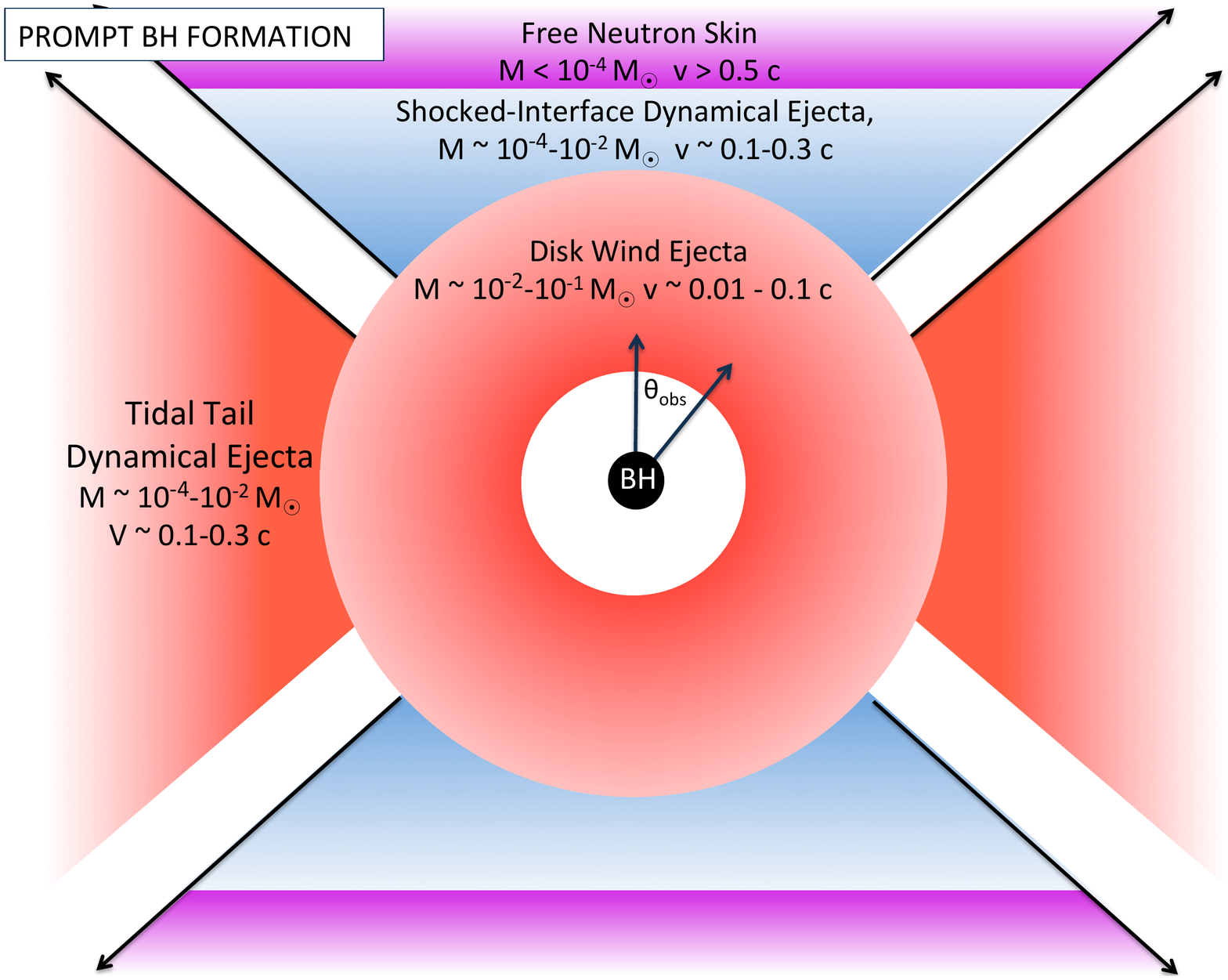}
\includegraphics[width=0.5\textwidth]{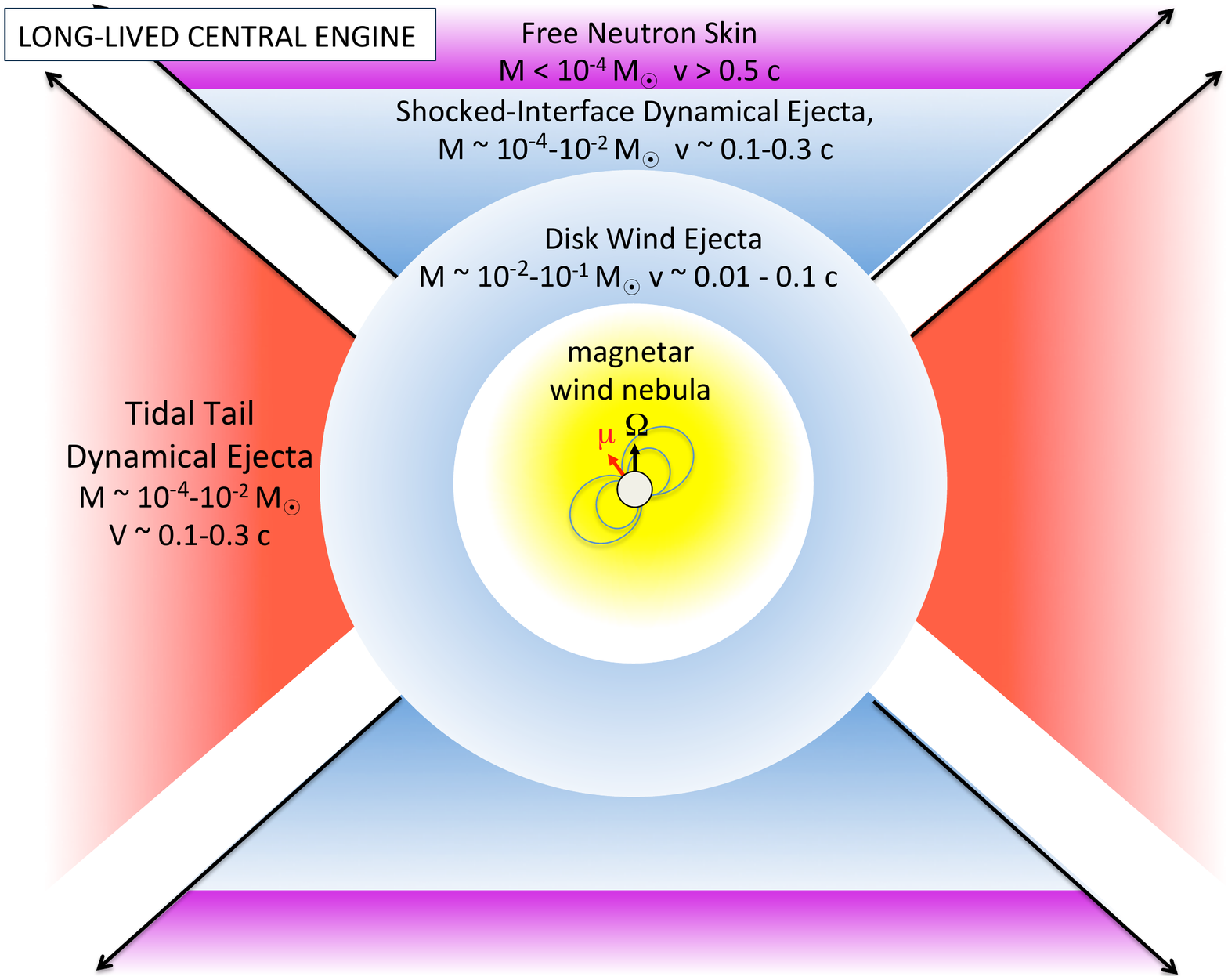}
\caption{Different components of the ejecta from NS-NS mergers and the dependence of their kilonova emission on the observer viewing angle, $\theta_{\rm obs}$, relative to the binary axis, in the case of prompt BH formation (top panel) and a long-lived magnetar remnant (bottom panel).  In both cases, the dynamical ejecta in the equatorial plane is highly neutron-rich ($Y_e \lesssim 0.1$), producing lanthanides and correspondingly ``red'' kilonova emission peaking at NIR wavelengths.  Mass ejected dynamically in the polar directions by shock heating may be sufficiently neutron-poor ($Y_e \gtrsim 0.3$; \citealp{Wanajo+14}) to preclude Lanthanide production, instead powering ``blue'' kilonova emission at optical wavelengths (although this component may be suppressed if BH formation is extremely prompt).  The outermost layers of the polar ejecta may contain free neutrons, the decay of which powers a UV transient lasting a few hours following the merger (Sect.~\ref{sec:neutrons}).  The innermost layers of the ejecta originate from accretion disk outflows, which are likely to emerge more isotropically.  When BH formation is prompt, this matter is also mainly neutron-rich, powering red kilonova emission \citep{Just+15,Wu+16}.  If the remnant is instead long-lived, then neutrinos from the NS remnant can increase the electron fraction of the disk outflows to suppress Lanthanide production and result in blue disk wind emission \citep{Metzger&Fernandez14,Perego+14,Martin+15}.  Energy input from the central accreting BH (top panel) or magnetar remnant (bottom panel) enhance the kilonova luminosity compared to the purely radioactive-powered case (Sect.~\ref{sec:engine}).    }
\label{fig:schematic}
\end{figure}

The mass ejected by the late disk wind can be comparable to or larger than that in the dynamical ejecta (e.g.~\citep{Wu+16}, their Fig.~1).  As the disk outflows emerge after the dynamical ejecta, they will be physically located behind the dynamical ejecta, and will possess a more isotropic geometry (Fig.~\ref{fig:schematic}).  

In addition to the dynamical and disk wind ejecta, additional neutrino or magnetically-driven outflows are expected from the long-lived NS remnant as it undergoes Kelvin-Helmholtz contraction to its final cold state, thus contributing an additional source of ejecta \citep{Dessart+09}.  Such outflows may be particularly important in cases when the remnant is supramassive or indefinitely stable.  The quantity and composition of this wind ejecta will be substantially different than that from `normal' (slowly-rotating, non-magnetized) proto-neutron star winds \citep{Qian&Woosley96} due to the effects of magneto-centrifugal acceleration \citep{Metzger+07,Vlasov+14} and, during early phases, winding of the magnetic field by latitudinal differential rotation \citep{Siegel+14}.

\subsection{Opacity}
\label{sec:opacity}

\begin{figure}[!t]
\includegraphics[width=1.0\textwidth]{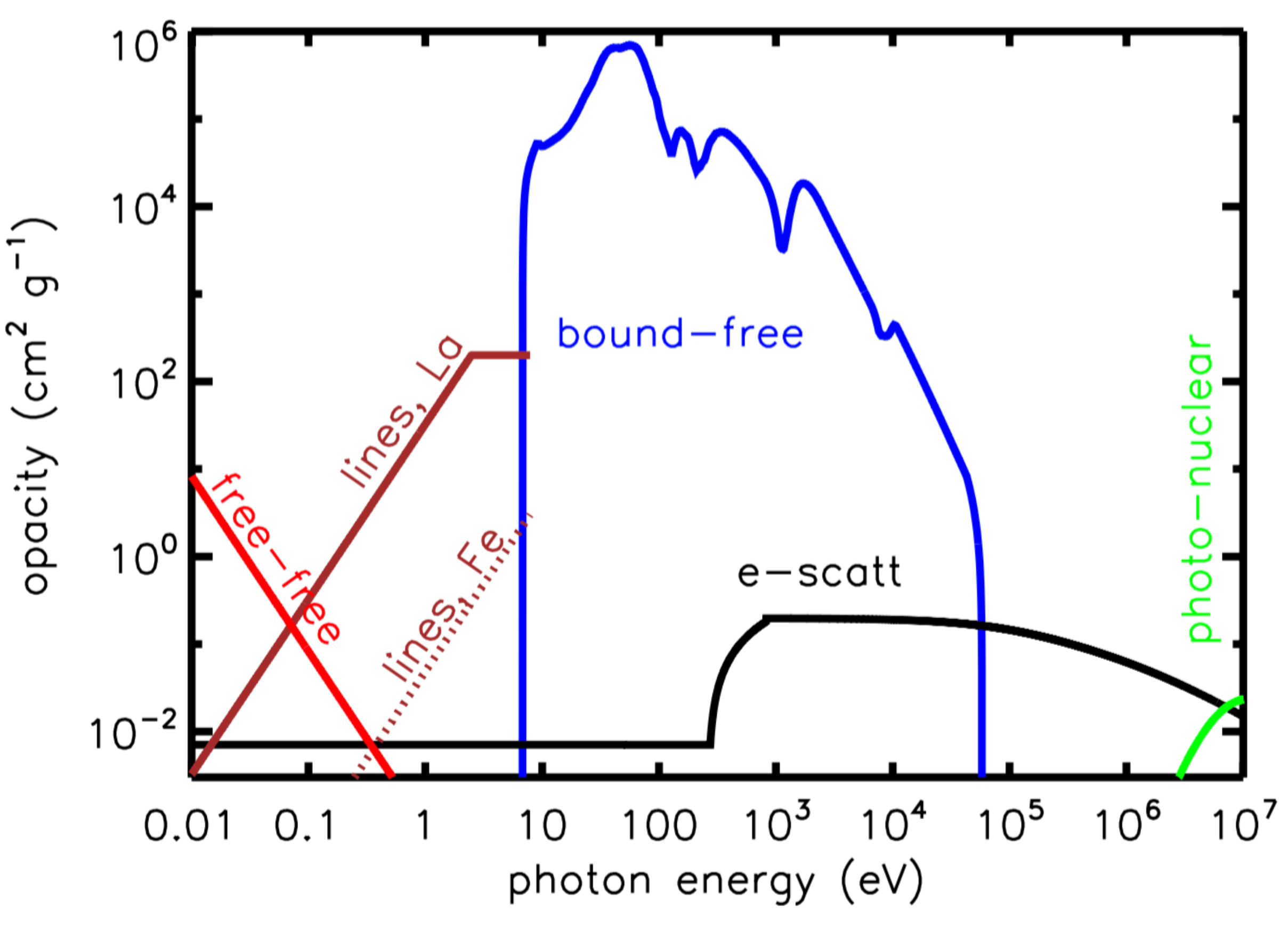}
\caption{Schematic illustration of the opacity of the NS merger ejecta as a function of photon energy near peak light.  The free-free opacity (red line) is calculated assuming singly-ionized ejecta of temperature $T = 2\times 10^{4}$ K and density $\rho = 10^{-14}$ g cm$^{-3}$, corresponding to the mean properties of 10$^{-2} M_{\odot}$ of ejecta expanding at $v = 0.1$ c at $t =$ 3 days.  Line opacities of Fe-like elements and lanthanide-rich elements are approximated from Figures 3 and 7 of \cite{Kasen+13}.  Bound-free opacities are estimated as that of neutral Fe \citep{Verner+96}, which we expect to crudely approximate the those of heavier $r$-process elements.  Electron scattering opacity accounts for the Klein-Nishina suppression at energies $\gg m_e c^{2}$ and (very schematically) for the rise in opacity that occurs above the keV energy scale due to all electrons (including those bound in atoms) contributing to the scattering opacity when the photon wavelength is smaller than the atomic scale.  At the highest energies, opacity is dominated by pair creation by gamma-rays interacting with the electric fields of nuclei in the ejecta (shown schematically for Xenon, $A = 131$, $Z = 54$).  Not included are possible contributions from $r$-process dust; or $\gamma-\gamma$ pair creation opacity at energies $\gg m_e c^{2}$, which is important for high compactness $\ell \gg 1$ (Eq.~\ref{eq:compactness}).  }
\label{fig:opacities}
\end{figure}

It's no coincidence that kilonova emission is centered in the optical/IR band, as this is among the first spectral windows through which the expanding merger ejecta becomes transparent.  Figure \ref{fig:opacities} provides a semi-quantitative illustration of the opacity of NS merger ejecta near peak light as a function of photon energy. 

 At the lowest frequencies (radio and far-IR), free-free absorption from ionized gas dominates, as shown with a red line in Fig.~\ref{fig:opacities}, and calculated for the approximate ejecta conditions three days post merger.  As the ejecta expands, the free-free opacity will decrease rapidly due to the decreasing density $\propto \rho \propto t^{-3}$ and the fewer number of free electrons as the ejecta cools and recombines.  The latter can be counteracted by photo-ionization from the central engine, which generally becomes more effective with time as the ejecta dilutes (see below).  

At near-IR/optical frequencies, the dominant source of opacity is a dense forrest of line (bound-bound) transitions.  The magnitude of this \emph{effective} continuum opacity is determined by the strengths and wavelength density of the lines, which in turn depend sensitively on the ejecta composition.  If the ejecta contains elements with relatively simple valence electron shell structures, such as iron, then the resulting opacity is comparatively low (dashed brown line), only moderately higher than the Fe-rich ejecta in Type Ia SNe \citep{Pinto&Eastman00}.  On the other hand, if the ejecta also contains even a modest fraction of elements with partially-filled f-shell valence shells, such as those in the lanthanide and actinide group, then the opacity can be an order of magnitude or more higher \citep{Kasen+13,Tanaka&Hotokezaka13,Fontes+16,Fontes+17}.  In both cases, the opacity rises steeply from the optical into the UV, due to the increasing line density moving to higher frequencies.  Based on Fig.~10 of \cite{Kasen+13}, we crudely approximate the Planck mean expansion opacity of lanthanide-bearing ejecta near the time of peak light as
\be
\kappa_r = \left\{
\begin{array}{lr}
200(T/4000\,{\rm K})^{5.5}\,\, {\rm cm}^{2} {\rm g}^{-1}
, &
{\rm 10^{3}\, K} < T < {\rm 4000\,K}\\
200\,\, {\rm cm}^{2} {\rm g}^{-1} &
 {\rm 4000\, K} < T < {\rm 10^{4}\, K} \\
\end{array}
\label{eq:kappar}
\right. ,
\ee

Considerable uncertainty remains in current calculations of the lanthanide/actinide opacities because the atomic states and line strengths of these complex elements are not measured experimentally.  Theoretically, such high$-Z$ atoms represent an unsolved problem in N-body quantum mechanics, with statistical models that must be calibrated to experimental data.  

Beyond identifying the line transitions themselves, there is considerably uncertainty in how to translate these data into an effective opacity.  The commonly employed ``line expansion opacity'' formalism \citep{Pinto&Eastman00}, based on the Sobolev approximation and applied to kilonovae by \cite{Barnes&Kasen13} and \cite{Tanaka&Hotokezaka13}, may break down if the line density is sufficiently high that the wavelength spacing of strong lines becomes comparable to the intrinsic thermal) width of the lines \citep{Kasen+13,Fontes+16,Fontes+17}.  Nevertheless, the qualitative dichotomy between the opacity of La/Ac-free and La/Ac-bearing ejecta is likely robust and may imprint diversity in the kilonova color evolution (Sect.~\ref{sec:blue}).        

Another uncertainty arises because, at low temperatures $\lesssim 10^{3}$ K, the ejecta may condense from gaseous to solid phase \citep{Takami+14}.  The formation of such `r-process dust' could act to either increase or decrease the optical/UV opacity, depending on uncertain details such as when the dust condenses and how quickly it grows.  Dust formation is already complex and poorly understood in less exotic astrophysical environments \citep{Cherneff&Dwek09,Lazzati&Heger16}.  

 Throughout the far UV and X-ray bands, bound-free transitions of the partially neutral ejecta dominates the opacity (blue line in Fig.~\ref{fig:opacities}, showing an approximation based on the opacity of iron).  This prevents radiation from escaping the ejecta at these frequencies, unless non-thermal radiation from the central magnetar or BH remnant remains luminous enough to re-ionize the ejecta (Sect.~\ref{sec:magnetar}).  The central engine luminosity $L_{\rm ion}$ required to ionize the ejecta at time $t$ is crudely estimated by balancing the rate of photo-ionization and radiative recombination.  Manipulating Eq.~(B9) of \cite{Metzger&Piro14}, we find
\be
L_{\rm ion} \approx 5\times 10^{45}\,{\rm erg\,s^{-1}}\left(\frac{t}{\rm 1 \,day}\right)^{-5}\left(\frac{M}{10^{-2}M_{\odot}}\right)^{3}\left(\frac{v}{\rm 0.3c}\right)^{-15/4}\left(\frac{T}{10^{5}\,{\rm K}}\right)^{-0.45},
\label{eq:Lion}
\ee
where $T$ is the electron temperature in the recombination layer.  

Extremely large luminosities are required to ionize the ejecta on timescales of days to a week near peak emission; however, the value of $L_{\rm ion}$ decreases rapidly with time, typically faster than the luminosity of the central engine (Fig.~\ref{fig:heating}, bottom panel), such that the ejecta may become ionized at late times.  Chances of such an ionization break-out are higher in the case of a long-lived magnetar remnant (Eq.~\ref{eq:Lsd}; \citealp{Metzger&Piro14}) than from fall-back accretion (Eq.~\ref{eq:Lxfb}).  In extreme cases in which the valence electrons of the lanthanide/actinide elements are completely ionized, this could substantially reduce the optical line opacity described above, suppressing or eliminating the hallmark infrared signal.  

At hard X-rays and gamma-ray energies, electron scattering, with Klein-Nishina corrections, provides an important opacity.  
At energies $\gg$ keV the scattering opacity is higher than at lower energies because, when the photon wavelength is smaller than the atomic scale ($\sim$ angstroms), one must include contributions from both free electrons and those bound into nuclei.  Also note that electron scattering becomes highly inelastic at energies approaching and exceeding $m_e c^{2}$.  The ejecta opacity to gamma-rays of energy $\sim$ MeV is important because it determines the efficiency with which radioactive $r$-process decay products thermalize their energy (Sect.~\ref{sec:rprocessheating}).

Ultra-high energy gamma-rays with $h\nu \gg m_e c^{2}$ can also interact with the more abundant lower energy optical or X-ray photons of energy $\ll m_e c^{2}$, leading to the creation of electron/positron pairs.  The importance of pair creation is quantified by the dimensionless ``compactness'' parameter,
\begin{eqnarray}
\ell \equiv \frac{U_{\rm rad}\sigma_T R}{m_e c^{2}} \simeq \frac{L \sigma_T}{4\pi R m_e c^{3}} 
 \approx 7\times 10^{-4}\left(\frac{L}{10^{41}\,{\rm erg\,s^{-1}}}\right)\left(\frac{v}{0.3\rm c}\right)^{-1}\left(\frac{t}{\rm 1\,day}\right)^{-1},
\label{eq:compactness}
\end{eqnarray}
which roughly defines the optical depth for pair creation assuming an order unity fraction of the engine luminosity is emitted in pair-creating photons.  Here $U_{\rm rad} \simeq L/(4\pi R^{2}c)$ is the energy density of seed photons, where $L$ is the central luminosity of seed photons.  For the most luminous kilonovae, powered by magnetar remnants with spin-down luminosities $L_X \gtrsim 10^{44}-10^{45}\mathrm{\ erg\ s}^{-1}$ (Sect.~\ref{sec:magnetar}), we can have $\ell \gg 1$ near the time of peak emission.  Pair creation is thus a potentially important agent, which can prevent the escape of the magnetar luminosity at very high photon energies.  

In addition to $\gamma-\gamma$ interactions, pair creation can occur due to the interaction of gamma-rays with nuclei in the ejecta.  For gamma-ray energies greatly exceeding the pair creation threshold of $\sim m_e c^{2}$, the opacity approaches a constant value $\kappa_{A\gamma} \approx \alpha_{\rm fs}\kappa_{T}(Z^{2}/A)$, where $\alpha_{\rm fs} \simeq 1/137$, and $A$ and $Z$ are the nuclear mass and charge, respectively.  For $r$-process nuclei with $Z^{2}/A \gtrsim 10-20$ we see that this dominates inelastic scattering at the highest gamma-ray energies.

\section{Unified Toy Model}
\label{sec:model}

Kilonova emission can be powered by a variety of different energy sources (Fig.~\ref{fig:heating}), including radioactivity and central engine activity.  This section describes a simple model for the evolution of the ejecta and its radiation, which we use to motivate the potential diversity of kilonova light curves.  Though ultimately no substitute for full, multi-dimensional, multi-group radiative transfer, this toy model does a reasonable job at the factor of a few level.  Some sacrifice in accuracy may be justified in order to facilitate a qualitative understanding, given the other uncertainties on the mass, heating rate, composition, and opacity of the ejecta.    

Following the merger, the ejecta velocity structure approaches one of homologous expansion, with the faster matter lying ahead of slower matter \citep{Rosswog+14}.  We approximation the distribution of mass with velocity greater than a value $v$ as a power-law,
\be
M_{v} = M(v/v_{\rm 0})^{-\beta},\,\,\,\, v \ge v_{0},
\label{eq:veldist}
\ee
where $M$ is the total mass, $v_{0} \approx 0.1$ c is the average ($\sim$ minimum) velocity.  We adopt a fiducial value of $\beta \approx 3$, motivated by a power-law fit to the dynamical ejecta in the numerical simulations of \citep{Bauswein+13}.  In general the velocity distribution derived from numerical simulations cannot be fit by a single power-law (e.g., Fig.~3 of \citealt{Piran+13}), but the following analysis can be readily extended to the case of an arbitrary velocity distribution.

In analogy with Eq.~(\ref{eq:tdiff}), radiation escapes from the mass layer $M_{v}$ on the diffusion timescale
\be
t_{d,v} \approx \frac{3 M_{v} \kappa_{v}}{4\pi \beta R c} \underset{R = vt}= \frac{M_{v}^{4/3}\kappa_{v}}{4\pi M^{1/3} v_{0} t c},
\label{eq:tdv}
\ee
where $\kappa_v$ is the opacity of the mass layer $v$ and in the second equality makes use of Eq.~(\ref{eq:veldist}) with $\beta = 3$.  Equating $t_{d,v} = t$ gives the mass depth from which radiation peaks for each time $t$,
\be
M_{v}(t)  = \left\{
\begin{array}{lr}
 M(t/t_{\rm peak})^{3/2}
, &
t < t_{\rm peak}\\
M &
t > t_{\rm peak} \\
\end{array}
\label{eq:Mv}
\right. ,
\ee
where $t_{\rm peak}$ is the peak time for diffusion out of the whole ejecta mass, e.g., Eq.~(\ref{eq:tpeak}) evaluated for $v = v_0$.  Emission from the outer layers (mass $M_v < M$) peaks first, while the luminosity of the innermost shell of mass $\sim M$ peaks at $t = t_{\rm peak}$.  The deepest layers usually set the peak luminosity of the total light curve, except when the heating rate and/or opacity are not constant with depth if the outer layers are free neutrons instead of $r$-process nuclei (Sect.~\ref{sec:neutrons}).

As the ejecta expands, the radius of each layer $M_{v}$ of mass $dM_{v}$ evolves according to 
\be \frac{dR_v}{dt} = v.
\ee
The thermal energy $E_v$ of the layer evolves according to
\be \frac{dE_v}{dt} = -\frac{E_v}{R_v}\frac{dR_v}{dt} - L_v + \dot{Q},
\label{eq:dEdt}
\ee
where the first term accounts for losses due to PdV expansion in the radiation-dominated ejecta.  The second term in Eq.~(\ref{eq:dEdt}), 
\be
L_{v} = \frac{E_v}{t_{d,v} + t_{lc,v}},
\ee 
accounts for radiative losses (the observed luminosity) and $t_{lc,v} = R_v/c$ limits the energy loss time to the light crossing time (this becomes important at late times when the layer is optically thin).  The third term in Eq.~(\ref{eq:dEdt}),
\be
\dot{Q}(t) = \dot{Q}_{r,v} + \dot{Q}_{\rm mag} + \dot{Q}_{\rm fb}
\ee
accounts for sources of heating, including radioactivity ($\dot{Q}_{r,v}$; Sect.~\ref{sec:rprocessheating}), a millisecond magnetar ($\dot{Q}_{\rm mag}$; Sect.~\ref{sec:magnetar}) or fall-back accretion ($\dot{Q}_{\rm fb}$; Sect.~\ref{sec:fallback}).  The radioactive heating rate, being intrinsic to the ejecta, will in general vary between different mass layers $v$.  In the case of magnetar or accretion heating, radiation must diffuse from the central cavity through the entire ejecta shell (Fig.~\ref{fig:schematic}, bottom panel).

One must in general also account for the time evolution of the ejecta velocity (Eq.~\ref{eq:veldist}) due to acceleration by pressure forces.  For radioactive heating, the total energy input $\int \dot{Q}_{r,v}dt$ is less than the initial kinetic energy of the ejecta \citep{Metzger+11,Rosswog+12}, in which case changes to the initial velocity distribution (Eq.~\ref{eq:veldist}) are safely ignored.  However, free expansion is not always a good assumption when there is substantial energy input from a central engine.  In such cases, the velocity of the central shell is evolved separately according to
\be
\frac{d}{dt}\left(\frac{M v_0^{2}}{2}\right) = Mv_0 \frac{dv_0}{dt} = \frac{E_{v_{0}}}{R_0}\frac{dR_0}{dt},
\label{eq:dvdt}
\ee
where the source term on the right hand side balances the PdV \emph{loss} term in the thermal energy equation (\ref{eq:dEdt}), and $R_0$ is the radius of the inner mass shell.  Equation (\ref{eq:dvdt}) neglects special relativistic effects, which are important for low ejecta masses $\lesssim 10^{-2}M_{\odot}$ and energetic engines, such as stable magnetars \citep{Zhang13,Gao+13,Siegel&Ciolfi16a,Siegel&Ciolfi16b}.\footnote{However, \cite{Metzger&Fernandez14} find that almost the entire mass of the remnant accretion disk ($\sim 0.1 M_{\odot}$) is ejected in the case of a long-lived remnant (Sect.~\ref{sec:ejecta}), in which case relativistic corrections remain relatively moderate even in this case.}

Assuming blackbody emission, the temperature of the thermal emission is 
\be T_{\rm eff} = \left(\frac{L_{\rm tot}}{4\pi \sigma R_{\rm ph}^{2}}\right)^{1/4},
\ee where $L_{\rm tot} = \Sigma (L_v dm_v)$ is the total luminosity (summed over all mass shells).  The radius of the photosphere $R_{\rm ph} (t)$ is defined as that of the mass shell at which the optical depth $\Sigma (\kappa_v dm_v) = 1$.  The flux density of the source at photon frequency $\nu$ is given by
\be
F_{\nu}(t) = \frac{2\pi h \nu^{3}}{c^{2}}\frac{1}{\exp\left[h\nu/kT_{\rm eff}(t)\right]-1}\frac{R_{\rm ph}^{2}(t)}{D^{2}},
\ee
where $D$ is the source distance (neglecting cosmological effects).  

The opacity $\kappa_v$ of each mass layer depends on its temperature, 
\be
T_{v} \simeq \left(\frac{3E_v}{4\pi a R_v^{3}}\right)^{1/4},
\ee
where we have assumed that the internal energy of the ejecta is dominated by radiation (easy to verify).  For Lanthanide-bearing ejecta  ($Y_e \lesssim 0.30$) we approximate the opacity using the approximate fit from Eq.~(\ref{eq:kappar}), based on \cite{Kasen+13}.  For Lanthanide-free ejecta  ($Y_e \gtrsim 0.30$) we adopt the same temperature dependence as in the Lanthanide case, but with a normalization which is 100 times smaller (Sect.~\ref{sec:opacity}). 

The full emission properties are determined by solving Eq.~(\ref{eq:dEdt}) for $E_v$, and hence $L_v$, for a densely sampled distribution of shells of mass $dM_v$ and velocity $v > v_0$.  When considering  radioactive heating acting alone, one can fix the velocity distribution (Eq.~\ref{eq:veldist}).  For an energetic engine, the velocity of the central shell is evolved simultaneously using Eq.~(\ref{eq:dvdt}).  As initial conditions at the ejection radius $R(t = 0) \approx 100$ km, it is reasonable to assume the thermal energy of the ejecta is comparable to the kinetic energy, $E_{v}(t = 0) \sim (1/2)dM_v v^2(t=0)$.  The emission properties at much later times near peak are insensitive to this assumption because the initial thermal energy is quickly removed by adiabatic expansion: one could take the initial thermal energy to be zero and obtain a similar result for the light curve near and after peak emission.  

\subsection{$R$-Process Heating}
\label{sec:rprocessheating}
At a minimum, the ejecta receives heating from the radioactive decay of heavy nuclei synthesized in the ejecta by the $r$-process.  This occurs at a rate
\be
\dot{Q}_{r,v} = dM_v X_{r,v} \dot{e}_r(t),
\label{eq:qdotr}
\ee
where $X_{r,v}$ is the r-process mass fraction in mass layer $M_v$ and $e_r$ is the specific heating rate.  For neutron-rich ejecta ($Y_e \lesssim 0.2$), the latter is reasonably approximated by the fitting formula \citep{Korobkin+12}
\be
\dot{e}_r  = 4\times 10^{18}\epsilon_{th,v} \left(0.5-\pi^{-1}\arctan[(t-t_0)/\sigma]\right)^{1.3}\,{\rm erg\,s^{-1}\,g^{-1}},
\label{eq:edotr}
\ee
where $t_0 = 1.3$ s and $\sigma = 0.11$ s are constants, and $\epsilon_{\rm th,m}$ is the thermalization efficiency.  Equation (\ref{eq:edotr}) predicts a constant heating rate for the first $\sim$ 1 second (while neutrons are being consumed during the $r$-process), followed by a $\propto t^{-1.3}$ decay at later times as nuclei decay back to stability \citep{Metzger+10, Roberts+11}; see Fig.~\ref{fig:heating}.  The time dependence is more complicated for higher $0.2 \lesssim Y_e \lesssim 0.4$, with `bumps' and `wiggles' caused by the heating rate being dominated by a few discrete nuclei instead of the large statistical ensemble present at low $Y_e$ \citep{Korobkin+12,Martin+15}.  However, when averaged over a realistic $Y_e$ distribution, the heating rate on timescales of days-weeks (of greatest relevance to the peak luminosity; Eq.~\ref{eq:Arnett}), is constant to within a factor of a few for $Y_e \lesssim 0.4$ \citep[][their Fig.~7]{Lippuner&Roberts15}.  The radioactive decay rate is also largely insensitive to uncertainties in the assumed nuclear masses, cross sections, and fission fragment distribution (although the $r$-process abundance pattern will be, \citealp{Eichler+15,Wu+16,Mumpower+16}).

Radioactive heating occurs through a combination of $\beta$-decays, $\alpha$-decays, and fission \citep{Metzger+10,Barnes+16,Hotokezaka+16b}.  The thermalization efficiency $\epsilon_{th,v}$ depends on how these decay products share their energy with the thermal plasma.  Neutrinos escape from the ejecta without interacting; $\sim$ MeV gamma-rays are trapped at early times ($\lesssim 1$ day), but they leak out at later times due to the comparatively low Klein-Nishina opacity (Fig.~\ref{fig:opacities}; \citealp{Hotokezaka+16b,Barnes+16}).  $\beta$-decay electrons, $\alpha-$particles, and fission fragments share their kinetic energy effectively with the ejecta via Coulomb collisions \citep{Metzger+10} and through ionization \citep{Barnes+16}.  However, for a fixed energy release rate, the thermalization efficiency is smallest for $\beta-$decay, higher for $\alpha-$decay, and the highest for fission fragments.  The thermalization efficiency of charged particles depends on the magnetic field orientation within the ejecta, since the particle Larmor radius is generally shorter than the Coulomb mean free path.  \cite{Barnes+16} find that the quantity of actinides produced around $A \sim 230$ varies significantly with the assumed nuclear mass model, such that the effective heating rate can vary by a factor of 2\,--\,6, depending on time.  This is because the actinides decay by alpha decay, which contribute more to the total effective heating than the energy released by beta decays.

\cite{Barnes+16} find that the combined efficiency from all of these processes typically decreases from $\epsilon_{ th,v} \sim 0.5$ on a timescale of 1 day to $\sim 0.1$ at $t \sim 1$ week (their Fig.~13).  In what follows, we adopt the fit provided in their Table 1,
\be
\epsilon_{th,v}(t) = 0.36\left[\exp(-a_v t_{\rm day})  + \frac{{\rm ln}(1+2b_v t_{\rm day}^{d_v})}{2b_v t_{\rm day}^{d_v}}\right],
\label{eq:eth}
\ee
where $t_{\rm day} = t/1$ day, and $\{a_v,b_v,d_v\}$ are constants that will in general depend on the mass and velocity of the layer under consideration.  For simplicity, we adopt fixed values of $a_v = 0.56, b_v = 0.17, c_v = 0.74$, corresponding to a layer with $M = 10^{-2}M_{\odot}$ and $v_0 = 0.1$ c.

\subsubsection{Red Kilonova: Lanthanide-Bearing Ejecta}
\label{sec:vanilla}

All NS-NS and BH-NS mergers capable of producing bright EM counterparts eject at least some highly neutron-rich matter ($Y_e < 0.30$), which will form heavy $r$-process nuclei.  This Lanthanide-bearing high-opacity material resides within tidal tails in the equatorial plane, or in more spherical outflows from the accretion disk in cases when BH formation is prompt or the HMNS phase is short-lived (Fig.~\ref{fig:schematic}, top panel).

\begin{figure}[!t]
\includegraphics[width=0.5\textwidth]{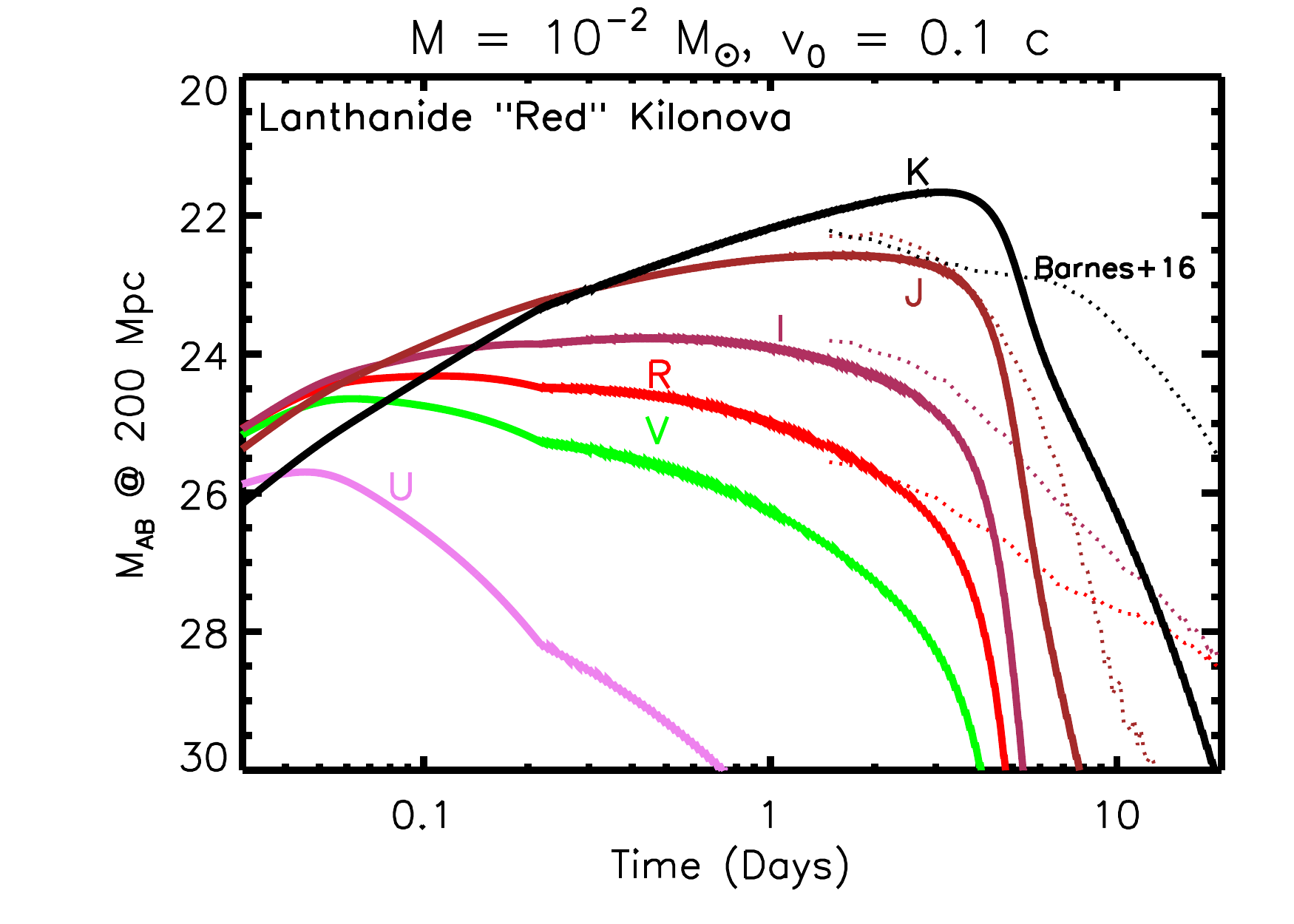}
\includegraphics[width=0.5\textwidth]{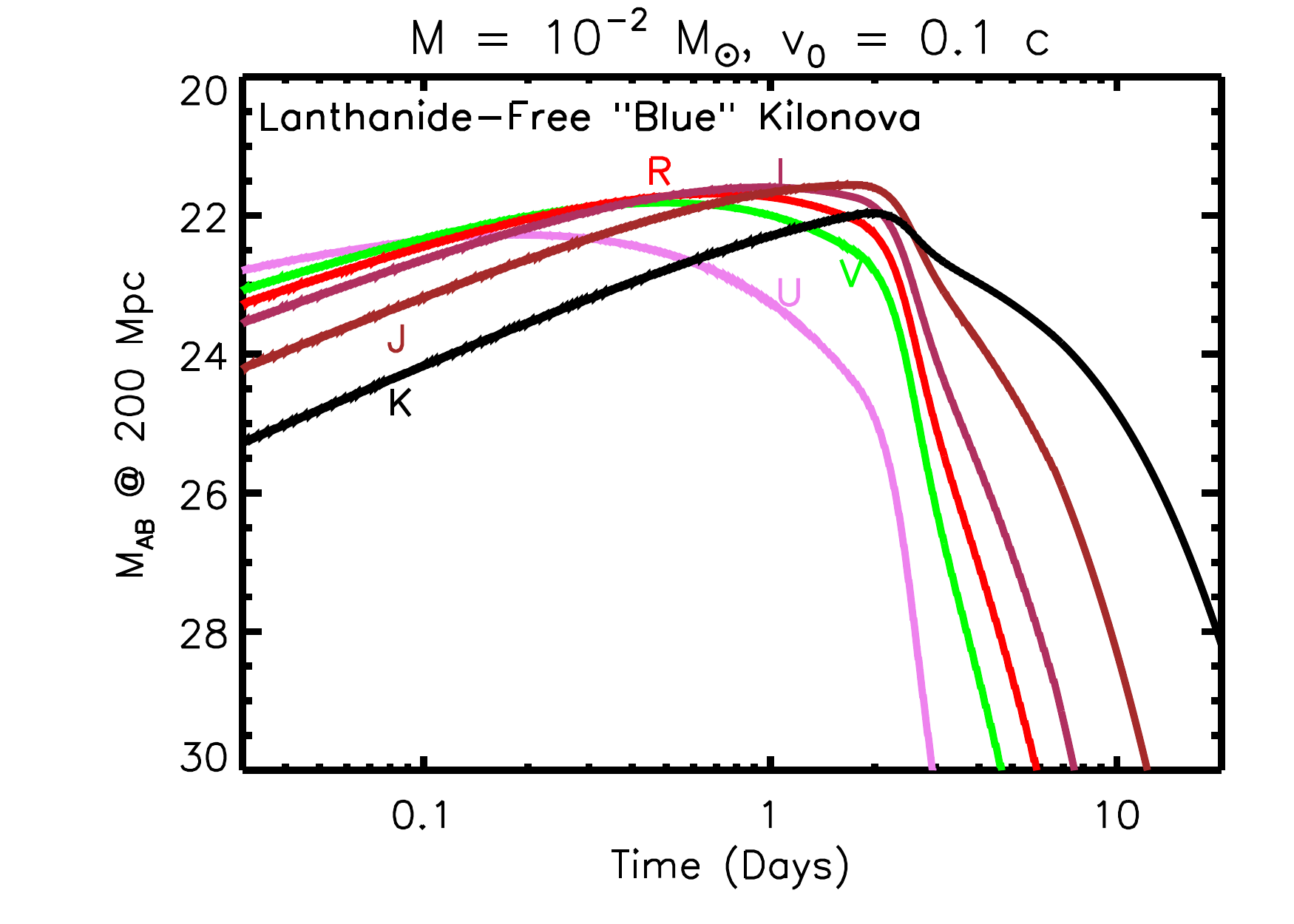}
\caption{Kilonova light curves in AB magnitudes for a source at 200 Mpc, calculated using the toy model presented in Sect.~\ref{sec:model}, assuming a total ejecta mass $M = 10^{-2}$ and minimum velocity $v_0 = 0.1$ c.  The top panel shows a standard ``red'' kilonova, corresponding to very neutron-rich ejecta with Lanthanide elements, while the bottom panel shows a ``blue'' kilonova produced by ejecta without Lanthanides.   Shown for comparison in the red kilonova case with dashed lines are models from \cite{Barnes+16} for $v = 0.1$ c and $M = 10^{-2}M_{\odot}$.  Depending on the viewing angle of the observer, both red and blue emission components may  be present in a single merger, if they originate from different locations in the ejecta (Fig.~\ref{fig:schematic}).}
\label{fig:vanilla}
\end{figure}

The top panel of Fig.~\ref{fig:vanilla} shows an example light curve of such a `red' kilonova, calculated from the toy model assuming an ejecta mass $M = 10^{-2}M_{\odot}$, minimum velocity $v_0 = 0.1$ c, and velocity index $\beta =3$.  For comparison, dashed lines show light curves calculated from \cite{Barnes+16}, based on a full 1D radiative transfer calculation, for similar parameters.  The emission is seen to peak are NIR wavelengths on a timescale of several days to a week at J and K bands (1.2 and 2.2 $\mu$m, respectively).  The significant suppression of the emission at optical wavebands $RVI$ due to the high opacity illustrates the great challenge to GW follow-up programs posed by kilonovae, at these for these most conservative (vanilla) models.    

The abrupt post-maximum light curve drop-off in our toy models disagrees with the smoother decline predicted by \cite{Barnes+16}.  These differences may result because of our toy model approximation of optically-thick blackbody emission, which breaks down at late times as the ejecta becomes optically-thin due to the strong temperature sensitivity of the assumed opacity.  In reality, other sources of post-maximum opacity, such as dust formation or additional electron scattering due to photo-ionization from the central engine, could also act to smooth the light curve decline as compared to the toy model predictions.

\subsubsection{Blue Kilonova: Lanthanide-Free Ejecta }
\label{sec:blue}

In addition to the highly neutron-rich ejecta ($Y_e \lesssim 0.30$), there are growing indications from simulations that some of the matter which is unbound from a NS-NS merger is less neutron rich ($Y_e \gtrsim 0.30$; e.g., \citealp{Wanajo+14,Goriely+15}) and thus will be free of Lanthanide group elements \citep{Metzger&Fernandez14}.  This low-opacity ejecta can reside either in the polar regions, due to dynamical ejection from the NS-NS merger interface, or in more isotropic outflows from the accretion disk in cases when BH formation is significantly delayed (Fig.~\ref{fig:schematic}, bottom panel).

The bottom panel of Fig.~\ref{fig:vanilla} shows an otherwise identical calculation to that presented in the last section, but assuming a lower opacity appropriate to Lanthanide-free ejecta.  The emission now peaks at the visual bands R and I, on a timescale of about 1 day at a level 2\,--\,3 magnitudes brighter than the Lanthanide-rich case.  In general, the total kilonova emission from a NS-NS merger will be a combination of `blue' and `red' components, as both high- and low-$Y_e$ ejecta components could be visible for viewing angles close to the binary rotation axis (Fig.~\ref{fig:schematic}).  For equatorial viewing angles, the blue emission is likely to be blocked by the higher opacity of the lanthanide-rich equatorial matter \citep{Kasen+15}.  Thus, although the week-long NIR transient is fairly generic, an early blue kilonova will be observed in only a fraction of mergers.  

\subsubsection{Free Neutron Precursor}
\label{sec:neutrons}

\begin{figure}[!t]
\includegraphics[width=0.5\textwidth]{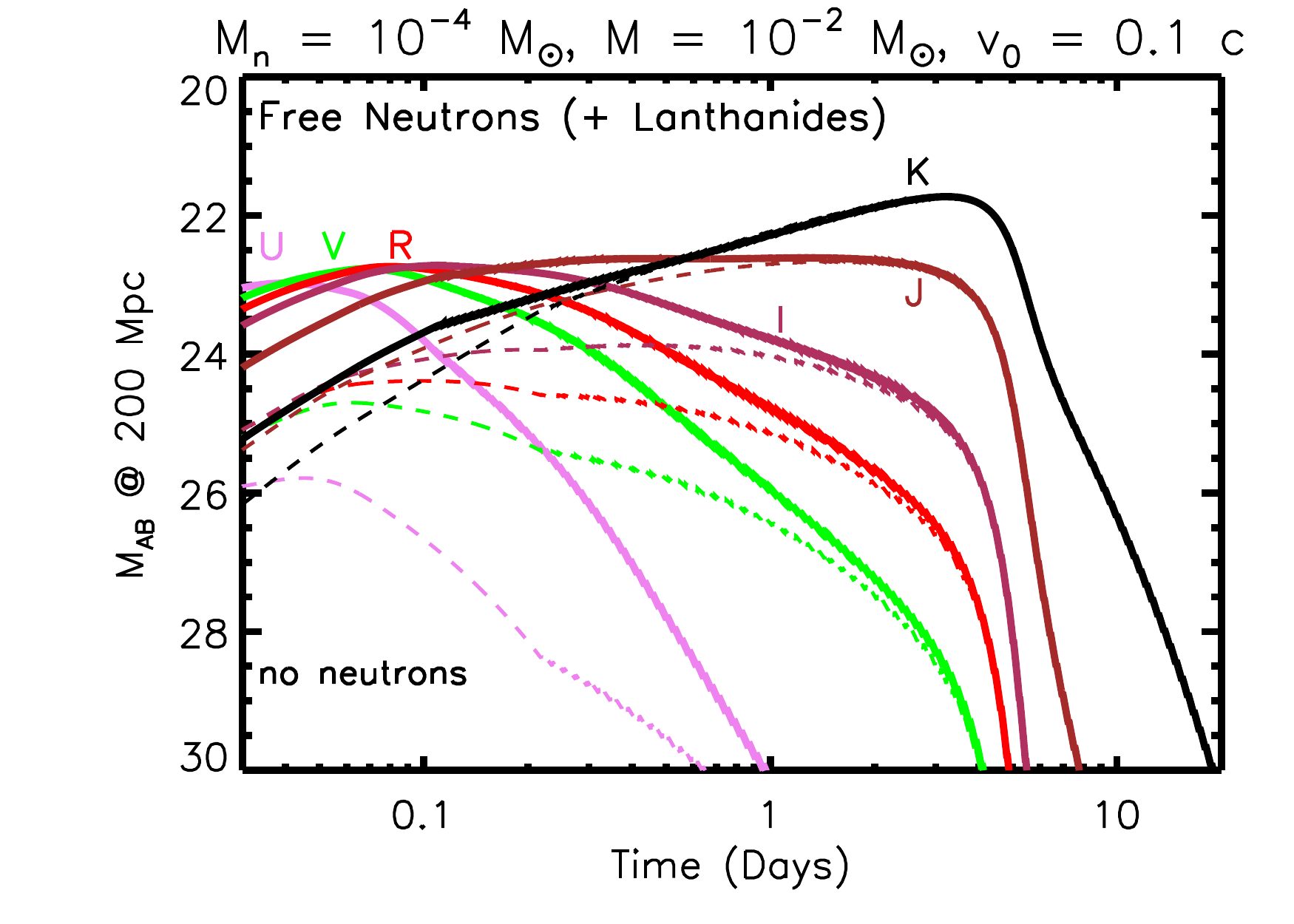}
\includegraphics[width=0.5\textwidth]{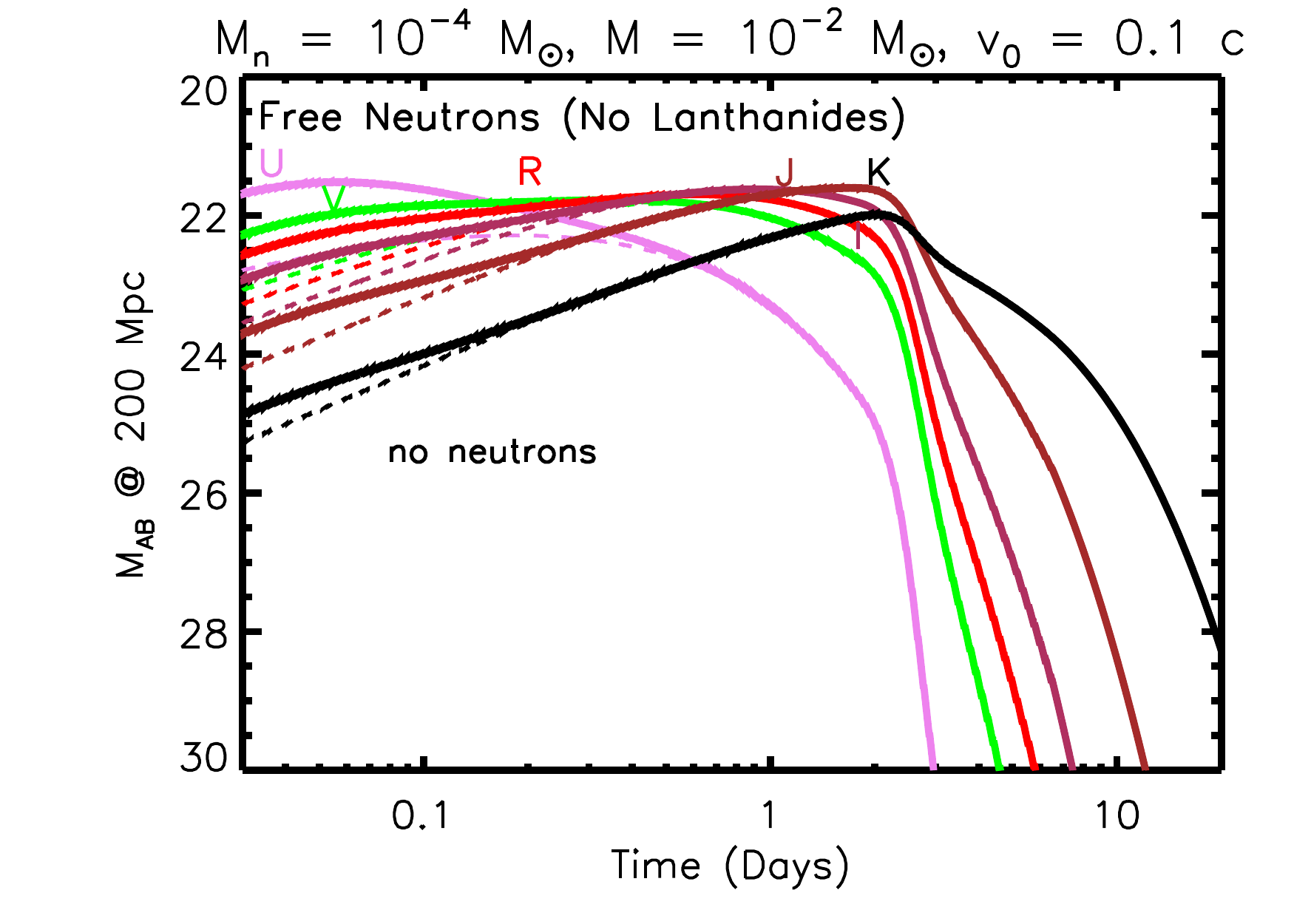}
\caption{Kilonova light curves, including the presence of free neutrons in the outer $M_{\rm n} = 10^{-4}M_{\odot}$ mass layers of the ejecta (``neutron precusors''), calculated for the same parameters of total ejecta mass $M = 10^{-2}$ and velocity $v_0 = 0.1$c  used in Fig.~\ref{fig:vanilla}.  The top panel shows a calculation with an opacity appropriate to lanthanide-bearing nuclei, while the bottom panel shows an opacity appropriate to lanthanide-free ejecta.  Models without a free neutron layer ($M_n = 0$; Fig.~\ref{fig:vanilla}) are shown for comparison with dashed lines.}
\label{fig:neutrons}
\end{figure}

The vast majority of the ejecta from a NS-NS merger remains sufficiently dense during its expansion that all neutrons are captured into nuclei during the $r$-process on a timescale of $\sim 1$ s.  However, recent NS-NS merger simulations show that a small fraction of the dynamical ejecta (typically a few percent, or $\sim 10^{-4}M_{\odot}$) expands sufficiently rapidly that the neutrons do not have time to be captured into nuclei  \citep{Bauswein+13}.  This fast expanding matter, which reaches asymptotic velocities $v \gtrsim 0.4-0.5$ c, originates from the shock-heated interface between the merging stars and resides on the outermost layers of the polar ejecta.  This `neutron skin' can super-heat the outer layers of the ejecta, enhancing the early kilonova emission \citep{Metzger+15,Lippuner&Roberts15}.  Before proceeding, it is important to emphasize that the presence of this fast-expanding matter is still highly speculative, and confirming or refuting its presence in actual merger events will require additional simulation work.  

Ejecta containing free neutrons experiences a radioactive heating rate of
\be
\dot{Q}_{r,v} = dM_v X_{n,v}\dot{e}_n(t),
\ee
where the initial mass fraction of neutrons,
\be 
X_{n,v} = \frac{2}{\pi} (1-Y_e)\arctan\left(\frac{M_{n}}{M_v}\right),
\label{eq:Xnv}
\ee
is interpolated in a smooth (but otherwise ad-hoc) manner between the neutron-free inner layers at $M \gg M_n$ and the neutron-rich outer layers $M \ll M_n$, which have a maximum mass fraction of $1- 2Y_e$.  The specific heating rate due to neutron $\beta-$decay (accounting for energy loss to neutrinos) is given by
\be
\dot{e}_n = 3.2\times 10^{14}\exp[-t/\tau_{n}]\,{\rm erg\,s^{-1}\,g^{-1}},
\label{eq:edotn}
\ee
where $\tau_n \approx 900$ s is the neutron half-life.  The rising fraction of free neutrons in the outermost layers produces a corresponding decreasing fraction of $r$-process nuclei in the outermost layers, i.e., $X_{r,v} = 1-X_{n,v}$ in calculating the $r$-process heating rate from Eq.~(\ref{eq:qdotr}).  

Figure \ref{fig:neutrons} shows kilonova light curves, including an outer layer of neutrons of mass $M_n = 10^{-4}M_{\odot}$ and electron fraction $Y_e = 0.1$.  In the top panel, we have assumed that the $r$-process nuclei which co-exist with the neutrons contain lanthanides, and hence would otherwise (absent the neutrons) produce a ``red'' kilonova.  Neutron heating acts to substantially increase the UVR luminosities on timescales of hours after the merger (the otherwise identical case without free neutrons is shown for comparison with a dashed line).  Even compared to the early emission from otherwise lanthanide-free ejecta (``blue kilonova''), the neutrons increase the luminosity during the first few hours by a magnitude or more, as shown in the bottom panel of Fig.~\ref{fig:neutrons}.  

It might seem counter-intuitive that heating from such a small layer of neutrons can have such a substantial impact on the light curve.  First, note that the \emph{specific} heating rate due to free neutrons $\dot{e}_n$ (Eq.~\ref{eq:edotn}) exceeds that due to $r$-process nuclei $\dot{e}_r$ (Eq.~\ref{eq:edotr}) by over an order of magnitude on timescales $\sim 0.1-1$ hr after the merger.  Coincidentally, this timescale is comparable to the photon diffusion depth from the inner edge of the neutron mass layer.  Indeed, setting $t_{\rm d,v} = t$ in Eq.~(\ref{eq:tdv}), the emission from mass layer $M_v$ peaks on a timescale
\begin{eqnarray}
&&t_{\rm peak,v} \approx \left(\frac{M_{v}^{4/3}\kappa_{v}}{4\pi M^{1/3} v_{0}  c}\right)^{1/2} \nonumber \\
 &\approx& 3.7\,{\rm hr}\left(\frac{M_v}{10^{-5}M_{\odot}}\right)^{2/3}\left(\frac{\kappa_v}{100\,{\rm cm^{2}\,g^{-1}}}\right)^{1/2}\left(\frac{v_0}{0.1\, \rm c}\right)^{-1/2}\left(\frac{M}{10^{-2}M_{\odot}}\right)^{-1/6} 
\label{eq:tpeakv}
\end{eqnarray}The total energy energy released by neutron-decay is $E_n \simeq \int \dot{e}_n M_{\rm n} dt \approx 6\times 10^{46}(M_{\rm n}/10^{-4}M_{\odot})\mathrm{\ erg}$ for $Y_e \ll 0.5$.  Following adiabatic losses, a fraction $\tau_{\rm n}/t_{\rm peak,v} \sim 0.01-0.1$ of this energy is available to be radiated over a timescale $\sim t_{\rm peak,v}$.  The peak luminosity of the neutron layer is thus approximately
\begin{eqnarray}
&& L_{\rm peak,n} \approx  \frac{E_n \tau_n}{t_{\rm peak,v}^{2}} \nonumber \\
&\approx& 3\times 10^{41}\,{\rm erg\,s^{-1}}\left(\frac{M_{v}}{10^{-5}M_{\odot}}\right)^{-1/3}\left(\frac{\kappa_v}{100\,{\rm cm^{2}\,g^{-1}}}\right)^{-1}\left(\frac{v_0}{0.1\, \rm c}\right)\left(\frac{M}{10^{-2}M_{\odot}}\right)^{1/3}, \nonumber\\
\label{eq:Lpeakn}
\end{eqnarray}
and hence is relatively insensitive to the mass of the neutron layer, $M_{v} = M_{\rm n}$.  As important as the peak luminosity itself, which is $\sim 10$ times higher than that of the main kilonova peak, is the high temperature of the ejecta during the first hours of the merger.  This may place the emission squarely in the optical/UV band, even in the presence of---or, in fact, partly thanks to---the high Lanthanide opacity.

Additional work is sorely needed to assess the robustness of the fast-moving ejecta and its abundance of free neutrons, which thus far has been seen in a single numerical code \citep{Bauswein+13}.\footnote{SPH simulations are generally better at following such a small quantity of ejecta, given the limitations imposed by the background density floor in grid-based codes.  Furthermore, general relativistic gravity appear to be critical in producing the high collision velocity between the merging NSs, which gives rise to the shock-heated polar ejecta responsible for the fast-expanding material in the simulations of \cite{Bauswein+13}.  On the other hand, SPH codes are only solving approximations to the hydrodynamical equation, and the simulations of \cite{Bauswein+13} include only the conformal flat approximation to GR.}  The freeze-out of the $r$-process, and the resulting abundance of free neutrons, is also sensitive to the expansion rate of the ejecta, which must currently be extrapolated from the merger simulations (running at most tens of milliseconds) to the much longer timescales of $\sim$ 1 second over which neutrons would nominally be captured into nuclei.  Figure \ref{fig:neutrons} and Eq.~(\ref{eq:Lpeakn}) also makes clear that the neutron emission is sensitive to the opacity of the ejecta at early stages, when the temperatures and ionization states are furthermore higher than those employed in kilonova calculations to date \citep{Kasen+13}.

\subsection{Engine Power}
\label{sec:engine}

The end product of a NS-NS or BH-NS merger is a central compact remnant, either a BH or a massive NS.  Sustained energy input from this remnant provides an additional potential source of heating and kilonova emission.  Though more speculative and uncertain than radioactive heating, this possibility is important to consider because a central engine has the potential to produce a much more luminous signal.

Much evidence exists for late-time central engine activity following short GRBs, on timescales from minutes to days.  A fraction $\approx 15-25\%$ of \textit{Swift} short bursts are followed by a prolonged `hump' of X-ray emission lasting for tens to hundreds of seconds \citep{nb06,Perley+09,Kagawa+15}.   The isotropic X-ray light curve of such `extended emission' in GRB 080503 is shown in the bottom panel of Fig.~\ref{fig:heating} (\cite{Perley+09}; Fig.~\ref{fig:magnetarjet}).  Other GRBs exhibit a temporary flattening or ``plateau'' in their X-ray afterglows lasting $\approx 10^2-10^3$ seconds \citep{nkg+06}.  X-ray flares have been observed on even later timescales of $\sim$few days \citep{Perley+09,fbm+14}.  The power output of the engine which is required to explain this emission depends on the radiative efficiency and uncertain beaming fraction of the (potentially jetted) X-ray emission, and hence is uncertain by several orders of magnitude.  A comparison of the top and bottom panels of Fig.~\ref{fig:heating} should nevertheless make clear that central engine heating, though subject to large uncertainties, could well dominate that of radioactivity.

\subsubsection{Fall-Back Accretion}
\label{sec:fallback}

\begin{figure}[!t]
\includegraphics[width=0.5\textwidth]{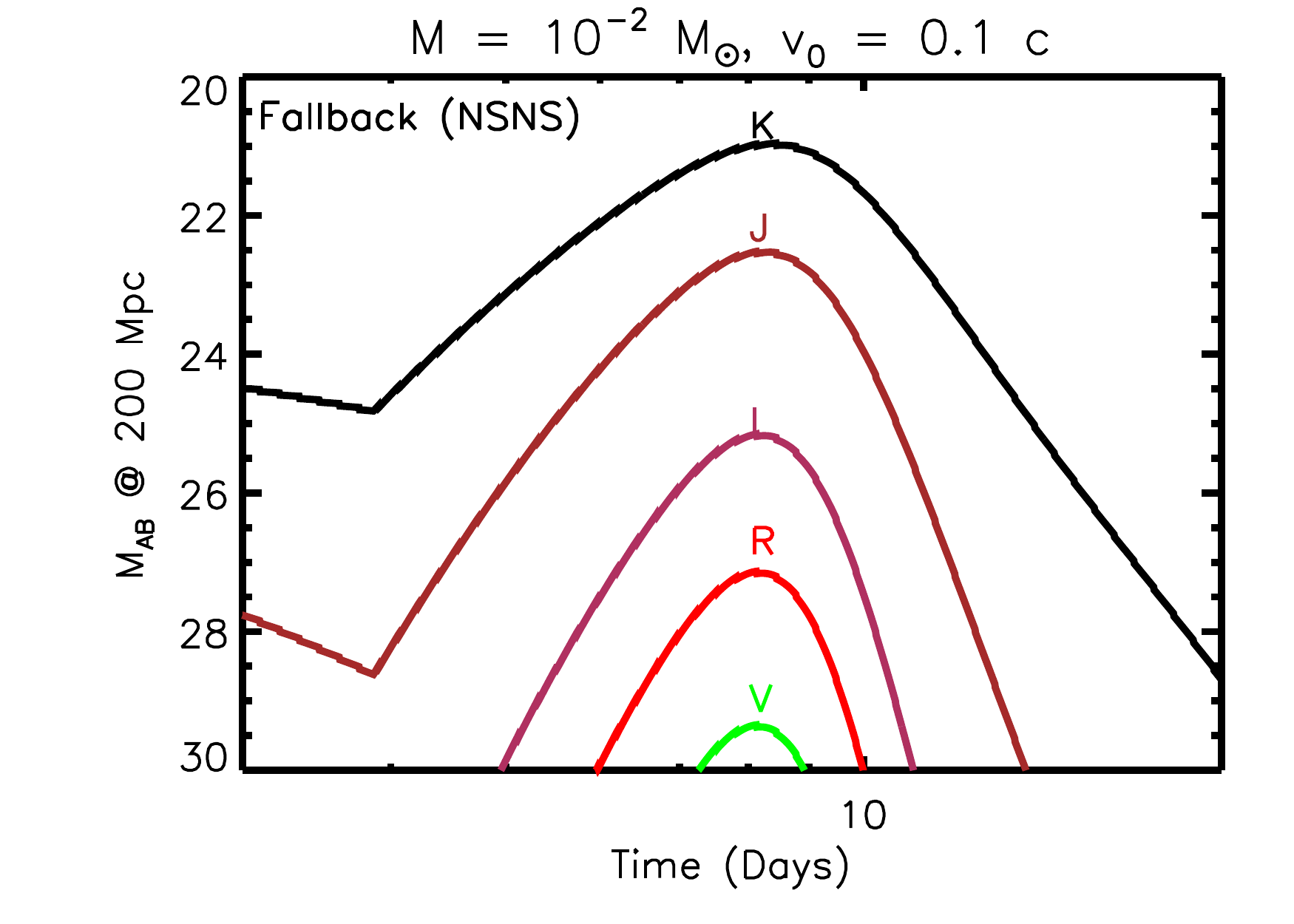}
\includegraphics[width=0.5\textwidth]{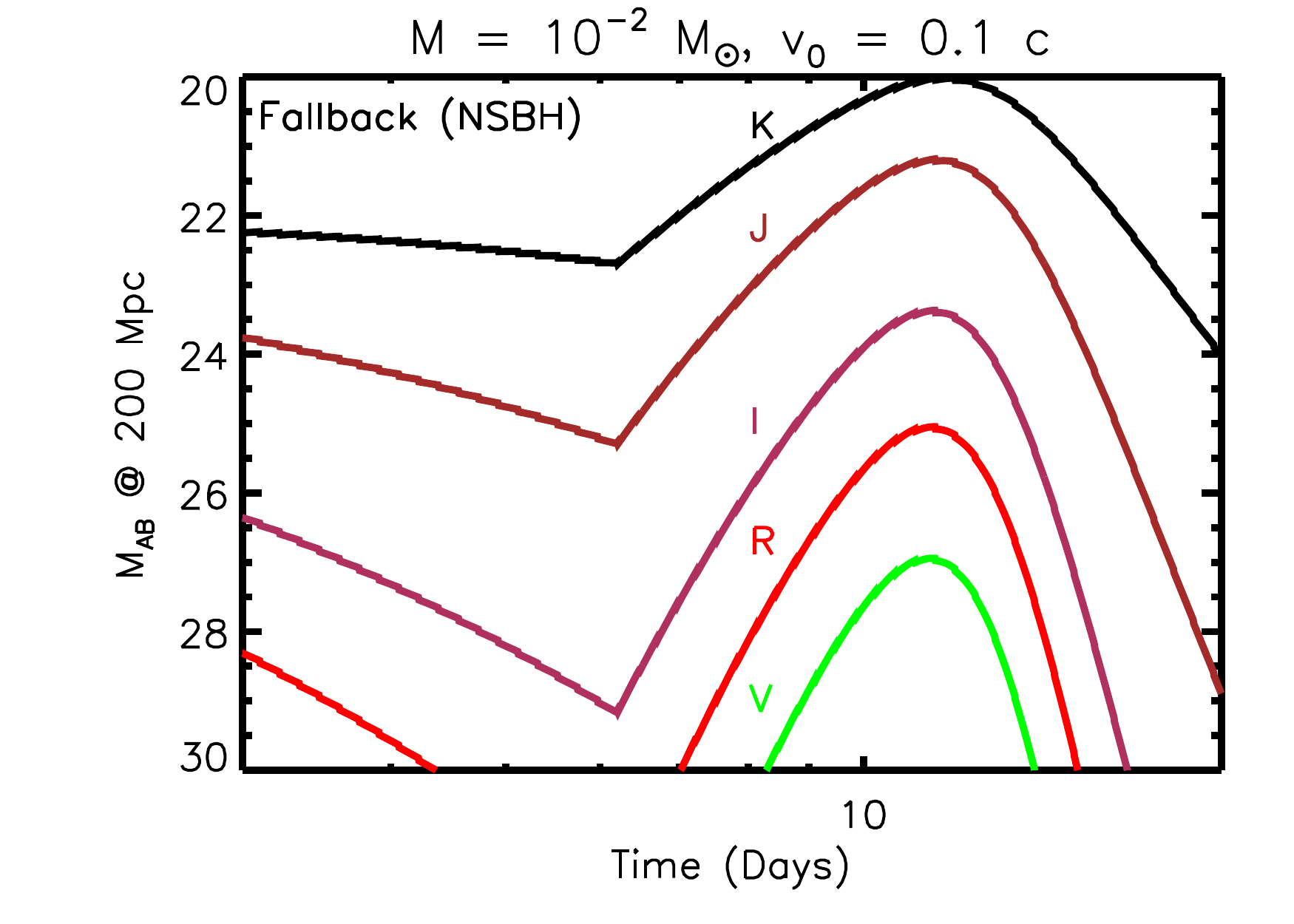}
\caption{Kilonova light curves powered by fall-back accretion, calculated for the same parameters of total ejecta mass $M = 10^{-2}$ and velocity $v_0 = 0.1$c used in Fig.~\ref{fig:vanilla}, and for an opacity appropriate to lanthanide-bearing nuclei.  We adopt an ejecta heating rate from Eq.~(\ref{eq:Lxfb}) for a fixed efficiency $\epsilon_{\rm j} = 0.1$.  We normalize the mass fall-back rate to a value of $\dot{M}_{\rm fb}(t = 0.1) = 10^{-3}M_{\odot}$ s$^{-1}$ in the case of NS-NS mergers (top panel), and to a value 10 times higher in BH-NS mergers (bottom panel), based on \cite{Rosswog07}.}
\label{fig:fallback}
\end{figure}

In addition to the ejecta which is unbound from a NS-NS/BH-NS merger, a comparable or greater quantity of mass remains gravitationally bound, falling back to the remnant over a range of timescales from seconds to days or longer \citep{Rosswog07,Rossi&Begelman09,Chawla+10,Kyutoku+15}.  At late times $t \gg 0.1$ s, the mass fall-back rate decays as a power-law
\be
\dot{M}_{\rm fb} \approx \left(\frac{\dot{M}_{\rm fb}(t = 0.1\,{\rm s})}{10^{-3}M_{\odot}\,s^{-1}}\right)\left(\frac{t}{0.1\,{\rm s}}\right)^{-5/3},
\label{eq:mdotfb}
\ee
where the normalization $\dot{M}_{\rm fb}(t = 0.1)$ at the reference time $t = 0.1$ s can vary from $\sim 10^{-3}M_{\odot}\,{\rm s}^{-1}$ in NS-NS mergers, to values up to an order of magnitude larger in BH-NS mergers \citep{Rosswog07}.  The decay exponent of 5/3 is only expected at late times if the mass distribution of the ejecta $dM/dE$ is constant with energy $E$ for marginally bound matter ($E \approx 0$; \citealt{Phinney89}).  This condition appears to be approximately satisfied for the dynamical ejecta from NS-NS \citep{Rosswog07} and NS-BH mergers \citep{Foucart+15}, though additional studies of the energy distribution of the ejecta warranted.

Hydrodynamical simulations of the interaction between fall-back accretion and the inner accretion flow show that disk winds are sufficiently powerful to stifle the fall-back material from reaching the BH on timescales $t \gtrsim 100$ ms \citep{Fernandez+15}.  Sustained heating due to the $r$-process over the first $\sim 1$ second can also unbind matter which is originally marginally-bound, causing a cut-off in the fall-back rate after a timescale of seconds or minutes \citep{Metzger+10c}.  However, it seems unlikely that fall-back will be completely suppressed on the much longer timescales of $t \sim$ days-weeks, which are most relevant to kilonovae.

If matter reaches the central compact object at the rate $\dot{M}_{\rm fb}$ (Eq.~\ref{eq:mdotfb}), then a fraction of the resulting accretion power $L_{\rm acc} \propto \dot{M}_{\rm fb}c^{2}$ could be available to heat the ejecta, enhancing the kilonova emission.  The still highly super-Eddington accretion flow could power a collimated ultra-relativistic jet, similar to that responsible for the earlier GRB.  At early times, such a jet is sufficiently powerful to propagate through the ejecta, producing high energy emission at larger radii powering the `extended X-ray emission' following the GRB).  However, as the jet power decreases in time, the jet is more likely to become unstable to the magnetic kink instability \citep{Bromberg&Tchekhovskoy16}, in which case its energy will instead be dissipated by magnetic reconnection, and ultimately as heat behind the ejecta.  The fall-back accretion flow may also power a mildly relativistic, wider-angle disk wind, which carries a substantial fraction of the accretion power.  This wider angle wind could collide with the (slower, but higher mass) ejecta shell, thermalizing a large fraction of its kinetic energy.  

In either of the cases described above, the heating rate of the ejecta due to fall-back accretion can be parametrized as follows,
\be
\dot{Q}_{\rm fb}= \epsilon_{j} \dot{M}_{\rm fb} c^{2} \approx 2\times 10^{51}\,{\rm erg\,s^{-1}}\left(\frac{\epsilon_{j}}{0.1}\right)\left(\frac{\dot{M}_{\rm fb}(0.1{\rm s}}{10^{-3}M_{\odot}\,s^{-1}}\right)\left(\frac{t}{\, \rm 0.1 s}\right)^{-5/3},
\label{eq:Lxfb}
\ee
where $\epsilon_{j}$ is a jet/disk wind efficiency factor.\footnote{It is typically assumed that the efficiency $\epsilon_j$ is constant in time.  However, if the jet power derives from the Blandford--Znajek process, its luminosity actually depends more sensitively on the magnetic flux threading the BH than the accretion rate, at least until the flux exceeds the critical value for which the jet power saturates ($\epsilon_j \approx 1$; \citealp{Tchekhovskoy+11}).  \citet{Kisaka&Ioka15} show that the topology of the magnetic field expected in the fallback debris could give rise to a complex temporal evolution of the jet power, which differs greatly from the $\propto t^{-5/3}$ decay predicted by Eq.~(\ref{eq:Lxfb}) for constant $\epsilon_j$.}
For characteristic values of $\epsilon_j \sim 0.01-0.1$, the fall-back heating rate is comparable to that from $r$-process radioactive heating on timescales of days to weeks (Fig.~\ref{fig:heating}).  

Based on the observed luminosity of the X-ray emission following GRB 130603B, \cite{Kisaka+16} argued that the NIR emission,  attributed to radioactive heating by \cite{Tanvir+13,Berger+13}, was instead powered by X-rays absorbed and re-emitted by the ejecta.  The viability of such a model depends on the assumption that the observed X-ray emission is isotropic, in contrast to the relativistically-beamed emission during the earlier GRB or typically expected for the synchrotron afterglow.

Figure \ref{fig:fallback} shows kilonova light curves, calculated from our toy model assuming the ejecta is heated exclusively by fall-back accretion according to Eq.~(\ref{eq:Lxfb}) for a (temporally constant) jet efficiency of $\epsilon_j = 0.1$, ejecta mass $M = 10^{-2}M_{\odot}$, and velocity $v_0 = 0.1$ c.  We normalize the fall-back rate to a value of $\dot{M}_{\rm fb}(t = 0.1) = 10^{-3}M_{\odot}$ s$^{-1}$ in the case of NS-NS mergers (top panel), and to a value 10 times higher in BH-NS mergers (bottom panel), motivated by the results of \cite{Rosswog07}.  The K-band peaks of 21 and 20 in NS-NS and BH-NS mergers, respectively, are 1\,--\,2 magnitudes higher than in the radioactive heating case (Figs.~\ref{fig:vanilla}, \ref{fig:neutrons}).  Because emission from the central engine must diffuse through the entire ejecta mass, and the outer mass layers contribute no emission, the light curve peak is more pronounced (`mountain shaped') than in the radioactive heating case.  However, the precise shape of the light curve should not be taken too literally, as it is particularly sensitive to our simplified opacity prescription (Eq.~\ref{eq:kappar}).

\subsubsection{Magnetar Remnants}
\label{sec:magnetar}

\begin{figure}[!t]
\includegraphics[width=1.0\textwidth]{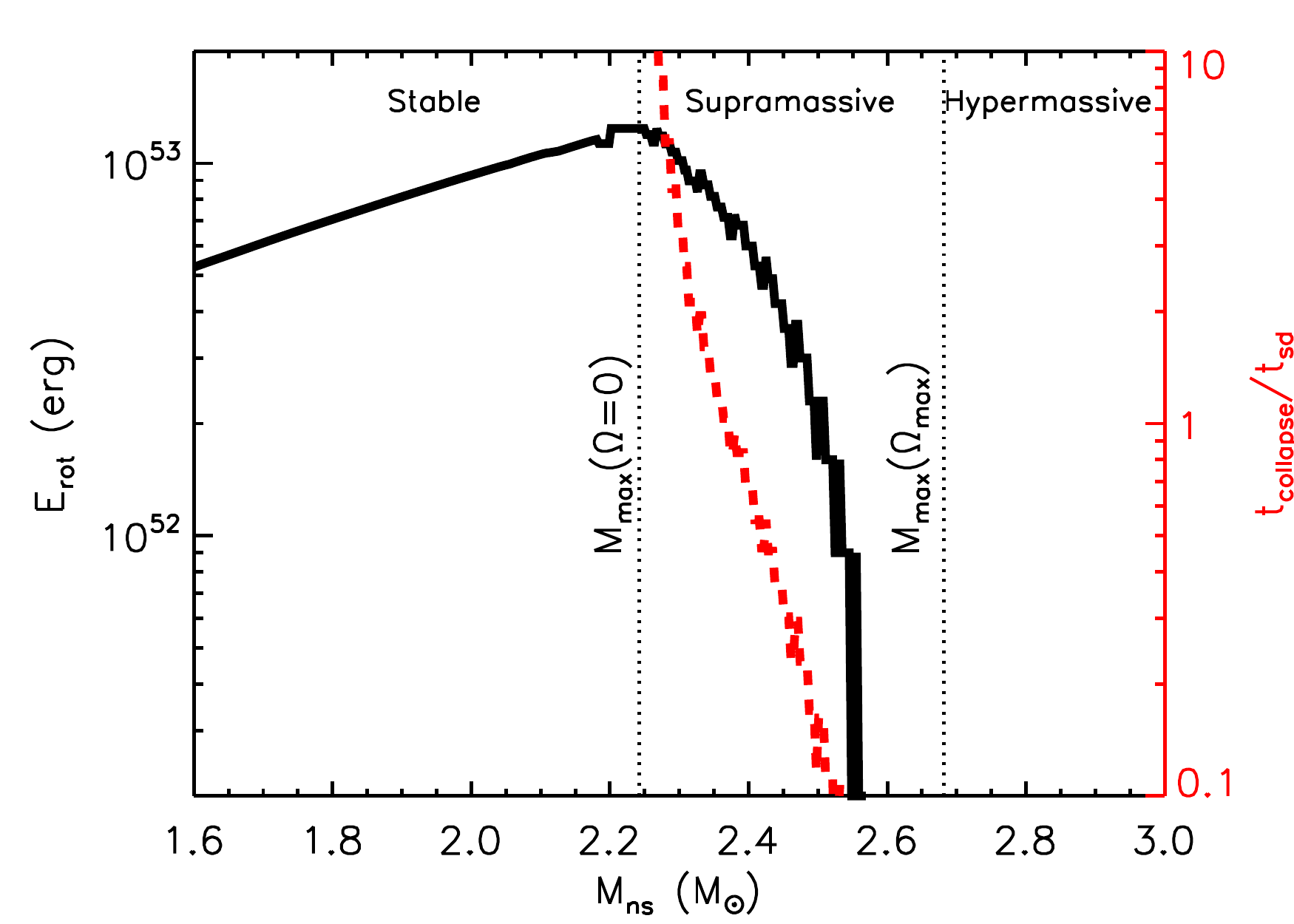}
\caption{Maximum extractable rotational energy from the magnetar remnant of a NS-NS merger as a function of its gravitational mass $M_{\rm ns}$ (black line, left axis).  Below the maximum mass of a non-rotating NS of $M_{\rm max}(\Omega = 0)$, this is just the rotational energy at the mass-shedding limit.  For $M_{\rm ns} \gtrsim M_{\rm max}(\Omega = 0)$, the extractable rotational energy is the difference between the mass-shedding limit at the rotational energy at the point of collapse into a black hole.  Also show with a red dashed line is the time to spin-down via magnetic dipole to the point of collapse, in units of the characteristic spin-down time $t_{\rm sd}$ (Eq.~\ref{eq:tsd}).  The remnant mass of a merger consisting of two NSs of mass $\approx 1.3-1.4M_{\odot}$ is typically $\approx 2.3-2.5M_{\odot}$, after accounting for neutrino losses and mass ejection \citep{Ruffert+97,Belczynski+08,Kaplan+14}.  The structure of the solid-body rotating NS is calculated using the {\tt rns} code \citep{sf95} assuming a parametrized piecewise polytropic EOS with an adiabatic index $\Gamma = 3$ above the break density of $\rho_{1} = 10^{14.7}$ g cm$^{-3}$ at a pressure of $P_{1} = 3.2\times 10^{34}$ dyn cm$^{-2}$ \citep{Margalit+15}.  The chosen EOS results in a 1.4$M_{\odot}$ NS radius of 10.6 km and maximum non-rotating mass of $M_{\rm max}(\Omega = 0) \approx 2.24M_{\odot}$.  This figure is modified from a related figure in \cite{Metzger+15b}.}
\label{fig:Erot}
\end{figure}

As described in Sect.~\ref{sec:ejecta}, the type of compact remnant produced by a NS-NS merger prompt BH formation, hypermassive NS, supramassive NS, or indefinitely stable NS) depends sensitively on the total mass of the binary relative to the maximum mass of a non-rotating NS, $M_{\rm max}(\Omega = 0)$.  The value of $M_{\rm max}(\Omega = 0)$ exceeds about 2$M_{\odot}$ \citep{Demorest+10,Antoniadis+13} but is otherwise unconstrained\footnote{High masses $M_{\rm max}(\Omega = 0) \gtrsim 2.1M_{\odot}$ are also suspected for the black widow binary pulsars; however, the precise mass estimates in these systems are unfortunately subject to large systematic uncertainties \citep{Romani+12}.  } by observations or theory up to the maximum value $\sim 3M_{\odot}$ set by the causality limit on the EOS.  A `typical' merger of two $\approx 1.3-1.4M_{\odot}$ NS results in a remnant gravitational mass of $\approx 2.3-2.5M_{\odot}$ \citep{Ruffert+97,Belczynski+08,Kaplan+14} after accounting for gravitational wave losses and neutrinos ($\approx 7.5\%$ of the mass according to \citep{Timmes+96}), although the precise range of values is uncertain.   If the value of $M_{\rm max}(\Omega = 0)$ is well below this value  $2.1-2.2M_{\odot}$), then most mergers will undergo prompt collapse or form hypermassive NSs with very short lifetimes.  On the other hand, if the value of $M_{\rm max}(\Omega = 0)$ is close to or exceeds $2.3-2.4M_{\odot}$, then a order unity fraction of NS-NS mergers could result in long-lived supramassive or indefinitely stable remnants. 

A massive NS rotating near the mass shedding limit possesses a rotational energy of
\be
E_{\rm rot} = \frac{1}{2}I\Omega^{2} \simeq 1\times 10^{53}\left(\frac{I}{I_{\rm LS}}\right)\left(\frac{M_{\rm ns}}{2.3 M_{\odot}}\right)^{3/2}\left(\frac{P}{\rm 0.7\rm ms}\right)^{-2}\,{\rm erg}
\ee
where $P = 2\pi/\Omega$ is the rotational period and $I$ is the NS moment of inertia, which we have normalized to an approximate value for a relatively wide class of nuclear equations of state $I_{\rm LS} \approx \simeq 1.3\times 10^{45}(M_{\rm ns}/1.4M_{\odot})^{3/2}\mathrm{\ g\ cm}^{2}$, motivated by  Fig.~1 of \citet{Lattimer&Schutz05}.  This energy reservoir is enormous compared to the kinetic energy of the merger ejecta ($\approx 10^{50}\mathrm{\ erg}$) or to the  energy released by radioactive decay.  

If $E_{\rm rot}$ could be extracted in non-GW channels on timescales of hours to years after the merger by electromagnetic torques, this could substantially enhance the EM emission from NS-NS mergers \citep{Gao+13,Metzger&Piro14,Gao+15,Siegel&Ciolfi16a}.  However, for NSs of mass $M_{\rm ns} \gtrsim M_{\rm max}(\Omega =0)$, only a fraction of $E_{\rm rot}$ is available to power EM emission, even in principle.  This is because the loss of angular momentum that accompanies spin-down results in the NS collapsing into a BH before all of its rotational energy $E_{\rm rot}$ is released.\footnote{The remaining rotational energy is stored in the spin of the BH.}  Figure \ref{fig:Erot} shows this \emph{extractable} rotational energy as a function of the remnant NS mass, calculated assuming a NS EOS which supports a maximum non-rotating mass of $M_{\rm max}(\Omega = 0) \approx 2.24M_{\odot}$.  The extractable energy of a stable remnant is $\sim 10^{53}$ erg, but $\Delta E_{\rm rot}$ decreases rapidly with increasing mass for $M_{\rm ns} \gtrsim M_{\rm max}(\Omega =0)$, reaching values of $\lesssim 10^{52}\mathrm{\ erg}$ for remnants with masses only $\gtrsim 10\%$ larger than $M_{\rm max}(\Omega =0)$, corresponding to $M_{\rm ns} \gtrsim 2.4M_{\odot}$ for the example shown in Fig.~\ref{fig:Erot}.

A strong magnetic field provides an agent for extracting rotational energy from the NS remnant via electromagnetic torques.  MHD simulations of NS-NS mergers show that the original magnetic field strengths of the NS are amplified to very large values, similar or exceeding the field strengths of $10^{15}-10^{16}$ G inferred for Galactic `magnetars' \citep{Price&Rosswog06,Zrake&MacFadyen13,Kiuchi+14}.  However, most of this amplification occurs on small spatial scales, and at early times when the NS is still differentially-rotating, resulting in a complex and time-dependent field geometry \citep{Siegel+14}.  However, once the NS comes into solid body rotation, as likely occurs hundreds of milliseconds or less following the merger due to the removal of differential rotation by internal magnetic stresses, one may speculate (and many have!) that the remnant will possess a dipole magnetic field of comparable strength, $B \sim 10^{15}-10^{16}$ G.  

The spin-down luminosity of an aligned dipole\footnote{Unlike vacuum dipole spin-down, the spin-down rate is not zero for an aligned rotator in the force-free case, which is of greatest relevance to the plasma-dense, post-merger environment.} rotator is given by \cite{Spitkovsky06,Philippov+15}
\be
L_{\rm sd}   = \left\{
\begin{array}{lr}
\frac{\mu^{2}\Omega^{4}}{c^{3}} =  7\times 10^{50}\,{\rm erg\,s^{-1}}\left(\frac{I}{I_{\rm LS}}\right)\left(\frac{B}{10^{15}\,{\rm G}}\right)^{2}\left(\frac{P_{\rm 0}}{\rm 0.7\,ms}\right)^{-4}\left(1 + \frac{t}{t_{\rm sd}}\right)^{-2}
, &
t < t_{\rm collapse}\\
0 &
t > t_{\rm collapse} \\
\end{array}
\label{eq:Lsd}
\right. ,
\ee
where $\mu = B R_{\rm ns}^{3}$ is the dipole moment, $R_{\rm ns} = 12\,{\rm km}$ is the NS radius, $B$ is the surface equatorial dipole field, 
\be
t_{\rm sd} = \left.\frac{E_{\rm rot}}{L_{\rm sd}}\right|_{t = 0}\simeq 150\,{\rm s}\left(\frac{I}{I_{\rm LS}}\right)\left(\frac{B}{10^{15}\,{\rm G}}\right)^{-2}\left(\frac{P_{\rm 0}}{\rm 0.7\,ms}\right)^{2}
\label{eq:tsd}
\ee
is the characteristic `spin-down timescale' over which an order unity fraction of the rotational energy is removed, where $P_{0}$ is the initial spin-period and we have assumed a remnant mass of $M = 2.3M_{\odot}$.  The latter is typically close to, or slightly exceeding, the mass-shedding limit of $P = 0.7$ ms.  If the remnant is born with a shorter period, mass shedding or non-axisymmetric instabilities set in which will result in much more rapid loss of angular momentum to GWs \citep{Shibata+00}, until the NS rotates at a rate close to $P_0 \gtrsim 0.7$ ms.

\begin{figure}[!t]
\includegraphics[width=0.5\textwidth]{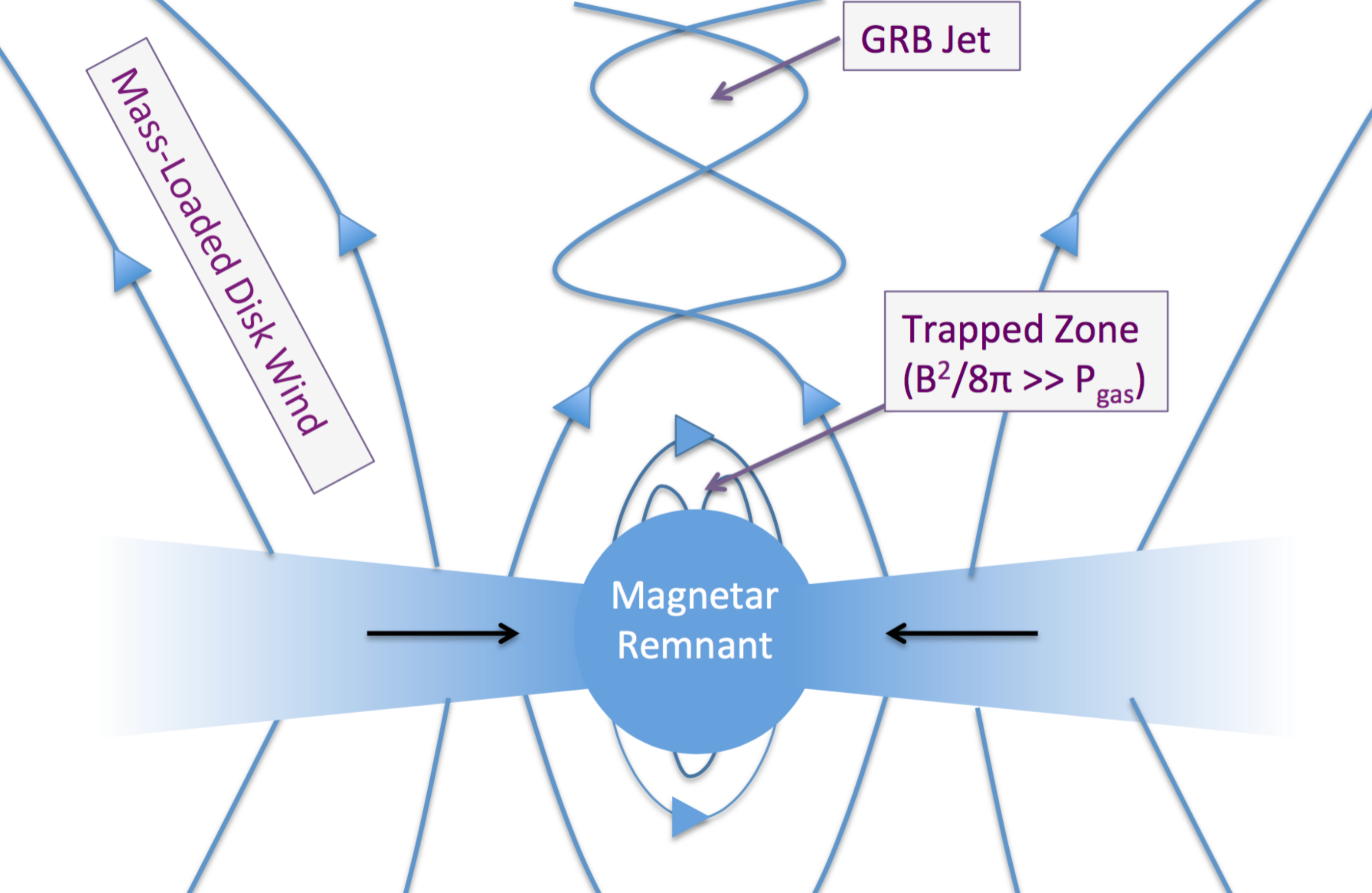}
\includegraphics[width=0.5\textwidth]{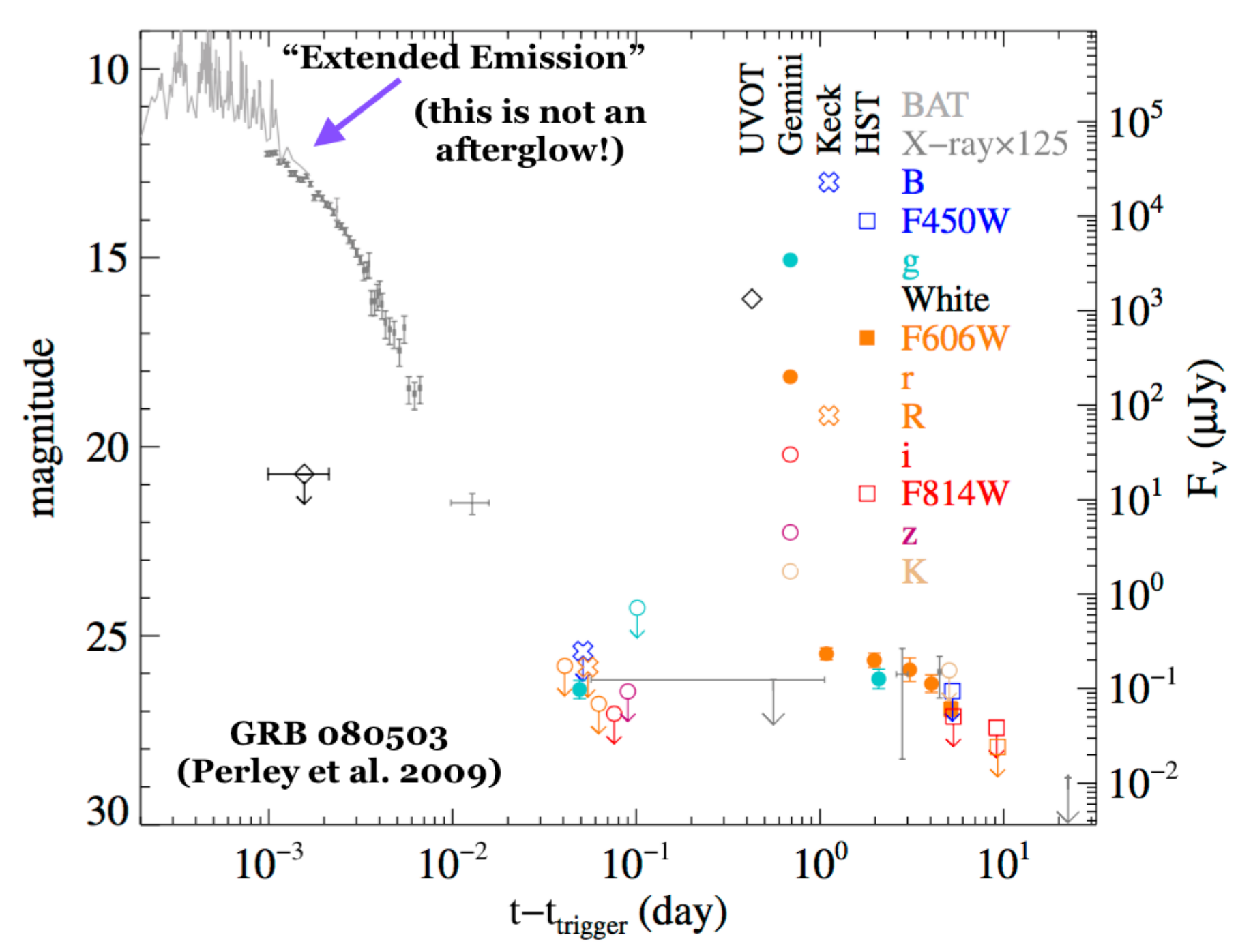}
\caption{LEFT: Schematic illustration of a possible scenario by which accretion onto the magnetar remnant of a NS-NS merger could power an ultra-relativistic short GRB jet.  Strong magnetic fields in the polar region confine the hot atmosphere of the proto-NS \citep{Thompson03}, preventing the formation of a steady neutrino-driven wind in this region.  Open magnetic field lines, which thread the accretion disk or shear boundary layer, carry the Poynting flux powering the GRB jet.  These field lines are relatively devoid of baryonic matter due to the large centrifugal barrier, enabling the outflow to accelerate to attain high asymptotic Lorentz factors.  At larger radii in the disk, outflows will be more heavily mass-loaded and form a potential collimating agent for the jet.  RIGHT: X-ray and optical light curves following the short duration GRB 080503.  Note the sharp drop, by over 6 orders of magnitude, in the X-ray flux, within hours following the burst.  This `steep decay phase', often observed following the prompt emission in long duration GRBs, is probably not related to the `afterglow' (forward or reverse shock created by the jet interacting with the circumburst medium), instead requiring ongoing central engine activity. Image reproduced with permission from \citet{Perley+09}, copyright by AAS.}
\label{fig:magnetarjet}
\end{figure}

The spin-down luminosity in Eq.~(\ref{eq:Lsd}) is assumed to go to zero\footnote{The collapse event itself has been speculated to produce a brief (sub-millisecond) electromagnetic flare \citep{Palenzuela+13} or a fast radio burst \citep{Falcke&Rezzolla13} from the detaching magnetosphere; however, no accretion disk, and hence long-lived transient, is likely to be produced \citep{Margalit+15}.} when the NS collapses to the BH at time $t_{\rm collapse}$.  For a stable remnant, $t_{\rm collapse} \rightarrow \infty$, but for supramassive remnants, the NS will collapse to a black hole after a finite time which can be estimated\footnote{We also assume that dipole spin-down exceeds gravitational wave losses, as is likely valid if the non-axisymmetric components of the interior field are $\lesssim 100$ times weaker than the external dipole field \citep{DallOsso+09}.  We have also neglected angular momentum losses due to $f$-mode instabilities \citep{Doneva+15}.} by equating $\int_0^{t_{\rm collapse}}L_{\rm sd}dt$ to the maximum extractable energy (black line in Fig.~\ref{fig:Erot}).  The value of $t_{\rm collapse}$, in units of the spin-down $t_{\rm sd}$ (Eq.~\ref{eq:tsd}), is showed by a dashed red line in Fig.~\ref{fig:Erot} as a function of the remnant NS mass, assuming that the NS is initially rotating near its mass-shedding limit.  The collapse time decreases rapidly with increasing NS mass above the stable mass.

Energy input from long-lived magnetar remnants of the type just described have been invoked to explain the prompt-like extended X-ray emission observed following short GRBs \citep{Gao&Fan06}\cite{Metzger+08magnetar,Bucciantini+12}.  Several studies have also fit phenomenological magnetar spin-down models to short GRBs with extended emission \citep{gow+13}, X-ray and optical plateaus \citep{Rowlinson+10,Rowlinson+13,gvo+15}, and late-time excess emission \citep{fyx+13,fbm+14}.  All of these models require magnetic fields of strength $B \sim 10^{15}-10^{16}$ G.

Magnetar models for the late-time activity after short GRBs have been criticized, based on the argument that it is not possible to produce a GRB until after BH formation, due to the high expected baryon pollution in the polar region above the surface of a remnant magnetar \citep{Murguia-Berthier+14,Murguia-Berthier+16}.  This led \cite{Ciolfi&Siegel15} and \cite{Rezzolla&Kumar15} to consider a `time reversal' scenario, in which the GRB is produced up to several minutes \emph{after} the merger, once the stable NS collapses to a BH.  This  enables the extended X-ray emission to be attributed to magnetar spin-down energy released prior to the collapse, which takes a finite time to diffuse out of the ejecta.  \cite{Margalit+15}, however, showed that the collapse of a solid body rotating NS is unlikely to leave an accretion disk for realistic NS structures, and so it is unclear how the GRB would be powered following such a long delay.  The disk required to power the GRB jet is also unlikely to originate from the merger itself and survive to late times, since such a disk or shear layers outside the NS remnant surface will be subject to angular momentum transport, if not by the MRI itself, then due to global acoustic waves at the star-disk interface \citep{Philippov+16}.  Even if a sizable disk remained following several minutes of viscous spreading evolution, its radial size would have become much too large to explain the durations of short GRBs.

Personally, I am not ready to concede that GRB formation requires BH formation.  Empirically, accreting NSs in our Galaxy produce ultra-relativistic jets (e.g., Circinus X-1; \citealp{Fender+04}).  In a purely hydrodynamical scenario, the region above the NS remnant will be polluted by a neutrino-driven wind on timescales of seconds following the merger \citep{Dessart+09,Murguia-Berthier+14,Murguia-Berthier+16}.  However, in the presence of the expected strong magnetic field $B \gg 10^{15}$ G, the plasma in this wind region could well be confined by small-scale magnetic flux bundles, which are dynamically dominant over the thermal or ram pressure of the nominal neutrino wind \citep{Thompson03}.  Within such quasi-hydrostatic regions, neutrino heating and cooling can balance with little or no outflow (see Fig.~\ref{fig:magnetarjet} for a schematic illustration).  The open magnetic field lines which carry the Poynting flux of the GRB jet could then originate from the accretion disk or shear interface, and would be relatively devoid of baryonic matter due to the centrifugal barrier.  Although shear due to the differential rotation of the magnetar surface will cause periodic openings of the polar field lines, one may speculate that such transient events might cause variability in the jet properties without substantially enhancing the time-averaged baryon pollution.\footnote{More speculatively, a baryon-free GRB outflow could be produced after the merger NS remnant undergoes a phase transitions from normal matter to deconfined-quark matter, since it is not possible to ablate baryons from the surface of a quark star \citep{Drago+16}.}  It is important to recall that the strongest motivation for magnetar activity after short GRBs is empirical: extended X-ray emission (Fig.~\ref{fig:magnetarjet}, bottom panel) which does not track the expected power-law decay of fall-back accretion rate (Eq.~\ref{eq:mdotfb}). Some have attributed early X-ray emission to the afterglow \citep{Holcomb+14}, but in most cases the observed variability and very rapid decay phase---seen also in long duration GRBs---decays too rapidly to be attributed to the forward or reverse shock.   

\cite{Yu+13} suggested\footnote{In fact, \cite{Kulkarni05} earlier had suggested energy input from a central pulsar as a power source in addition to radioacitivty, though he did not develop the idea in detail.} that magnetic spin-down power, injected by the magnetar behind the merger ejecta over a timescale of days, could enhance the kilonova emission (the termed such events ``merger-novae''; see also \citealp{Gao+15}).  Their model was motivated by similar ideas applied to super-luminous supernovae \citep{Kasen&Bildsten10,Woosley10,Metzger+14} and is similar in spirit to the `fall-back powered' emission described in Sect.~\ref{sec:fallback}.  Although the spin-down luminosity implied by Eq.~(\ref{eq:Lsd}) is substantial on timescales of hours to days, the fraction of this energy which will actually be thermalized by the ejecta, and hence available to power kilonova emission, may be much smaller.

As in the Crab Nebula, pulsar winds inject a relativistic wind of electron/positron pairs.  This wind is generally assumed to undergo shock dissipation or magnetic reconnection near or outside a termination shock, inflating a nascent `magnetar wind nebula' of relativistic particles \citep{Kennel&Coroniti84}.  Given the high energy densities of the post-NS-NS merger environment, these heated pairs cool extremely rapidly via synchrotron and inverse Compton emission inside the nebula \citep{Metzger+14, Siegel&Ciolfi16a, Siegel&Ciolfi16b}, producing broadband radiation from the radio to gamma-rays (again similar to conventional pulsar wind nebulae; e.g., \citealp{Gaensler&Slane06}).  A fraction of this non-thermal radiation, in particular that at UV and soft X-ray frequencies, will be absorbed by the neutral ejecta walls and reprocessed to lower, optical/IR frequencies \citep{Metzger+14}, where the lower opacity allows the energy to escape, powering luminous kilonova-like emission.

On the other hand, this non-thermal nebular radiation may also escape directly from the ejecta without being thermalized through spectral windows in the opacity (Fig.~\ref{fig:opacities}).  This can occur for hard X-ray energies above the bound-free opacity, or for high energy $ \gg$ MeV gamma-rays between the decreasing Klein--Nishina cross section and the $\gamma-\gamma$ opacity (once the nebula compactness $\ell$ has decreased sufficiently; Eq.~\ref{eq:compactness}).  Furthermore, if the ejecta mass is sufficiently low $\lesssim 10^{-2}M_{\odot}$, the ejecta can become ionized, allowing radiation to freely escape also from the far UV and softer X-ray bands where bound-free opacity normally dominates (Eq.~\ref{eq:Lion}).  Such X-ray leakage itself provides a potential isotropic high energy counterpart to the merger \citep{Metzger&Piro14,Siegel&Ciolfi16a, Siegel&Ciolfi16b,Wang+16}.  However, it also reduces the fraction of the magnetar spin-down luminosity which thermalizes and is available to power optical-band radiation, where wide-field telescopes are typically most sensitive.  Magnetar energy could also escape without thermalizing in the form of a relativistic jet \citep{Bucciantini+12} or due to hydrodynamic instabilities (e.g., Rayleigh--Taylor) that occur as the hot bubble of relativistic particles accelerates the relatively modest amount of mass to high energies \citep{Chen+16}.

\begin{figure}[!t]
\includegraphics[width=0.5\textwidth]{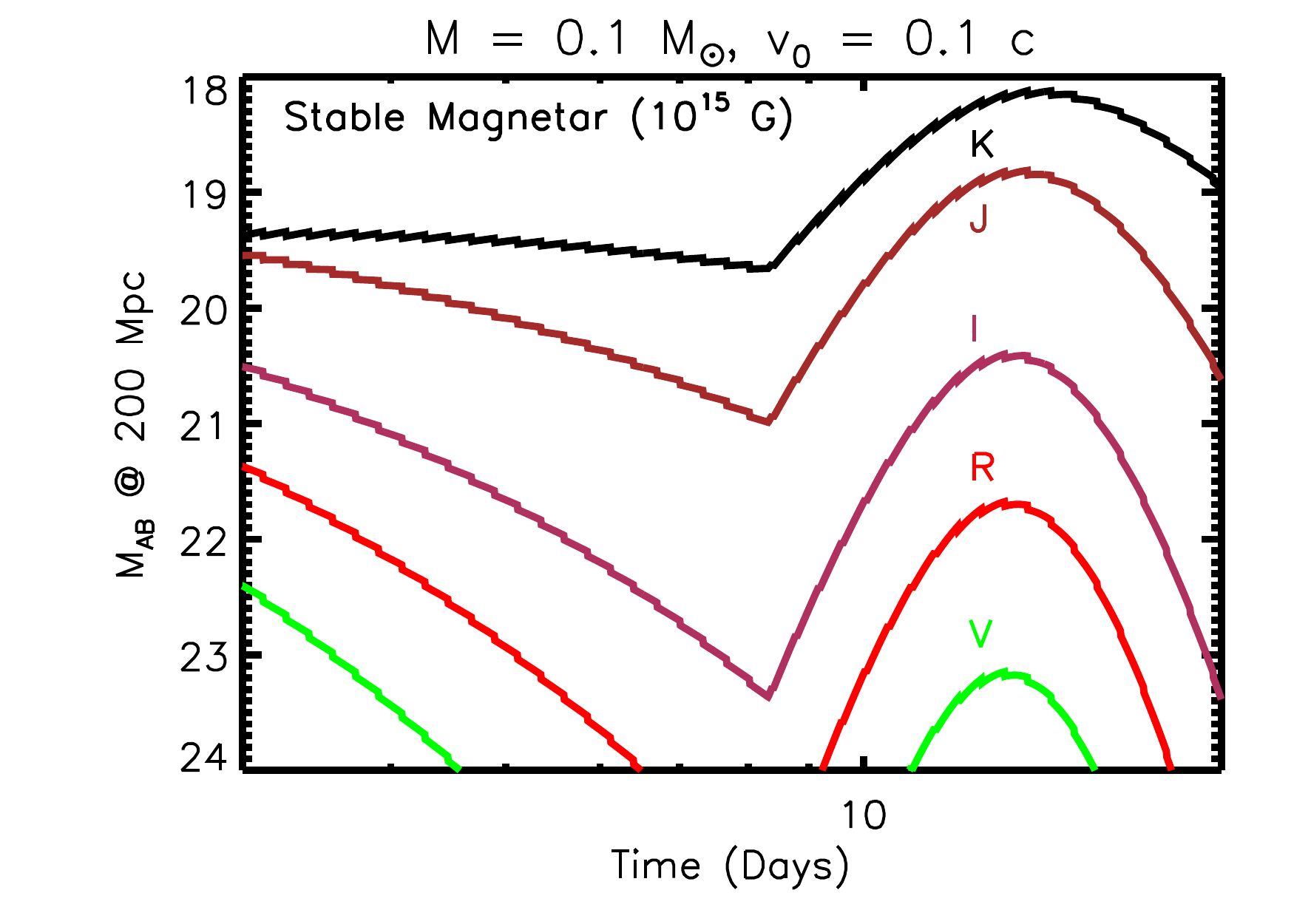}
\includegraphics[width=0.5\textwidth]{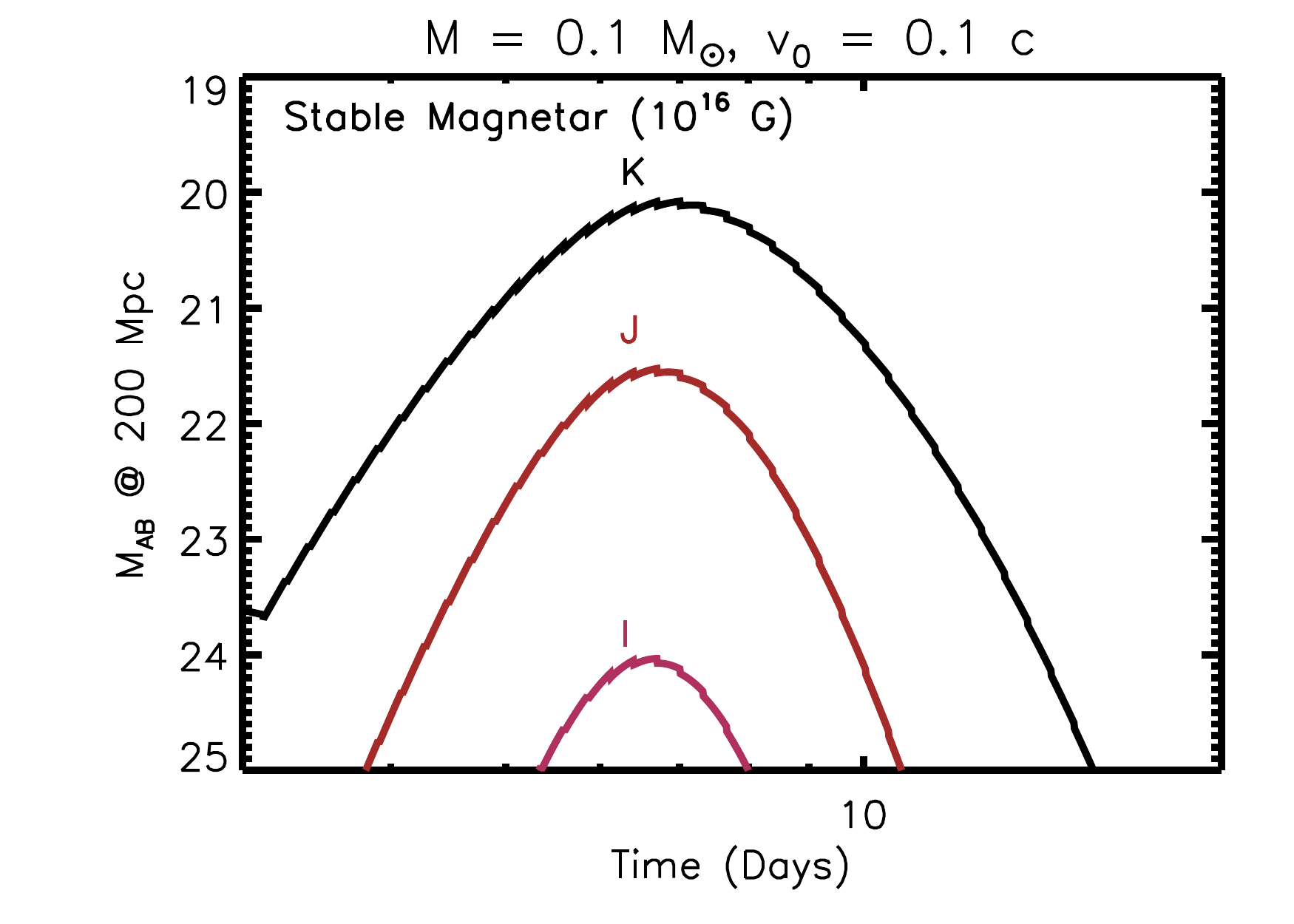}
\caption{Kilonova light curves, boosted by spin-down energy from an indefinitely stable magnetar ($t_{\rm collapse} = \infty$).  We assume an ejecta mass $M = 0.1M_{\odot}$ \citep{Metzger&Fernandez14}, initial magnetar spin period $P_0 = 0.7$ ms,  thermalization efficiency $\epsilon_{\rm th} = 1$ and magnetic dipole field strength of $10^{15}$ G (left panel) or $10^{16}$ G (right panel).}
\label{fig:magnetar15}
\end{figure}

\begin{figure}[!t]
\includegraphics[width=0.5\textwidth]{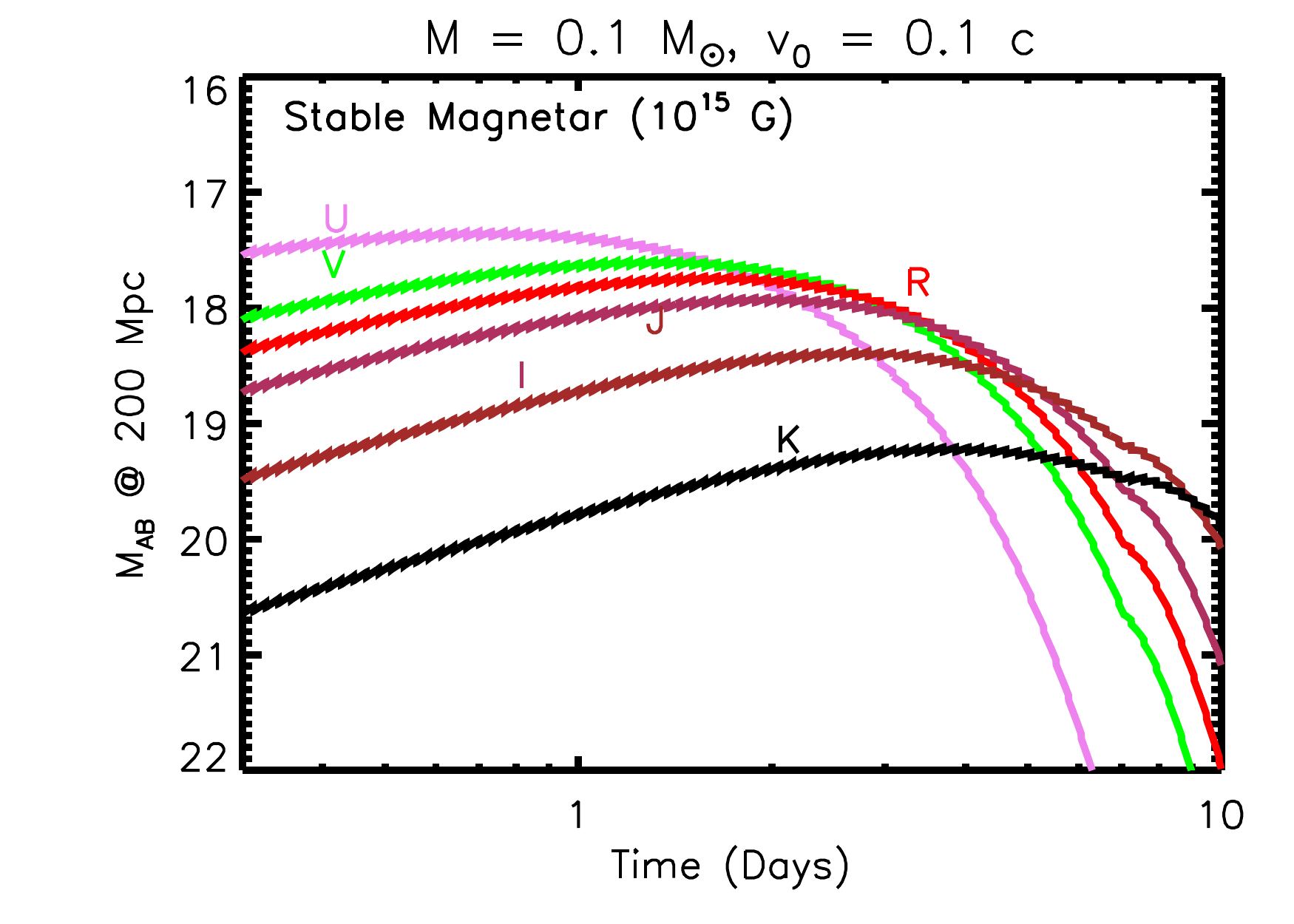}
\includegraphics[width=0.5\textwidth]{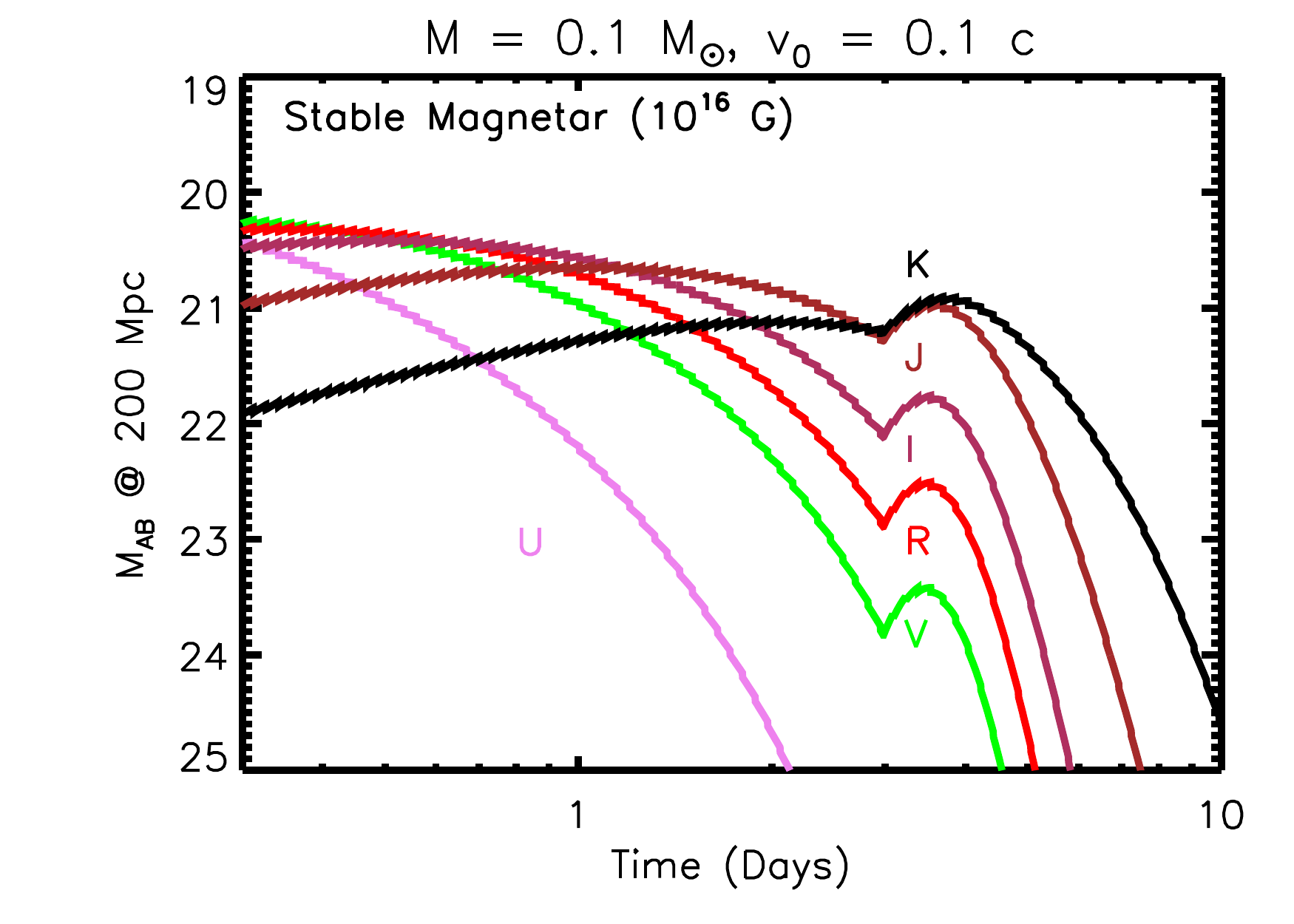}
\caption{Same as Fig.~\ref{fig:magnetar15}, but calculated for ejecta opacities corresponding to lanthanide-free matter.}
\label{fig:magnetar16}
\end{figure}

We can parameterize the magnetar spin-down contribution to the ejecta heating as
\be
\dot{Q}_{\rm sd} = \epsilon_{\rm th}L_{\rm sd},
\label{eq:qdotsd}
\ee  
where, as in the fall-back case (Eq.~\ref{eq:Lxfb}), $\epsilon_{\rm th}$ is the thermalization efficiency.  We expect $\epsilon_{\rm th} \sim 1$ at early times when the ejecta is opaque (unless significant energy escapes in a jet), but the value of $\epsilon_{\rm th}$ will decrease as the optical depth of the expanding ejecta decreases.

\cite{Metzger&Piro14} point out another inefficiency, which, unlike radiation leakage, is most severe at early times.  High energy $\gtrsim$ MeV gamma-rays in the nebula behind the ejecta produce copious electron/positron pairs when the compactness is high.  These pairs in turn are created with enough energy to Compton upscatter additional seed photons to sufficient energies to produce another generation of pairs (and so on...).  For high compactness $\ell \gg 1$, this process repeats multiple times, resulting in a `pair cascade' which acts to transform a significant fraction $Y \sim 0.01-0.1$ of the pulsar spin-down power $L_{\rm sd}$ into the rest mass of electron/positron pairs \citep{Svensson87,Lightman+87}.  Crucially, in order for non-thermal radiation from the central nebula to reach the ejecta and thermalize, \emph{it must diffuse radially through this pair cloud}, during which time it experiences adiabatic PdV losses.  Because at early times the Thomson optical depth of the pair cloud, $\tau_{\rm es}^{\rm n}$, actually exceeds the optical depth through the ejecta itself, this suppresses the fraction of the magnetar spin-down power which is available to thermalize and power the emission.

Following \cite{Metzger&Piro14} and \cite{Kasen+15}, we account in an approximate manner for the effect of the pair cloud by suppressing the observed luminosity according to,
\be
L_{\rm obs} = \frac{L}{1 + (t_{\rm life}/t)}
\label{eq:Lobs}
\ee
where $L$ is the luminosity of the kilonova, calculated as usual from the energy equation (\ref{eq:dEdt}) using the magnetar heat source (Eq.~$\ref{eq:qdotsd}$), and
\be
\frac{t_{\rm life}}{t} = \frac{\tau_{\rm es}^{\rm n}v}{c(1-A)} \approx \frac{0.6}{1-A}\left(\frac{Y}{0.1}\right)^{1/2}\left(\frac{L_{\rm sd}}{10^{45}\, {\rm erg\,s^{-1}}}\right)^{1/2}\left(\frac{v}{0.3\,\rm c}\right)^{1/2}\left(\frac{t}{\rm 1\, day}\right)^{-1/2}
\ee
is the characteristic `lifetime' of a non-thermal photon in the nebula relative to the ejecta expansion timescale, where $A$ is the (frequency-averaged) albedo of the ejecta.  In what follows we assume $A = 0.5$.

For high spin-down power and early times ($t_{\rm life} \gg t$), pair trapping acts to reduce the thermalization efficiency of nebular photons, reducing the effective luminosity of the magnetar-powered kilonova by several orders of magnitude compared to its value were this effect neglected.  The bottom panel of Fig.~\ref{fig:heating} shows the spin-down luminosity $L_{\rm sd}$ for stable magnetars with $P_0 = 0.7$ ms and $B = 10^{15}, 10^{16}$ G.  We also show the spin-down power, `corrected' by the factor $(1 + t_{\rm life}/t)^{-1}$, as in Eq.~(\ref{eq:Lobs}) for $Y = 0.1$.  We emphasize, however, that when one is actually calculating the light curve, the pair suppression (Eq.~\ref{eq:Lobs}) should be applied \emph{after} the luminosity has been calculated using the full spin-down power as the heating source (Eq.~\ref{eq:qdotsd}).  This is because the non-thermal radiation trapped by pairs is also available to do PdV work on the ejecta, accelerating it according to Eq.~(\ref{eq:dvdt}).

\begin{figure}[!t]
\includegraphics[width=0.5\textwidth]{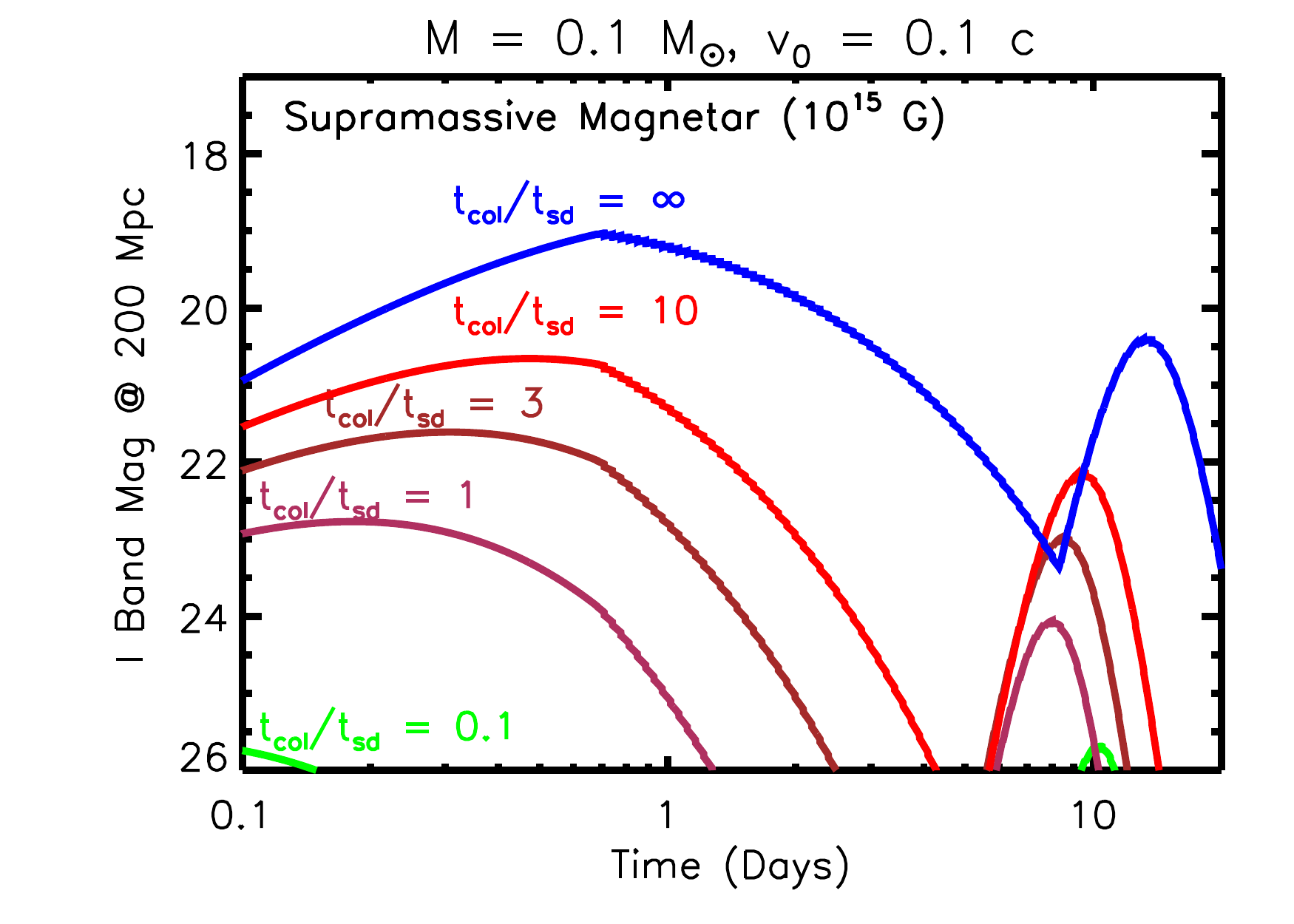}
\includegraphics[width=0.5\textwidth]{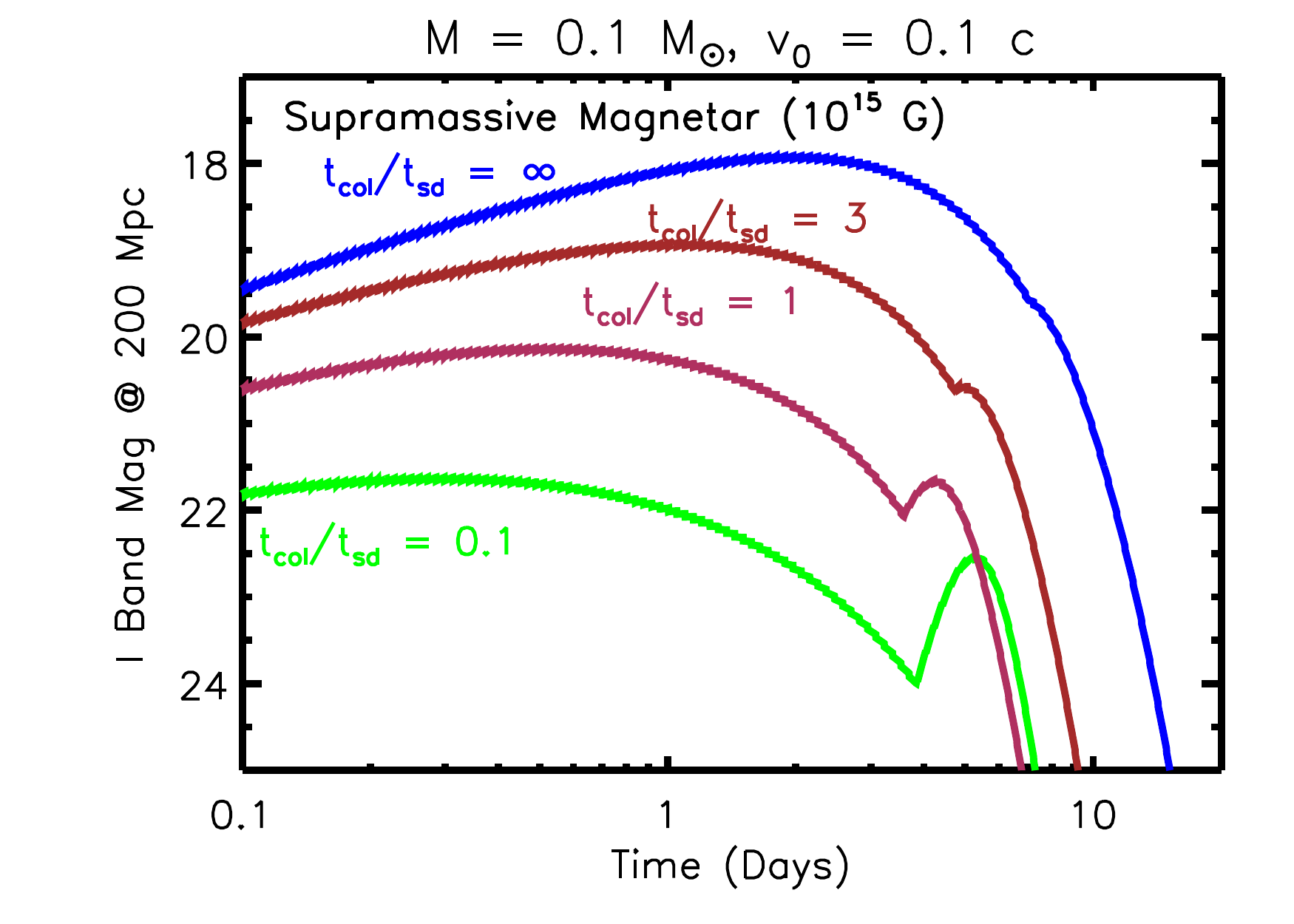}
\caption{I band light curves for a magnetar with a field strength of $10^{15}$ G and a variable lifetime $t_{\rm col}$ in units of the spin-down timescale $t_{\rm sd} \simeq 150$ s (shown as different colored line).  Other ejecta parameters are identical to those in Figs.~\ref{fig:magnetar15}, \ref{fig:magnetar16}.  The calculation shown in the left panel assumes opacities appropriate to Lanthanide matter, while the right panel assumes Lanthanide-free ejecta.}
\label{fig:tcollapse}
\end{figure}

Figure \ref{fig:magnetar15} shows kilonova light curves powered by rotational energy input from a stable magnetar with assumed dipole field strengths of $B = 10^{15}$ G (top panel) and $10^{16}$ G (bottom panel), calculated for an opacity appropriate for lanthanide-bearing ejecta.  Although the emission is still red in color and peaks on a timescale $\sim 1\mbox{\,--\,}2$ weeks (similar to Fig.~\ref{fig:vanilla}), the peak luminosity is greatly enhanced compared to the radioactive case, with peak magnitudes of $K \approx 18\mbox{\,--\,}20$.  Figure \ref{fig:magnetar16} shows a similar calculation assuming an opacity for lanthanide-free ejecta, as might apply to a polar observer.  In this case, the UVR luminosities reach similar magnitudes of $\sim 18\mbox{\,--\,}20$, but on an earlier timescale of $\sim 1$ day, as expected given the lower opacity. 

Figs.~\ref{fig:magnetar15}  and \ref{fig:magnetar16} were calculated assuming an indefinitely stable magnetar, as would likely be produced only for a very low mass binary.  Somewhat more massive binaries will produce supramassive NSs that collapse to BHs, terminating their energy input at a finite time (dashed line in Fig.~\ref{fig:Erot}.  Figure~\ref{fig:tcollapse} compares the I band (800 nm) light curves, calculated for a $B = 10^{15}$ G magnetar for different assumed collapse times (as accounted for by the termination in heating rate after the collapse time $t \gtrsim t_{\rm collapse}$ in Eq.~(\ref{eq:Lsd})).  As expected, the peak luminosity and peak timescale decrease rapidly for earlier collapse times, corresponding to more massive NS remnants and/or softer nuclear EOS.

\section{Implications}
\label{sec:discussion}

\subsection{Kilonova candidates following short GRBs}
\label{sec:candidates}

If short duration GRBs originate from NS-NS or NS-BH mergers, then one way to constrain kilonova models is via optical and NIR follow-up observations of nearby short bursts on timescales of hours to a week.  All else being equal, the closest GRBs provide the most stringent constraints; however, the non-thermal afterglow emission---the strength of which can vary from burst to burst---must also be relatively weak, so that it does not outshine the thermal kilonova.

The NIR excess observed following GRB 130603B \citep{Berger+13,Tanvir+13}, if powered by the radioactive decay of $r$-process nuclei, required a total ejecta mass of lanthanide-bearing matter of $\approx 0.05\mbox{\,--\,}0.1M_{\odot}$ \citep{Barnes+16}.  This is generally too high to be explained by the dynamical ejecta from a NS-NS merger, possibly implicating a BH-NS merger \citep{Hotokezaka+13b,Tanaka+14,Kawaguchi+16}.  However, NS-NS mergers may also produce such high ejecta masses if a large fraction of the remnant accretion disk of mass $\gtrsim 0.1M_{\odot}$ is unbound in disk winds \citep{Metzger&Fernandez14}, as occurs in the case of a very long-lived or stable NS remnant (see also \citealp{Kasen+15}).  Alternatively, the unexpectedly high luminosity of this event could attributed to energy input from a central engine rather than radioactivity \citep{Kisaka+16}, which for fall-back accretion indeed produces the correct luminosity to within an order-of-magnitude (Fig.~\ref{fig:fallback}).  

\cite{Yang+15,Jin+15,Jin+16} found evidence for NIR emission in excess of the expected afterglow following the short GRBs 050709 and 060614, indicative of possible kilonova emission.  In light of the above discussion, it is thus noteworthy that both bursts produced extended emission \citep{Fox+05,Gehrels+06}, indicating the presence of a sustained late-time central engine.  The short GRB 080503 \citep{Perley+09} showed an optical peak on a timescale of $\sim 1$ day (Fig.~\ref{fig:magnetarjet}, bottom panel), potentially consistent with a blue kilonova powered by $r$-process heating \citep{Metzger&Fernandez14,Kasen+15} or a central engine \citep{Metzger&Piro14,Gao+15}.  These possibilities cannot be distinguished because the host galaxy (and hence distance) of GRB080503 was not identified, resulting in its luminosity being unconstrained.  A rebrightening in the X-ray luminosity, coincident with the optical brightening, was also observed following GRB 080503.  This was used to argue against a $r$-process kilonova origin \citep{Perley+09}, but it might also potentially be consistent with non-thermal emission from a central engine \citep{Metzger&Piro14,Gao+15,Siegel&Ciolfi16a,Siegel&Ciolfi16b}.  

Additional upper limits on kilonova counterparts were obtained for GRB050509b \citep{Bloom+06} and GRB150101B \citep{Fong+16b}.  In GRB050509b, the R-band limit $M_R \gtrsim -16$ at $t \sim 1$ day corresponds to $R \gtrsim 20.5$ at 200 Mpc, thus ruling out only stable magnetar models (Figs.~\ref{fig:magnetar15},\ref{fig:magnetar16}).  In GRB150101B, \cite{Fong+16b} placed constraints on a source at 200 Mpc of $R \gtrsim 22$ at $t \approx 11$ days, $J \gtrsim 20$ at $t \approx 2.7$ days, and $H \gtrsim 21$ at $t \approx 15$ days.  These again rule out stable magnetars, but do not constrain less luminous blue or red $r$-process-powered emission.  Even with deep observations of a particularly nearby burst, \cite{Fong+16b} emphasize how challenging it is to constrain non-extreme kilonova models with ground follow-up of GRBs.  This highlights the crucial role played by the Hubble Space Telescope, and in the future by the James Webb Space Telescope (JSWT) and Wide Field Infrared Survey Telescope (WFIRST), in such efforts.  Fortunately, a typical NS-NS merger detected by Advanced LIGO at 200 Mpc (redshift $z = 0.045$) is roughly three times closer (2.5 magnitudes brighter) than the nearest short GRBs.

\begin{figure}[!t]
\includegraphics[width=1.0\textwidth]{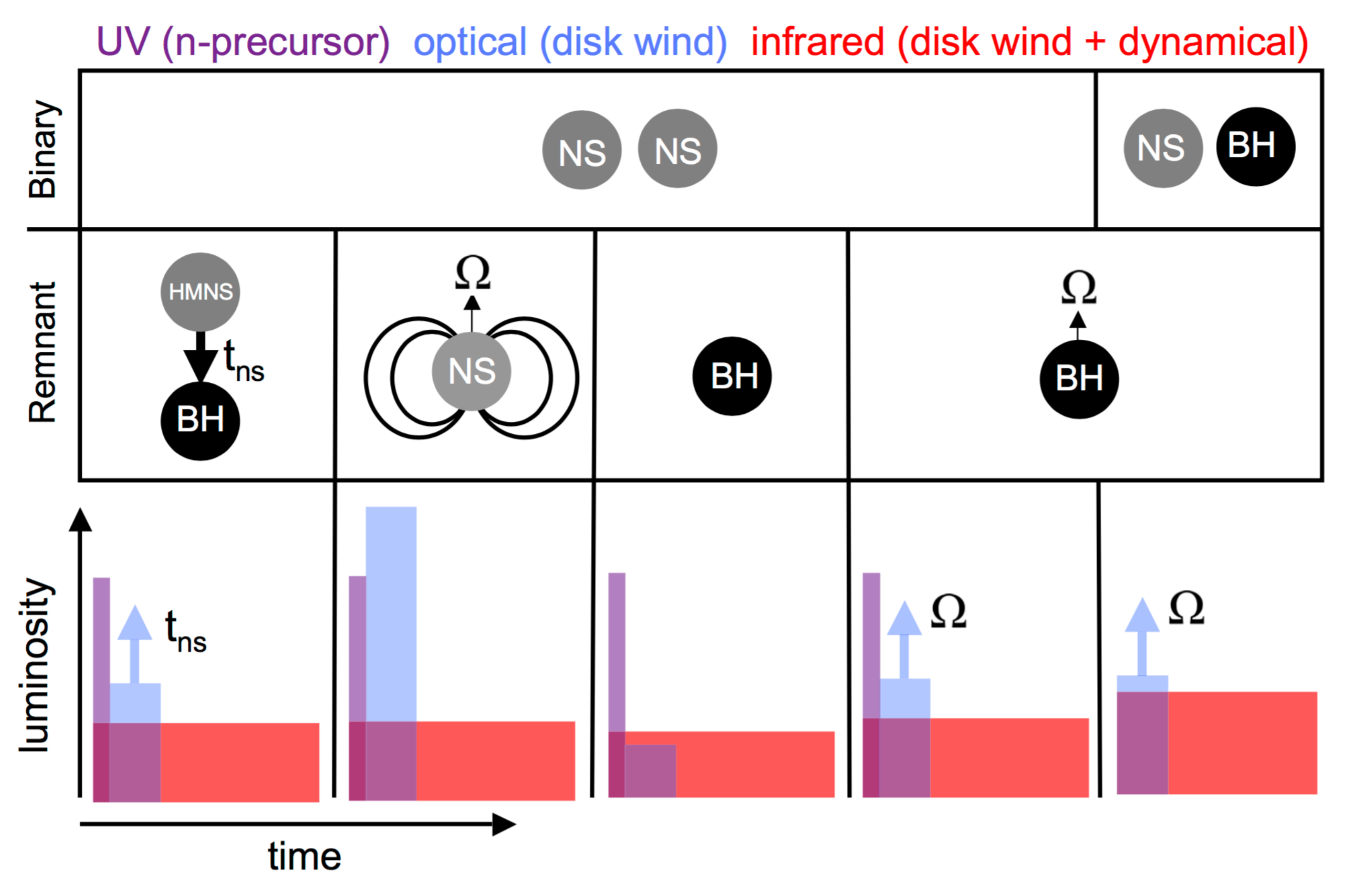}
\caption{Schematic illustration of the mapping between mergers and kilonova light curves. The top panel shows the progenitor system, either an NS-NS or an NS-BH binary, while the middle plane shows the final merger remnant (from left to right: an HMNS that collapses to a BH after time $t_{\rm collapse}$, a spinning magnetized NS, a non-spinning BH and a rapidly spinning BH). The bottom panel illustrates the relative amount of UV/blue emission from an neutron precursor (purple), optical emission from lanthanide-free material (blue) and IR emission from lanthanide containing ejecta (red).  We caution that the case of a NS-NS merger leading to a slowly spinning black hole is very unlikely, given that at a minimum the remnant will acquire the angular momentum of the original binary orbit.  Image reproduced with permission from from \cite{Kasen+15}, copyright by the authors.}
\label{fig:outcomes}
\end{figure}

\subsection{GW follow-up: prospects and strategies}
\label{sec:detection}

Optical follow-up observations were performed following the GW trigger from the BH-BH merger event GW150914, as summarized in \cite{Abbott+16}  (\citealp{SoaresSantos16,Lipunov+16}; see also \citealp{Cowperthwaite+16} and \citealp{Yoshida+17} for follow-up of the second confirmed BH-BH merger, GW151226).  This remarkable campaign covered a range of wavelengths from the near UV to the NIR ($u-z$ bands), on timescales ranging from hours to weeks after the merger.  There were shallow observations covering most of the error region ($R \approx 20.4$ for $i$PTF) and deeper observations ($i \approx 22.5$ by DECam and $i \approx 24.6$ for Subaru Hyper Suprime-Cam [HSC]) covering a narrower portion.  Based on this trial run, we can  assess how effectively future similar searches will be at detecting or constraining the presence of kilonova emission, assuming these events had been NS-NS or BH-NS mergers at the fiducial distance of 200 Mpc, instead of BH-BH mergers.

This review has hopefully made clear that kilonovae may not be homogeneous in their properties, with potentially significant differences in their colors and peak luminosity, depending on the binary inclination and the type of merging system (Fig.~\ref{fig:outcomes} for a summary).  This is especially true in the case of NS-NS mergers, where the signal depends on the remnant lifetime, which in turn is a sensitive function of the mass of the binary relative to the maximum mass of a non-rotating NS.  When BH formation is prompt, the ejecta mass is lower, and radioactivity, and potentially fall-back accretion, provide the only heating sources.  For viewing angles with only the lanthanide-rich ejecta being visible, the $r$-process kilonova is extremely red, peaking at $K \sim 22/24$ at $K$ and $R$ bands, respectively, on a timescale of several days (Fig.~\ref{fig:vanilla}, top panel).  If GW150914 or GW151226 had been a NS-NS binary, only DECam and Subaru HSC \citep{Yoshida+17} would have been able to place meaningful constraints on the ejecta mass in such a scenario, and only provided it could cover enough of the GW sky error region to identify the counterpart.

For a longer-lived NS-NS remnant which survives $\gtrsim 100$ ms \citep{Metzger&Fernandez14,Kasen+15},  or given the presence of high-$Y_e$ dynamical ejecta \citep{Wanajo+14}, the resulting lanthanide-free ejecta may power `blue' kilonova emission peaking at $URI \sim 22$ (Fig.~\ref{fig:vanilla}, bottom panel) on a timescale of several hours to a few days.  Even if this blue emission is not present, for instance due to it being blocked by lanthanide-rich matter, the source may still reach UVR magnitudes of $\sim 22\mbox{\,--\,}23$ on a timescale of hours if the outer layers of the ejecta contain free neutrons (Fig.~\ref{fig:neutrons}).  Although not much brighter in magnitude than the NIR peak at later times, the blue kilonova may be the most promising counterpart for the majority of follow-up telescopes, for which the greatest sensitivity at optical wavelengths.  In such cases, it is essential that follow-up begin within hours to one day following the GW trigger.

In the case of a stable or long-lived magnetar, the peak magnitude can reach up to $\approx 18$ (Figs.~\ref{fig:magnetar15}, \ref{fig:magnetar16}), depending on several uncertain factors: the dipole magnetic field strength of the remnant NS, the thermalization efficiency (Eq.~\ref{eq:qdotsd}), and the NS collapse time (which in turn depends on the binary mass and the nuclear EOS; Fig.~\ref{fig:Erot}).  Shallower follow-up observations, such as those conducted by smaller robotic telescopes, are thus still relevant to kilonova follow-up, insofar as they could be sufficient to detect the kilonova in these extreme cases.  They could also be sufficient to detect the on-axis GRB afterglow for face-on mergers.  Late-time radio observations of short GRBs on timescales of years to decades are now placing stringent constraints on the presence of stable or long-lived supramassive NSs in these systems \citep{Metzger&Bower13, Horesh+16,Fong+16}.  However, we should be cautious that, even if short GRBs result from NS-NS mergers, they may represent a biased subset of the entire merger population, as would be the case if only mergers resulting in prompt BH formation produce successful GRB jets \citep{Belczynski+08,Lawrence+15,Fryer+15, Murguia-Berthier+14,Murguia-Berthier+16}.  Given their extremely bright radio emission, upcoming wide-field radio surveys will also measure or constrain the fraction of mergers which produce stable magnetars, regardless of whether they produce GRBs \citep{Metzger+15c}.

In BH-NS mergers, the quantity of dynamical ejecta can be larger than in NS-NS mergers, by a typical factor of $\sim 10$ (Sect.~\ref{sec:ejecta}); all else being equal, this results in the kilonova peaking one magnitude brighter in BH-NS mergers.  Likewise, the rate of mass fallback in BH-NS mergers can be up to $\sim 10$ times higher than in NS-NS mergers, enhancing potential accretion-powered contributions to the kilonova emission (Fig.~\ref{fig:fallback}, bottom panel).  However the amount of high-$Y_e$ ejecta is potentially less in BH-NS mergers than NS-NS mergers due to the lack of shock-heated ejecta, and for the same reason no neutron precursor is anticipated.  Thus, the kilonova emission from BH-NS mergers is likely dominated by the NIR component, although moderate amounts of high-$Y_e$ matter and blue emission could still be produced by the disk winds \citep{Just+15,Fernandez+15}.  Furthermore, the benefits of the higher ejecta mass on the light curve luminosity may be more than offset by the larger expected source distance, which will typically be $\approx 2\mbox{\,--\,}3$ times greater than the 200 Mpc horizon characteristic of NS-NS mergers for an otherwise equal GW event detection rates.  

Several works have explored the optimal EM follow-up strategies of GW sources, or ways to achieve lower latency GW triggers \citep{Metzger&Berger12,Cowperthwaite&Berger15,Gehrels+16,Ghosh+16,Howell+16,Rana+16}.  Extremely low latency \citep{Cannon+12,Chen&Holz15}, though crucial to searching for a potential low-frequency radio burst \citep{Kaplan+16}, is generally not essential for kilonova follow-up.  One possible exception is the speculative neutron precursor (Sect.~\ref{sec:neutrons}), which peaks hours after the merger.  However, in this case, the greatest advantage is arguably to instead locate the follow-up telescope in North America, producing a better chance of the source being directly overhead of the LIGO detectors where their sensitivity is greatest \citep{Kasliwal&Nissanke13}.  More essential for kilonova follow-up is providing accurate sky error regions on a timescale of several hours to a day.  Of secondary importance is providing information on whether the merging binary is a BH-NS or NS-NS system.  The inclination and total binary mass, though challenging to measure to high precision via the strain data, are the parameters which most strongly affect the kilonova signal in the NS-NS case.  
 
The generally greater sensitivity of telescopes at optical wavelengths, as compared to the infrared, motivates a general strategy by which candidate targets are first identified by wide-field optical telescopes on a timescale of days, and then followed-up spectroscopically or photometrically in the NIR over a longer timescale of $\sim 1$ week.  \cite{Cowperthwaite&Berger15} show that no other known or predicted astrophysical transients are as red and evolve as quickly as kilonovae, thus reducing the number of optical false positives to a manageable level.  Follow-up observations of candidates at wavelengths of a few microns could be accomplished, for instance, by the James Webb Space Telescope \citep{Bartos+16}, WFIRST \citep{Gehrels&Spergel15}, or a dedicated GW follow-up telescope with better target-of-opportunity capabilities.  

The ultimate `smoking gun' confirmation of kilonova emission would be a spectroscopic measurement of absorption lines from $r$-process elements.  Individual lines are are unlikely to be identifiable for the simple reason that their precise wavelengths are usually not known.  However, the very strange spectrum, unlike any SN detected to date, would confirm the exotic composition of the ejecta.  Absorption lines will be Doppler-broadened near peak due to the substantial velocities $v \sim 0.1\mbox{\,--\,}$ c of the ejecta, but the line-widths will become narrower post-maximum as the photosphere recedes to lower velocity coordinates through the ejecta and nebular lines appear.  Spectroscopic IR observations of such dim targets is a compelling science case for future 30-meter telescopes.  For instance, the planned Infrared Imaging Spectrograph (IRIS) on the Thirty Meter Telescope \citep{Skidmore+15} will obtain a signal to noise ratio of 10 per wavelength channel (spectral resolution $R = 4000$) for a $K = 25$ mag point source.

\subsection{The GW/EM Horizon Ahead}

We now perform the dangerous exercise of looking ahead a decade or more, once the Advanced LIGO/Virgo/KAGRA detector network has received further upgrades, and possibly includes a third generation GW detector like the Einstein Telescope.  We can imagine an era when kilonovae and short GRB detections are commonplace in coincidence with high SNR GW chirps from NS-NS/BH-NS mergers out to redshifts $z \gtrsim 0.5$.  Intentionally ignoring issues of measurement precision, we consider (`dream') the types of scientific questions that could be addressed from a large sample of high quality events in the multi-messenger era.  

At present, the geometric structure of GRB jets is poorly constrained, with at most the half-opening angle measured or constrained in a few short GRBs \citep{Fong+15}.  However, by combining the GRB prompt emission and afterglow properties of a sample of mergers with GW-measured inclinations (for which the distance-inclination degeneracy has been broken by a kilonova or afterglow-enabled redshift measurement), we will obtain detailed information on the angular structure of the luminosity and Lorentz factor of GRB jets.  

Perhaps a class of `dirty fireballs' will be discovered, for which a low-frequency non-thermal afterglow is observed but no GRB is produced, despite an optimal face-on orientation.  For NS-NS mergers, we will connect this subclass of GRB-less mergers to those binaries with particularly low masses (as inferred from the GW signal), which produce stable magnetar remnants instead of BHs and thus are incapable of producing high Lorentz factor jets (despite the objects to this simple picture raised in Sect.~\ref{sec:magnetar}).  Nevertheless, this class of sources are found to be extremely bright in radio synchrotron emission, due to the high kinetic energies of their afterglows caused by the injection of magnetar rotational energy.  

For BH-NS mergers, perhaps the GRB-less events will be connected to a misalignment between the BH spin and the binary angular momentum, which causes substantial (GW-measured) precession of the binary orbit and hence of the nominal jet \citep{Stone+13}.  Of course, a comparable or higher fraction of BH-NS mergers produce no GRB for the simple reason that the NS is swallowed whole before being tidally disrupted, something that is readily observed by the lack of a truncation of the GW chirp at higher frequencies \citep{Pannarale+15}.  By measuring this transition point precisely, we obtain constraints on the NS radius, which are consistent with those determined by tidal effects on the waveforms in the case of NS-NS mergers \citep{Read+13,Hinderer+16}.

Another diagnostic of the angular structure of the ejecta from NS-NS/NS-BH mergers comes by comparing the relative strength of the `blue' and `red' components of the kilonova emission for binaries with different inclination angles relative to the line of sight (but otherwise similar masses and mass ratios).  With a large sample of NS-NS mergers, an inverse correlation between the blue/red fraction and the total binary mass is established, which is later confirmed to result from the merger remnant lifetime based on a measurement of the strength of oscillations in the HMNS for a particularly nearby event (\cite{Bauswein+16}, and references therein).     
  
The strength of the blue and red kilonova emission components in a number of events is used to obtain measurements of the yield of both heavy (lanthanide-bearing) and lighter (lanthanide-free) $r$-process elements.  Combining these with the measured GW event rates, we obtain a quantitative assessment of the total contribution of NS mergers to the $r$-process yields of the galaxy (Eq.~\ref{eq:Mr}).  Based on the observed positions of the GW-detected mergers in or around their host galaxies, we learn about the spatial distribution of the pollution events.  Spectroscopic measurements at their positions provide information about the metallicity distribution of the pollution environments.  A few events are identified as occurring in globular clusters, establishing that the delay time of at least some events is extremely long ($\gtrsim 10$ Gyr).  

Low mass NS-NS binaries may produce stable magnetars that substantially enhance the kilonova luminosity, while higher mass binaries, the magnetar lifetime is very short and its impact on the kilonova is negligible.  The transition between these cases is very abrupt, as seen clearly by the extractable rotational energy curve in Fig.~\ref{fig:Erot}.  It is thus conceivable that follow-up observations of NS-NS mergers reveal two distinct classes of events in terms of their isotropic emission, `EM bright' and `EM dim' (although, as discussed above, the `EM bright' events may not necessarily be those accompanied by powerful GRBs).  The ratio of observed bright to dim events would be then be expected to be a decreasing function of the maximum stable mass of the NS, since for a fixed binary population this will control the relative number of mergers that result in short versus long-lived merger remnants.  Synchrotron radio emission from the interaction of the merger ejecta with the circumburst medium will also delineate this dichotomy, by placing constraints on the ejecta kinetic energy \citep{Nakar&Piran11,Metzger&Bower14,Margalit&Piran15}.  `EM bright' events are observed to occur preferentially within the midplanes of their host galaxies, due to the smaller natal kicks expected to accompany the lowest mass NSs if the latter are formed in electron capture SNe or from the accretion-induced collapse of a white dwarf (and hence which are also more likely to produce long-lived remnants once they merge).\footnote{In fact, there is growing evidence for the existence of a class of so-called ``ultra-stripped" supernovae (\citealt{Drout+14,Kleiser&Kasen14,Dessart&Hillier15}).  These originate from stars stripped of most of the outer envelopes by mass loss interaction and tight binaries, and which are expected to give rise to NS-NS binaries with small natal kicks due to the small ejecta mass \citep{Yoon+10,Tauris+15}. }

\section{Final Thoughts}
\label{sec:conclusions}

As a student entering this field in the mid/late 2000s, it was clear to me that the optical transients proposed by \cite{Li&Paczynski98} were not connected in most people's mind with the $r$-process.  \cite{Rosswog05} in principle had all the information needed to calculate the radioactive heating rate of the ejecta based on the earlier \cite{Freiburghaus+99} calculations, and thus to determine the true luminosity scale of these merger transients well before \cite{Metzger+10}.  I make this point not to cast blame, but simply to point out that the concept, now taken for granted, that the radioactive heating rate was something that could actually be calculated with any precision, came as a revelation, at least to a student of the available literature.  

When I first started to inquire about how to go about performing such a calculation, I was informed by a well-respected researcher (who will remain anonymous) that the $r$-process was endothermic, and thus could not release heat, because it involved forming isotopes with masses above the peak of the nuclear binding curve.\footnote{In fact, the $r$-process is exothermic because, although $r$-process nuclei have lower specific binding energy than seed nuclei, this is more than compensated by the energy released as free neutrons are incorporated into heavy nuclei.}   Fortunately, I was introduced to Gabriel Mart{\'{\i}}nez-Pinedo and Almudena Arcones, colleagues who had developed the nuclear reaction network and assembled the microphysics needed to calculate the late-time radioactive heating, and who were  enthusiastic about reviving the relic idea of \cite{Burbidge+56} of an `$r$-process-powered supernova'.   

Given the rapid evolution of this field in recent years, it is natural to question the robustness of current kilonova models.  What would it mean if kilonova emission is ruled out following a NS-NS merger, even to stringent limits?  First, it should be recognized that---unlike, for instance, a GRB afterglow---kilonovae are largely thermal phenomena.  The ejection of neutron-rich matter during a NS-NS merger at about ten percent of the speed of light appears to be a robust consequence of the hydrodynamics of such events, which all modern simulations agree upon.  Likewise, the fact that decompressing nuclear-density matter will synthesize heavy neutron rich isotopes is also robust \citep{Meyer89,Goriely+11}.  The properties of individual nuclei well off of the stable valley are not well understood, although that will improve soon due to measurements with the new Facility for Rare Isotope Beams \citep{Balantekin+14}.  However, the combined radioactive heating rate from a large ensemble of decaying nuclei is largely statistical in nature and hence is also relatively robust, even if individual isotopes are not; furthermore, most of the isotopes which contribute to the heating on the timescale of days to weeks most relevant to the kilonova peak are stable enough that their masses and half-lives are experimentally measured.  Although the thermalization efficiency of the decay products requires careful consideration \citep{Barnes+16}, this probably represents at most a factor of a few uncertainty on the peak luminosity.    

The largest remaining uncertainty in kilonova emission relates to the wavelength-dependent opacity of the ejecta, in particular when it includes lanthanide/actinides isotopes with partially-filled f-shell valence shells \citep{Kasen+13,Tanaka&Hotokezaka13,Fontes+16}.  As discussed in Sect.~\ref{sec:opacity}, the wavelengths and strengths of the enormous number of lines of these elements and ionization states are not experimentally measured and are impossible to calculate from first principles from multi-body quantum mechanics with current computational capabilities.  Furthermore, how to handle radiative transport in cases when the density of strong lines becomes so large that the usual expansion opacity formalism breaks down deserves further consideration and simulation work.  

From the standpoint of numerical advances, all published simulations of the long-term disk evolution to date are hydrodynamical, i.e., they adopt an $\alpha$-viscosity in place of a self-consistent physical mechanism for angular momentum transport, e.g., locally by the magnetorotational instability or by global torques driven by one-arm spiral instabilities in the hypermassive NS \citep{East+16}.  Future work should explore the impact of MHD or non-axisymmetric torques from the central NS on the disk outflows and their compositions.  Another issue which deserves prompt attention is the robustness of the presence of free neutrons in the outermost layers of the ejecta, given their potentially large impact on the very early-time kilonova optical emission (Sect.~\ref{sec:neutrons}).  With an ongoing dedicated effort, as more detections or constraints on kilonovae become possible over the next few years, we will be in an excellent position to use these observations to probe the physics of binary NS mergers, their remnants, and their role as an origin of the $r$-process. 

\begin{acknowledgements}
I want to acknowledge my many collaborators on binary neutron star mergers, who helped shape many of the ideas expressed in this article.  These include, but are not limited to, Almudena Arcones, Andrei Beloborodov, Edo Berger, Josh Bloom, Geoff Bower, Niccolo Bucciantini, Alessandro Drago, Rodrigo Fern\'andez, Wen-Fai Fong, Daniel Kasen, Ben Margalit, Gabriel Mart\'inez-Pinedo, Daniel Perley, Tony Piro, Eliot Quataert, Antonia Rowlinson, Daniel Siegel, Todd Thompson, and Meng-Ru Wu.  I gratefully acknowledge support from the National Science Foundation (AST-1410950, AST-1615084), NASA through the Astrophysics Theory Program (NNX16AB30G) and the Fermi Guest Investigator Program (NNX15AU77G, NNX16AR73G), the Research Corporation for Science Advancement Scialog Program (RCSA 23810), and the Alfred P.~Sloan Foundation. 
This work benefited from discussions at ``r-process nucleosynthesis: connecting FRIB with the cosmos workshop'', supported by the National Science Foundation under Grant No. PHY-1430152 (JINA Center for the Evolution of the Elements).
\end{acknowledgements}


\bibliographystyle{spbasic}      
\bibliography{refs}   

\end{document}